\documentclass[a4paper]{article}
\usepackage[widelayout,hyperrefblack]{./research18}
\usepackage{ulem}
\usepackage{subfigure}
\usepackage{threeparttable}
\usepackage{diagbox}
\usepackage{array}
\usepackage{boldline}
\usepackage{caption}
\usepackage{bbm}
\usepackage{algorithmicx}
\usepackage{algpseudocode}

\allowdisplaybreaks
\usetikzlibrary{datavisualization} 

\usepackage{pgfplots}
\usepgfplotslibrary{units}
\usetikzlibrary{spy,backgrounds}
\usetikzlibrary{decorations.markings}
\usepackage{pgfplotstable}
\usepackage{appendix}

\usepackage{setspace}
\usepackage{amsmath}
\usepackage{subfigure}
\usepackage{mathrsfs}
\usepackage{amsfonts}
\usepackage{amssymb}
\usepackage{psfrag}
\usepackage{cite}
\usepackage{color}
\usepackage{graphicx}
\usepackage{url}
\usepackage{bm}
\usepackage{booktabs}
\usepackage{diagbox}
\usepackage{multirow}
\usepackage{stfloats}
\usepackage{lettrine}
\usepackage{makecell}
\usepackage{listings}
\newcommand{\paperstatus}{Submitted to \textit{IEEE Transactions on Information Theory}}

\providecommand{\keywords}[1]{\textbf{Index terms ---} #1}
\makeatletter
\def\thanks#1{\protected@xdef\@thanks{\@thanks
		\protect\footnotetext{#1}}}
\makeatother

\newcommand{\smatlabaxislabel}[1]{\fontsize{12}{\f@baselineskip}%
\textsf{#1}}
\newcommand{\matlabaxislabel}[1]{\fontsize{14.4}{\f@baselineskip}%
\textsf{#1}}
\newcommand{\mmatlabaxislabel}[1]{\fontsize{17.28}{\f@baselineskip}%
\textsf{#1}}
\newcommand{\bmatlabaxislabel}[1]{\fontsize{20.74}{\f@baselineskip}%
\textsf{#1}}
\newcommand{\bbmatlabaxislabel}[1]{\fontsize{24.88}{\f@baselineskip}%
\textsf{#1}} \makeatother

\allowdisplaybreaks

\newcommand{\llg}[1]{{\color{black}#1}}

\newcommand{\half}%
{\raisebox{2.5pt}{\scriptsize 1}{\small/}\raisebox{-1pt}{\scriptsize 2}}

\newcommand{\beq}{\begin{equation}}
\newcommand{\eeq}{\end{equation}}

\title{On the Capacity of MISO Optical Intensity Channels With Per-Antenna {\color{black}Intensity} Constraints}

\author{Ru-Han Chen, Longguang Li$^\dagger$, Jian Zhang, Wenyi Zhang, and Jing Zhou
\thanks{This work is supported {\color{black}in part} by the National Natural Science Foundation of China under Grant No. 62071489 {\color{black}and 62101192, and in part by} Shanghai Sailing Program under Grant No. 21YF1411000. \textit{(Corresponding author: L. Li{\color{black}.})}}
\thanks{R. H. Chen and J. Zhang are with National Digital Switching System Engineering and Technological Research Center, Zhengzhou, China (e-mail: RyanChen1210@163.com,  zhang\_xinda@126.com).} 
\thanks{L. Li is with Dept. Communication and Electronic Engineering, East China Normal University, Shanghai, China  (e-mail: lgli@cee.ecnu.edu.cn). }
\thanks{W. Zhang {\color{black}is} with the CAS Key Laboratory of Wireless-Optical Communications, University of Science and Technology of China, Hefei, China (e-mail: wenyizha@ustc.edu.cn).}
\thanks{{\color{black}J. Zhou is with Shaoxing University, Shaoxing, China (e-mail: jzhou@usx.edu.cn). He was with the CAS Key Laboratory of Wireless-Optical Communications, University of Science and Technology of China, Hefei, China.}
}}

\begin{document}
\maketitle

\begin{abstract}
This paper investigates the capacity of general multiple-input single-output (MISO)  optical intensity channels {\color{black}(OICs)} under per-antenna peak- and average-{\color{black}intensity} constraints. We first consider the {\color{black}MISO equal-cost constrained OIC (EC-OIC)}, where, apart from the peak-{\color{black}intensity} constraint, {\color{black}average intensities} of inputs are {\it{equal to}} arbitrarily preassigned constants. The second model of our interest is the {\color{black} MISO bounded-cost constrained  OIC (BC-OIC)}, where, as compared with the {\color{black}EC-OIC}, {\color{black}average intensities} of inputs are {\it{no larger than}} arbitrarily \llg{preassigned} constants. 
By {\color{black}leveraging tools from} quantile functions, stop-loss transform and convex ordering of nonnegative random variables, we prove {\color{black}two decomposition theorems} for bounded and nonnegative random variables, {\color{black}based on which} we equivalently transform both the {\color{black}EC-OIC} and the {\color{black}BC-OIC} into {\color{black}respective} single-input single-output channels under a peak-{\color{black}intensity} and several {\color{black}stop-loss mean} constraints. {\color{black}Capacity} lower and upper bounds for both channels are established, based on which the asymptotic capacity at high and low signal-to-noise-ratio are determined.

\end{abstract}

\section{Introduction}
\label{sec:introduction}
Optical wireless communication (OWC) is a promising technique for future wireless communication due to its abundant bandwidth, license-free deployment, and absence of interference with existing radio frequency (RF) systems~\cite{khalighiuysal14_1, karunatilakazafarkalavallyparthiban15_1,
	uysalnouri14_1}. {\color{black}From the consideration of implementation complexity,} a particularly appealing transmission scheme in current OWC systems is intensity-modulation direct-detection (IM/DD). In such a scheme, the transmitter modulates the intensity of optical signals coming from light emitting diodes (LEDs), and the receiver applies photodetectors to measure incoming optical intensities \cite{gagliardikarp76_1,Kahn1997Wireless}. As a consequence of this operation, the transmit{\color{black}ted} signal is real and nonnegative, which is fundamentally different from that of its RF {\color{black}counterpart}. Furthermore, considering safety reasons and hardware limitations, the peak and the average {\color{black}intensities} of transmit{\color{black}ted} signals typically have to be restricted. 
	
The optical intensity channel (OIC) is the most commonly used model for linear IM/DD systems impaired by additive white Gaussian noise (AWGN), just like the AWGN channel in RF communications. During the last few decades, OICs have been extensively studied in the literature from two closely related aspects \cite{ hranilovickschischang04_1,lapidothmoserwigger09_7,faridhranilovic09_1,faridhranilovic10_1,chaabanrezkialouini17_1,
	chaabanrezkialouini18_1,
	chaabanrezkialouini18_2,moserwangwigger18_3,limoserwangwigger20_1,li2018miso,monteiro2014design,letzepishollandcowley08_1,Lowery2020Survey,Armstrong2006EL,Wang2015OE,
	yesilkayabasarmiramirkhanipanayirciuysalhaas17_1,leerandaelbreyerkoonen09_1,Steve2003TIT,zhang18_1,Chen2019FFT}. One aspect is on capacity analysis, which describes fundamental limits of OICs {\color{black}for reliable communication}.  While exact characterization of {\color{black}channel capacity} under a first-order moment or an amplitude constraint is still an open problem \cite{lapidothmoserwigger09_7,faridhranilovic09_1,faridhranilovic10_1,smith71_1,mckellips04_1,
	thangarajkramerbocherer17_1, rassouliclerckx16_1,dytsogoldenbaumshamaipoor17_1}, many capacity bounds and asymptotic results have been derived; see, e.g.,~\cite{chaaban2020capacity} for a comprehensive survey. For single-input single-output (SISO) OICs under a peak- and/or an average-{\color{black}intensity constraint}, capacity bounds and asymptotic results have been established~\cite{lapidothmoserwigger09_7,faridhranilovic09_1,faridhranilovic10_1}. {\color{black}In~\cite{chaabanrezkialouini17_1}, capacity results of parallel SISO OICs with a total average-intensity constraint have been presented, as well as a low-complexity intensity allocation algorithm.} {\color{black}Various capacity slopes at low signal-to-noise-ratios (SNRs) have been characterized in~\cite{chaabanrezkialouini18_1} for general MIMO OICs, while high-SNR asymptotic capacities have been investigated in \cite{chaabanrezkialouini18_2} for MIMO OICs of full column rank and under a total average-intensity constraint, or per-antenna peak-intensity constraints, or both.}
For general multiple-input single-output (MISO) and multiple-input multiple-output (MIMO) OICs with {\color{black}per-antenna peak-intensity constraints and a total average-intensity constraint}, the optimal signaling strategy and capacity results have also been derived {\color{black}in~\cite{moserwangwigger18_3} and \cite{limoserwangwigger20_1}, respectively}. The other aspect is on modulation and coding, related to system designs for optimizing various performance metrics, such as bit-error rates, block-error rates and the {\color{black}minimum} Euclidean distance of constellations.  Modulation schemes including on-off keying,  color-shift keying~\cite{monteiro2014design}, pulse-position modulation \cite{letzepishollandcowley08_1}, different variants of unipolar orthogonal frequency-division multiplexing \cite{Lowery2020Survey,Armstrong2006EL,Wang2015OE,
	yesilkayabasarmiramirkhanipanayirciuysalhaas17_1} have been studied, and code constructions described in \cite{Steve2003TIT,zhang18_1,Chen2019FFT}.

This paper mainly focuses on the capacity analysis of general MISO OICs under per-antenna {\color{black}peak- and average-intensity} constraints.
We first consider the {\color{black}MISO equal-cost constrained OIC (EC-OIC)}, where, apart from the peak constraints,  expectations (i.e., average {\color{black}intensities}) of inputs need to be \llg{arbitrarily} preassigned constants. The prime motivation for considering {\color{black}EC-OICs} is arising requirements of more granular illumination adjustment and chromaticity control in visible light communications~\cite{gancarz2013impact,gong2015power}. Then we extend the results to the {\color{black}MISO bounded-cost constrained OIC (BC-OIC)}, where the average-{\color{black}intensity} constraints are relaxed to be no larger than{\color{black}, but not necessarily equal to,} arbitrarily preassigned constants as compared to the {\color{black}EC-OIC}.
An important motivation for {\color{black}the BC-OIC} is the increasing attention to the distributed design of active antennas~\cite{vu11_1,loyka_17}, where each antenna has its own amplifier with limited power. Furthermore, it will be shown that the per-antenna average-{\color{black}intensity inequality} constraints are {\color{black}in general not} equivalent to {\color{black}equality} constraints for MISO OICs in terms of channel capacity{\color{black},} as those in the SISO OIC {\color{black}with the ratio of the maximum allowed average intensity to the maximum allowed peak intensity being no larger than ${1}/{2}$}. For this reason, theoretical investigation on MISO OICs under such two types of {\color{black}intensity} constraints is meaningful. The results concerning the above channels may also shed light on performance analysis of OWC networks, in which users usually have {\color{black}individual intensity} constraints \cite{zhou19_1,chaaban16_1}.


In previous works, the average{\color{black}-intensity} constraint in MISO or MIMO channels is mostly imposed on the total average {\color{black}intensity} of all transmit antennas. In those scenarios {\color{black}intensity} allocation among all transmit antennas is allowed, and by using the optimal signaling scheme, capacity analysis can be accomplished \cite{moserwangwigger18_3,li2018miso,limoserwangwigger20_1}. Per-antenna {\color{black}intensity} constraints spontaneously make the total average {\color{black}intensity} of inputs limited, and hence, impose a much stronger restriction on the choice of {\color{black}intensity} allocation. Consequently, the optimal signaling scheme in the total average-{\color{black}intensity} constrained scenario may no longer be {\color{black}admissible} to the per-antenna {\color{black}intensity}-constrained scenario. To the best of our knowledge, the existing relevant capacity results concerning {\color{black}multi-transmitter} OICs under per-antenna {\color{black}intensity} constraints are only found in  \cite{chaabanrezkialouini18_2}, which mainly focuses on MIMO OICs at low SNR, and shows that the low-SNR capacity slope can be achieved by a certain maximally-correlated input. 

As we will see, the mathematical techniques involved in analyzing a multi-transmitter system with per-antenna constraints are related to {\color{black}decomposing a random variable into several possibly dependent random variables with constraints on their supports and expectations.} As opposite to the usually studied problem of analyzing behaviors of the sum of random variables, its inverse problem, i.e., decomposition of a random variable, is less studied. {\color{black}Moreover}, most relevant mathematical results are limited to {\color{black}the case where summands are independent random variables}~\cite{linnik1977decomposition}. 
{\color{black}To address the challenges brought by the support constraint, individual first-moment constraints and statistical dependence on the inputs}, we introduce concepts of the stop-loss transform and convex ordering, which have been commonly used in economics and insurance mathematics. {\color{black}Our} main technical contribution is {\color{black}necessary and sufficient} conditions for {\color{black}two types of} decomposability of a bounded and {\color{black}nonnegative} random variable.
 Based on the proposed decomposition theorem{\color{black}s}, we {\color{black}develop} several capacity results {\color{black}on the} MISO {\color{black}OIC} under per-antenna {\color{black}intensity} constraints. The main results of this paper are summarized as follows:
\begin{enumerate}
	\item \textit{{\color{black}Decomposition results}:} We establish {\color{black}necessary and sufficient} conditions for {\color{black}two types of decomposition} of a bounded and nonnegative random variable into a convex summation in a particular form; see {\color{black}Theorems~\ref{main} and~\ref{main2}}. {\color{black}In the constructive proof of Theorem~\ref{main}, a novel approach, called greedy decomposition, of mapping the sum random variable onto its corresponding components is presented; see Sec.~\ref{sec:proof_main}. Furthermore, an iteration-based and a partition-based algorithm for the greedy decomposition are proposed; see Algorithms~\ref{Algorithm:iteration} and~\ref{Algorithm:partition}.}
	\item \textit{Equivalent capacity expression{\color{black}s}:}  {\color{black}Based on the decomposition theorems, we prove that either the MISO EC-OIC and the MISO BC-OIC is equivalent to some SISO channel with an amplitude constraint and several moment constraints (defined by the stop-loss transform); see Corollaries~\ref{corollary:ecc} and \ref{corollary:bcmc}}.

	\item \textit{{\color{black}Capacity bounds}:} For the {\color{black}EC-OIC} and the {\color{black}BC-OIC}, {\color{black}capacity lower} bounds are derived by maximizing the differential entropy of the input for the equivalent SISO channel and applying Entropy Power Inequality (EPI){\color{black}, respectively}; see Theorems~\ref{ecmc:lbd} and~\ref{bcmc:lbd}. {\color{black}A} type of upper bounds is derived by maximizing the variance {\color{black}of the equivalent input and  applying the ``Gaussian maximizes differential entropy'' argument}; see Theorems~\ref{ecmc:max-var-bnd} and~\ref{bcmc:max-var-bnd}. Another type of {\color{black}capacity} upper bounds is obtained by using {\color{black}the} duality-bounding technique; see Theorems~\ref{ubd:ecc} and~\ref{ubd:bcc}. 
	\item \textit{Asymptotic capacity {\color{black}characterizations}:} {\color{black}Low-SNR capacity slopes of both the {\color{black}EC-OIC} and the {\color{black}BC-OIC} are determined via a maximum-variance argument; see Theorem \ref{thm:lowsnr}. High-SNR asymptotic capacities of both channels are {\color{black}determined} by showing that derived capacity upper and lower bounds match asymptotically; see Theorem~\ref{thm:highsnr}.}

\end{enumerate}

{\color{black}The remaining part of this paper} is organized as follows. We end the introduction with {\color{black}Table~\ref{tab:notation} that summarizes notational conventions in the paper}.  {\color{black}Sec.}~\ref{sec:model} introduces {\color{black} the channel models, formulates and simplifies the problem of interest, and provides an outline of the approach used in the paper.} {\color{black}Sec.}~\ref{sec:preliminaries} gives preliminaries in this paper. In Sec.~\ref{sec:decomposability}, {\color{black}conditions for two types of decomposition of a bounded and nonnegative random variable and the resulting equivalent capacity expressions are presented.} {\color{black}Sec.}~\ref{sec:cap-res-ecmc} presents lower bounds, upper bounds, and asymptotic results on capacities. {\color{black}Derivation of results on decomposition is postponed until Sec.~\ref{sec:derivation}.} {\color{black}Sec. \ref{sec: signaling} focuses on the decomposition algorithms and signaling issues.} The paper is concluded in {\color{black}Sec.}~\ref{sec:conclusion}.

\begin{table}[!htbp] 
	\caption{{\color{black}Summary of notations used throughout the paper.}}
	\centering
	\begin{tabular}{|c||l|} 
		\hline
		$\mathbb{P}\{\cdot\}$&Probability of an event\\
		\hline
		$\mathbb{E}\left[\cdot\right]$&Expectation operator\\
		\hline
		$\mathcal{N}(0,\sigma^2)$&Gaussian distribution with zero mean and variance $\sigma^2$\\
		\hline
		$\mathcal{Q}(\cdot)$&Gaussian Q-function\\
		\hline
		$\mathsf{supp}\, X$&Support of a random variable $X$ \\
		\hline
		$\mathscr{P}_{\! X}$&Probability measure of $X$ \\
		\hline
		$Q_{\! X}(\cdot)$&Quantile function of $X$\\ 
		\hline
		$\hh(X)$&Differential entropy of $X$ \\
		\hline
		$\II(X;Y)$&Mutual information between $X$ and $Y$\\
		\hline
		$\bm{1}_{n}$&$n$-dimensional all-one column vector\\
				\hline
		$\left[n\right]$&Index set $\{ 1,2,\cdots,n\}$ for a positive integer $n$\\  
		\hline
		$\mathcal{A}^n$&$n$-fold Cartesian product of the set $\mathcal{A}$ \\
		\hline
		$|\mathcal{A}|$&Cardinality of $\mathcal{A}$\\
		\hline
		$\min \mathcal{A}$, $\max \mathcal{A}$&Minimum and maximum of a totally ordered set $\mathcal{A}$ \\
		\hline
		$(a)_+$& $\max \{a,0\}$ for a real number $a$ \\
		\hline
		\multirow{2}{*}{$\min\{\mathbf{a},\mathbf{b}\}$}&Element-wise minimum $\trans{(\min\{a_1,b_1\}, \cdots, \min\{a_n,b_n\})}$ \\
		~&  for $n$-dimensional column vectors $\mathbf{a}$ and $\mathbf{b}$\\ 
		\hline
		$\mathbf{a} \preccurlyeq \mathbf{b}$& Difference $\mathbf{b}-\mathbf{a}$ is nonnegative \\
		\hline
		\multirow{2}{*}{$\mathbf{a}_{\mathcal{J}}$}& $\trans{(a_{j_1},\cdots,a_{j_{|\mathcal{J}|}})}$, where $j_1,\cdots,j_{{|\mathcal{J}|}}$ are all elements of the set $\mathcal{J}\subseteq \left[n\right]$  \\
		~ &in the ascending order and $\mathbf{a}$ is an $n$-dimensional column vector\\
		\hline
	\end{tabular}\label{tab:notation}
\end{table}

\section{{\color{black}Channel Model, Problem Formulation, and Summary of Results}}\label{sec:model}
\subsection{{\color{black}Channel Model}}\label{subsec:model}
Consider an $\nt \times 1$ ($\nt\ge2$)  MISO {\color{black}OIC}
\begin{equation}\label{eqn:model1}
	Y=\trans{\mathbf{h}}{\color{black}\mathbf{X}}+Z,
\end{equation}
where the $\nt$-dimensional vector {\color{black}$\mathbf{X}=\trans{\left(X_1,\cdots,X_{\nt}\right)}$} denotes the channel input, whose entries denote optical intensities emitted from LED transmitters, and hence, are nonnegative:
$${\color{black}X_k} \ge 0, \quad \forall   k\in {\color{black}\left[\nt\right]};$$ 
where the $\nt$-dimensional vector ${\mathbf{h}}=\trans{\left(h_1,\cdots,h_{\nt}\right)}$ denotes the constant channel gain vector with positive entries, taking into account LED conversion factors, path loss, and the photodetector responsivity, etc., the sum of which is normalized to be {\color{black}unity}, i.e., 
$$\sum_{i=1}^{\nt} h_i=1;$$
where $Z$ denotes the overall effect of the ambient shot noise induced by background radiations and the thermal noise of electronic devices, which is approximated to follow the Gaussian distribution with zero mean and variance $\sigma^2$, i.e.,
$$Z \sim \mathcal{N}(0,\sigma^2),$$
and is independent of {\color{black}$\mathbf{X}$}; and where $Y$ denotes the photocurrent output. It should be noted that the output $Y$ can be negative even if the input ${\color{black}\mathbf{X}}$ is always nonnegative.

In this paper, we assume that full channel knowledge is available at both {\color{black}the transmitter and the receiver}. Considering the limited dynamic range of LED devices and the requirement of illumination quality or energy consumption, the input $\vect{X}$ is subject to one of the following two types of per-antenna {\color{black}intensity} constraints:
\begin{enumerate}
	\item Equal-cost constraints: the channel input is subject to per-antenna peak-{\color{black}intensity constraints} and average-{\color{black}intensity} equality constraints, i.e.,
	\begin{equation}\label{eqn:ecc}
		\begin{aligned}
			&{\color{black}0 \le X_k \le 1},\\
			&\mathbb{E}\left[X_k\right] = \alpha_k, \quad \forall   k\in {\color{black}\left[\nt\right]}, 
		\end{aligned}
	\end{equation}
where the constant $\alpha_k\in \left[0,1\right]$ denotes the ratio of the average {\color{black}intensity} to the maximum allowed peak {\color{black}intensity} of the $k$-th transmitter.
	\item Bounded-cost constraints: the channel input is subject to per-antenna peak-{\color{black}intensity} constraints and average-{\color{black}intensity} inequality constraints, i.e.,
	\begin{equation}\label{eqn:bcc}
		\begin{aligned}
			&{\color{black}0 \le X_k \le 1},\\
			&\mathbb{E}\left[X_k\right]\le \alpha_k,~\quad \forall   k\in {\color{black}\left[\nt\right]},
		\end{aligned}
	\end{equation}
where the constant $\alpha_k\in\left[0,1\right]$ is the ratio of the maximum allowed average {\color{black}intensity} to the maximum allowed peak {\color{black}intensity} of the $k$-th transmitter.
\end{enumerate}
For brevity of expression, {\color{black}we} let ${\bm{\alpha}} = \trans{\left(\alpha_1,\alpha_2,\ldots,\alpha_{\nt}\right)}$. Without loss of generality, we assume the entries of ${\bm{\alpha}}$ are ordered as $\alpha_1\ge \alpha_2 \ge \cdots \ge \alpha_{\nt}$. In the rest of the paper, we call the {\color{black}OIC} under input constraints~(\ref{eqn:ecc}) as the {\color{black}EC-OIC}, and under constraints~(\ref{eqn:bcc}) as the {\color{black}BC-OIC}. 

We rewrite the MISO channel \eqref{eqn:model1} as 
\begin{equation}\label{model:siso}
	Y= S +Z,
\end{equation}
where the equivalent input is
\begin{eqnarray}\label{eqn:model2}
	\begin{aligned}
		S&\triangleq \trans{\mathbf{h}}\mathbf{X}=\sum_{k=1}^{\nt}h_kX_k.\\
	\end{aligned}
\end{eqnarray}
Then $S$ must satisfy
\begin{flalign}\label{eq:supp}
	{\color{black}0 \le S \le 1},
\end{flalign}
and
\begin{flalign}\label{eq:ap1}
	\mathbb{E}\left[S\right]={\color{black}\trans{\mathbf{h}}\bm{\alpha}}
\end{flalign}
for the {\color{black}EC-OIC} or
\begin{flalign}\label{eq:ap2}
	\mathbb{E}\left[S\right]\le {\color{black}\trans{\mathbf{h}}\bm{\alpha}}
\end{flalign}
for the {\color{black}BC-OIC}.

\subsection{{\color{black}Problem Formulation and Simplification}}\label{subsec:problem}
In this paper, we {\color{black}are concerned about capacity results on} the MISO {\color{black}OIC}~(\ref{eqn:model1}) under {\color{black}the} two different per-antenna {\color{black}intensity} constraints (\ref{eqn:ecc}) or (\ref{eqn:bcc}). The single-letter capacity expression for the {\color{black}EC-OIC} is given by
\begin{flalign}
	\mathsf{C}_{\text{E}}\left(\vect{h},\bm{\alpha},\sigma\right) = \sup_{{\color{black}\mathscr{P}_{\mathbf{X}}} \textnormal{ satisfying }  \eqref{eqn:ecc}} \II\left(\mathbf{X};Y\right),
\end{flalign}
{\color{black}and} similarly, for the {\color{black}BC-OIC, by}
\begin{flalign}
	\mathsf{C}_{\text{B}}\left(\vect{h},\bm{\alpha},\sigma\right) = \sup_{{\color{black}\mathscr{P}_{\mathbf{X}}} \textnormal{ satisfying }  \eqref{eqn:bcc}} \II\left(\mathbf{X};Y\right).
\end{flalign}

By flipping the input as $\bm{1}_{\nt}-\mathbf{X}$, we have 
\begin{flalign}\label{eq:symmetry}
		\mathsf{C}_{\textnormal{E }}\left(\vect{h},\bm{\alpha},\sigma\right) = \mathsf{C}_{\textnormal{E  }}\left(\vect{h},   {\bm{1}}_{\nt}-{\bm{\alpha}},\sigma\right) {\color{black},~\forall \, \bm{0} \preccurlyeq \bm{\alpha}  \preccurlyeq \bm{1}_{\nt},}
\end{flalign}
for the {\color{black}EC-OIC}. Note that all coordinates of ${\bm{1}}_{\nt}-\bm{\alpha}$ are no larger than $\frac{1}{2}$ if $\alpha_{\nt} \geq \frac{1}{2}$. 

For the {\color{black}BC-OIC}, by following the proof of~\cite[Proposition $1$]{limoserwangwigger20_1}, we can easily show that 
	\begin{flalign}\label{eq:12}
		\mathsf{C}_{\textnormal{B  }}\left(\vect{h},\bm{\alpha},\sigma\right) = \mathsf{C}_{\textnormal{B}}\left(\vect{h}, \frac{1}{2}{\bm{1}}_{\nt},\sigma\right),~\text{if}~\alpha_{\nt} \geq \frac{1}{2}.
	\end{flalign}

{\color{black}Moreover, the} following proposition shows that we can further restrict our attention to the case where entries in $\bm{\alpha}$ follow a strictly decreasing order, i.e., $\alpha_1>\cdots>\alpha_{\nt}>0$. 
\begin{proposition} \label{lemma:mergeability}
	Given an $\nt\times 1$ {\color{black}EC-OIC} (or {\color{black}BC-OIC}) with a channel gain vector $\mathbf{h}$ and an average-{\color{black}intensity} constraint vector $\bm{\alpha}$, if there exists an index $ i \in {\color{black}\left[\nt-1\right]}$ satisfying ${\alpha}_i={\alpha}_{i+1}$, then
	\begin{flalign}
		\mathsf{C}_{\textnormal{E (or B)}}\left(\mathbf{h},\bm{\alpha},\sigma\right)=\mathsf{C}_{\textnormal{E (or B)}}\left(\mathbf{h}',\bm{\alpha}_{{\color{black}\left[\nt\right]}\setminus \{i\}},\sigma\right),
	\end{flalign}  
	with the $(\nt-1)$-dimensional vectors $$\mathbf{h}' =\trans{ (h_1,h_2,\ldots,h_{i-1},h_i+h_{i+1},h_{i+2},\ldots,h_{\nt})} $$ 
	and 
	$$\bm{\alpha}_{{\color{black}\left[\nt\right]}\setminus \{i\}} =\trans{ (\alpha_1,\alpha_2,\ldots,\alpha_{i-1},\alpha_{i+1},\alpha_{i+2},\ldots,\alpha_{\nt})}.$$
\end{proposition}
\begin{IEEEproof}
	Let $\mathbf{X}$ be an arbitrary feasible input.\footnote{By the convention of optimization theory, we use the term ``feasible'' to mean that the {\color{black}input $\mathbf{X}$ satisfies} the constraints (\ref{eqn:ecc}) or (\ref{eqn:bcc}) for the {\color{black}EC-OIC or the BC-OIC}, respectively.} Note that the equivalent input $S=\trans{\mathbf{h}}\mathbf{X}$ is a function of the channel input $\mathbf{X}$, and $\mathbf{X} \rightarrow S \rightarrow Y$ forms a Markov chain. Then $\II \left(\mathbf{X};Y\right)=\II \left(S;Y\right)$. 
	
	Construct a new input $\widetilde{\mathbf{X}}$ by letting  $\widetilde{X}_i=\widetilde{X}_{i+1}=\frac{h_iX_i+h_{i+1}X_{i+1}}{h_{i}+h_{i+1}}$ and $\widetilde{X}_j=X_j, \, \forall j \in {\color{black}\left[\nt\right]}\backslash\{i,i+1\}$. It is obvious that the constructed input $\widetilde{\mathbf{X}}$ is also feasible, and satisfies $\trans{\mathbf{h}}\widetilde{\mathbf{X}}=\trans{\mathbf{h}}\mathbf{X}$. Hence for any feasible input, we have 
	$$\II \left(\widetilde{\mathbf{X}};\trans{\mathbf{h}}\widetilde{\mathbf{X}}+Z\right)=\II \left(\mathbf{X};\trans{\mathbf{h}}\mathbf{X}+Z\right).$$
	Since $\trans{\mathbf{h}}\widetilde{\mathbf{X}}$ is also a feasible equivalent input (induced by the input $\widetilde{\mathbf{X}}_{{\color{black}\left[\nt\right]}\setminus \{i\}}$) to the $(\nt-1)\times 1$ channel with $\mathbf{h}'$ and $\bm{\alpha}_{{\color{black}\left[\nt\right]}\setminus \{i\}}$,  we have 
	$$\mathsf{C}_{\text{E (or B)}}\left(\mathbf{h},\bm{\alpha},\sigma\right)\le\mathsf{C}_{\text{E (or B)}}\left(\mathbf{h}',\bm{\alpha}_{{\color{black}\left[\nt\right]}\setminus \{i\}},\sigma\right).$$ 
	In the reverse direction, by restricting $X_i=X_{i+1}$, the original $\nt \times 1$ {\color{black}EC-OIC} (or {\color{black}BC-OIC}) is degenerated to {\color{black}an} $(\nt-1)\times 1$ channel with $\mathbf{h}'$ and $\bm{\alpha}_{{\color{black}\left[\nt\right]}\setminus \{i\}}$, which implies 
	$$\mathsf{C}_{\text{E (or B)}}\left(\mathbf{h},\bm{\alpha},\sigma\right)\ge\mathsf{C}_{\text{E (or B)}}\left(\mathbf{h}',\bm{\alpha}_{{\color{black}\left[\nt\right]}\setminus \{i\}},\sigma\right).$$ 
	The proof is completed.
\end{IEEEproof}
\vspace{0.2cm}

	Proposition~\ref{lemma:mergeability} shows the mergeability of transmitters with identical normalized per-antenna {\color{black}intensity} constraints for both the {\color{black}EC-OIC and the BC-OIC}, and spatial repetition across those transmitters is optimal in the sense of channel capacity. Specifically, if the average {\color{black}intensities} of all transmitters are constrained by an identical ratio, i.e., $\bm{\alpha}=\alpha\bm{1}$, simply sending identical signals and treating the considered MISO OIC as a SISO one {\color{black}do not} induce any loss. {\color{black}For example, according to~\eqref{eq:12} the optimal {\color{black}intensity} allocation for the {\color{black}BC-OIC} with $\alpha_{\nt}\ge \frac{1}{2}$ is exactly $\frac{1}{2}\bm{1}_{\nt}$, which implies that the channel can be regarded as a peak-limited SISO OIC with {\color{black}an inactive average intensity} constraint. }  	

{\color{black}In light of Proposition 1, in the rest of the paper, we assume $\alpha_1>\cdots>\alpha_{\nt}>0$ and $\alpha_{\nt} \leq \frac{1}{2}$  without loss of generality.}
%

\begin{remark} {\color{black}In terms of capacity, the above model also applies to channels with unnormalized parameters.} For a general MISO {\color{black}OIC} with per-antenna {\color{black}maximum allowed peak intensity} $ {\boldsymbol{\amp} } = \trans{(\amp_1,\amp_2,\ldots,\amp_{\nt})}$, average {\color{black}intensity} (or maximum allowed average {\color{black}intensity}) ${ \boldsymbol{\const{P}}}  = \trans{ (\alpha_1\amp_1,\alpha_2\amp_2,\ldots,\alpha_{\nt}\amp_{\nt})}$, unnormalized channel gain vector ${\widetilde{\mathbf{h}}} =\trans{(\tilde{h}_1,\tilde{h}_2,\ldots,\tilde{h}_{\nt})} $, and the AWGN with a standard deviation $\tilde{\sigma}$, its capacity can be easily shown to be
	$  \mathsf{C}_{\textnormal{E (or B)}}\left(\vect{h},\bm{\alpha},\sigma\right)$ via parameter normalization, where $\sigma=\tilde{\sigma}/\left(\sum_{i=1}^{\nt}\tilde{h}_i \amp_i\right)$ and $\vect{h}$ is a normalized vector with entries $h_k = \tilde{h}_k \amp_k/\left(\sum_{i=1}^{\nt}\tilde{h}_i \amp_i\right), \, \forall k\in {\color{black}[\nt]}$.
\end{remark}
{\color{black}
\subsection{Outline of Approach}\label{subsec:outline}
Herein, without introducing any new terminology, we provide an outline of the approach in the paper, which will help the reader grasp our basic idea.

\begin{itemize}
	\item \textit{Stage 1:} We have already shown that it is sufficient to consider the capacity problem in the case where $\alpha_1 > \alpha_2 > \cdots > \alpha_{\nt}$ and $\alpha_{\nt}\le \frac{1}{2}$; see Sec.~\ref{subsec:problem}.
	\item \textit{Stage 2:} Then we turn to the problem of for what distributions can a random variable $S$, satisfying $0 \le S \le 1$ and $\mathbb{E}[S]=\trans{\mathbf{h}}\bm{\alpha}$, be decomposed as $S=\sum_{k=1}^{\nt}h_k X_k$ with $0 \le X_k \le 1$ and $\mathbb{E}[X_k]=\alpha_k$. This problem is tackled in two steps.
	\begin{itemize}
		\item \textit{Necessity:} If $S$ can be decomposed in the above form, then $ \mathbb{E}\left[\left(S- \left(\sum_{k\in \mathcal{J}}h_k \right) \right)_+\right] \le \sum_{k\in\mathcal{J}^{\rm c}} h_k\alpha_k$ for all $2^{\nt}$ subsets $\mathcal{J}\subseteq [\nt]$. 
		\item \textit{Sufficiency:} If the above expectation condition is satisfied for $\nt-1$ subsets $\mathcal{J}=[1]$, $[2]$, $\cdots$, and $[\nt-1]$, then $S$ indeed has such a decomposition. For this, we give a constructive proof, where a procedure for generating the desired decomposition is presented.
	\end{itemize} 
\item \textit{Stage 3:}  Next, we answer the similar problem of under which conditions the random variable $S$, satisfying $0 \le S \le 1 $ and $\mathbb{E}[S]\le\trans{\mathbf{h}}\bm{\alpha}$, can be decomposed as $S=\sum_{k=1}^{\nt}h_k X_k$ with $0 \le X_k \le 1$ and $\mathbb{E}[X_k]\le \alpha_k$. It is shown that if sufficient conditions in Stage 2 are fulfilled, there exists such type of decomposition for $S$ as well. 
\item \textit{Stage 4:} The above results on decomposition lead to alternative capacity expressions for the MISO EC-OIC and the MISO BC-OIC. These expressions suggest that either channel is equivalent to some SISO OIC with certain corresponding constraints. Therefore, the subsequent capacity analysis can be significantly simplified.
\item \textit{Stage 5:} By relaxing some of the input constraints to be satisfied and using existing information-theoretic techniques, several capacity upper and lower bounds are derived, and then asymptotic capacities are obtained by showing tightness of some bounds.
\end{itemize}

}
\section{Preliminaries}
\label{sec:preliminaries}
This section is devoted to presenting some definitions and propositions that we will use.
\subsection{Distribution and Quantile Functions}
Unless otherwise stated, the supports of random variables under our consideration are subsets of the interval $[0,1]$.  

The cumulative distribution function $F_{\!X}(x)$ of a random variable $X$ with $\mathsf{supp} \, X \subseteq [0,1]$ is defined as
\begin{equation}
	F_{\!X}(x)=\mathbb{P}\left\{X\le x\right\},
\end{equation}
which implies that $F_{ \!X}(x)$ is non-decreasing and right-continuous, and has at most countable discontinuous points on the interval $x\in [0,1]$.

The generalized inverse function of $F_{\!X}(x)$, namely, \textit{quantile function}, is defined
as
\begin{flalign}\label{def:quantile}
	Q_{\! X}(p) \triangleq \inf \left\{x\in [0,1]:\, p\le F_{\! X} (x) \right\},
\end{flalign}
for $0 < p \le 1$,\footnote{{\color{black}Since the random variable $X$ is bounded from above, we define $p$ on $\left( 0,1\right]$ instead of the conventional open interval $\left( 0,1\right)$.}} which correspondingly is non-decreasing and left-continuous. The following well-known property of quantile functions can be immediately derived by the definition (\ref{def:quantile}).

\begin{proposition}[{\color{black}{\cite[p.~304]{van2000asymptotic}}}] \label{prop:galois}
	The quantile function {\color{black}$Q_{\! X}(p)$ satisfies} the Galois inequality:
	\begin{eqnarray}\label{eq:Galois}
		F_{\! X}(x)\ge p~\textnormal{if and only if}~  x\ge Q_{\! X}(p),
	\end{eqnarray}
		for any $p\in(0,1]$.
\end{proposition}

\begin{IEEEproof}
{\color{black}Omitted.	
	}
\end{IEEEproof}
\vspace{0.2cm}
A consequence of the Galois inequality is the following proposition.
\begin{proposition}\label{thm:quantile}
	Let $U$ be a random variable uniformly distributed on the interval $\left(0,1\right]$. For any random variable $X$, we have
	\begin{flalign}\label{eq:17}
		X\overset{\text{d}}= Q_{\! X}(U),
	\end{flalign}
	and furthermore,
	\begin{flalign}\label{eq:18}
		\mathbb{E}[X]=\int_0^1 Q_{\!X}(p) \dd p,
	\end{flalign}
	where $\overset{\text{d}}=$ stands for ``equality in {\color{black}cumulative} distribution''. Conversely, if a random variable $X\overset{\text{d}}= f(U)$, where the function $f(u)$ is nondecreasing, left-continuous {\color{black}and satisfies $0  \le f(u) \le 1$} for $u\in (0,1]$, then 
	\begin{flalign}
		Q_{\! X}(p)=f(p),~p\in\left(0,1\right].
	\end{flalign}
\end{proposition}

\begin{IEEEproof}
{\color{black}Note that~\eqref{eq:17} is an immediate consequence of the Galois inequality~\eqref{eq:Galois} and~\eqref{eq:18} follows~\eqref{eq:17}. If $X\overset{\text{d}}= f(U)$, we have
\begin{flalign}
	F_{\! X}(x)&=\mathbb{P}\left\{f(U)\le x\right\} \nonumber \\
	&=\sup \left\{p\in (0,1]:\,f(p)\le x \right\} \label{eq:20},
\end{flalign}
where~\eqref{eq:20} follows from the fact that $f(p)$ is nondecreasing and left-continuous. It is easy to show that the function $f(p)$ satisfies the Galois inequality~\eqref{eq:Galois} as well. Since $F_{\! X} (x)$ is nondecreasing and right-continuous, we have $f(p)=\inf \left\{x\in [0,1]:\, p\le F_{\! X} (x) \right\}=Q_{\! X}(p)$.	
}
\end{IEEEproof}

%

%

\subsection{Stop-Loss Transform}
Since a monotonic function on a closed interval must be Riemann integrable, we can define the stop-loss transform (SLT) of a random variable as follows.

\begin{definition}[\cite{Muller1996IME}]\label{Def1}
	The stop-loss transform of a random variable $X$ with $\mathsf{supp}\,X\subseteq [0,1]$ is 
	\begin{flalign}
		\pi_{\scriptscriptstyle \! X} (t)=\int_t^1 \left( 1-F_{ \!X}(x) \right) \dd x
	\end{flalign}
	for $t\in\left[0,1\right]$.
\end{definition}

The SLT is a well-known concept in actuarial science for ordering risks~\cite{goovaerts1990effective}.
{\color{black}Several} properties of SLT are listed as follows without proof; see, e.g.,~\cite{Muller1996IME} for more details.
\begin{enumerate}
	\item [(P1)]  $\pi_{\scriptscriptstyle \! X} (t)$ is nonnegative and nonincreasing;
	\item [(P2)] $\pi_{\scriptscriptstyle \! X} (t)$ is convex;
	\item [(P3)] $0\le \pi_{\scriptscriptstyle \! X} (t) \le 1-t$ and thus $\pi_{\scriptscriptstyle \! X} (1)=0$;
	\item [(P4)] $\pi_{\scriptscriptstyle \! X} (t)=\mathbb{E}\left[ (X-t)_+ \right]$ and thus $\pi_{\scriptscriptstyle \! X} (0)=\mathbb{E}\left[X\right]$.
\end{enumerate}

%
%

Here we present an alternative geometrical interpretation of the SLT, {\color{black}revealing a connection between the SLT and} the quantile $Q_{\!X}(p)$.


\begin{proposition}\label{prop:area}
	For a random variable $X$ with $\mathsf{supp}\, X \subseteq [0,1]$, let the region $\mathcal{R}^{  \scriptscriptstyle \! X}_t \subseteq \mathbb{R}^2$ be
	\begin{flalign}
		\mathcal{R}^{  \scriptscriptstyle \! X}_t= \left\{ (p,y)\in \mathbb{R}^2: t \le y < Q_{\!X}(p),~p\in(0,1] \right\},~t\in[0,1].
	\end{flalign}
	Then the area of $\mathcal{R}^{  \scriptscriptstyle \! X}_t$ satisfies
	\begin{flalign}
		\textnormal{Area}~ \mathcal{R}^{  \scriptscriptstyle \! X}_t
		&=\pi_{\scriptscriptstyle \! X} (t),~t\in\left[0,1\right].
	\end{flalign}	
\end{proposition}
\begin{IEEEproof}
	The contrapositive of Proposition~\ref{prop:galois} reveals that 
	\begin{flalign}
		\mathcal{R}^{  \scriptscriptstyle \! X}_t= \left\{ (p,y)\in \mathbb{R}^2: y\ge t, ~0 \le 1-p< 1- F_{\! X}(y) \right\}.
	\end{flalign}
	Then
	\begin{flalign}
		&\textnormal{Area}~ \mathcal{R}^{  \scriptscriptstyle \! X}_t \nonumber \\
		=& \textnormal{Area}~ \left\{ (y,\bar{p})\in \mathbb{R}^2: y\ge t, ~0 \le \bar{p}< 1- F_{\! X}(y) \right\} \nonumber \\
		=&\int_{t}^{1}
		\left( 1- F_{\! X}(y)\right)
		\dd y \nonumber \\
		=&\pi_{\scriptscriptstyle \! X} (t)
	\end{flalign}
	for $t\in\left[0,1\right]$.
\end{IEEEproof}

\subsection{Convex Ordering}
The convex order is a classical stochastic order that compares two random variables with an equal mean and commonly known in economics and insurance mathematics~\cite{shakedshanthikumar94_1,Muller1996IME}.
\begin{definition}[\cite{Muller1996IME}]
	Given two nonnegative random variables $X$ and $Y$ with finite mean, $X$ is said to precede $Y$ in the convex order sense, denoted as $X\le_{\text{cx}}Y$, if 
	either one of the following two equivalent propositions holds:
	\begin{itemize}
		\item $
		\mathbb{E}\left[f(X)\right]\le \mathbb{E}\left[f(Y)\right]
		$ for all convex functions $f$ such that $\mathbb{E}\left[f(X)\right]$ and $\mathbb{E}\left[f(Y)\right]$ exist;
		\item $
		\pi_{\scriptscriptstyle \! X}(t)\le \pi_{\scriptscriptstyle \! Y}(t)~\text{for all } t\ge 0,~\text{and}~\pi_{\scriptscriptstyle \! X}(0)=\pi_{\scriptscriptstyle \! Y}(0)
		$.
	\end{itemize}
\end{definition}

{\color{black}Due to $\pi_{\scriptscriptstyle \! X}(0)=\mathbb{E}\left[X\right]$, the relation $X\le_{\text{cx}}Y$ leads to $\mathbb{E}\left[X\right]=\mathbb{E}\left[Y\right]$.} Furthermore, by letting $f(x)=(x-\mathbb{E}\left[X\right])^2$, we attain the following corollary, which reveals that the convex order implies the variance order.
\begin{corollary}\label{var_order}
	If $X\le_{\text{cx}}Y$, then the following variance order holds
	\begin{flalign}
		\mathbb{E}\left[ \left(X-\mathbb{E}\left[X\right]\right)^2\right]\le  \mathbb{E}\left[ \left(Y-\mathbb{E}\left[Y\right]\right)^2\right] .
	\end{flalign} 
\end{corollary}
\begin{IEEEproof}
	Omitted.
\end{IEEEproof}

\subsection{Comonotonic Distribution}
Research on comonotonicity arises naturally in risk analysis since the upper bound (in the sense of convex ordering) of the sum of several random variables with given marginals is achieved by a comonotonic distribution~\cite{deelstra2011overview,cheung2013bounds}. In the following, we introduce two relevant definitions.
\begin{definition}[\cite{cheung2013bounds}]\label{Def1}
A set $\mathcal{X}\subseteq \mathbb{R}^n$ is comonotonic if $\mathbf{x} \preccurlyeq \mathbf{x}^{\prime}$ or $\mathbf{x}^{\prime} \preccurlyeq \mathbf{x}$ holds for any $\mathbf{x}$ {\color{black}and} $\mathbf{x}^\prime \in \mathcal{X}$.
\end{definition}

\begin{definition}[\cite{deelstra2011overview,cheung2013bounds}]\label{Def1}
	 A random vector $\mathbf{X}=\trans{\left(X_1,\cdots,X_n\right)}$ is comonotonic if either one of following equivalent propositions holds:
	\begin{itemize}
		\item there is a comonotonic set $\mathcal{X}\subseteq \mathbb{R}^n$ such that $\mathbb{P}\left\{\mathbf{X}\in \mathcal{X} \right\}=1$;
		\item $\mathbf{X}\overset{\text{d}}= \trans{{\color{black}\left(Q_{\! X_{\! 1}}(U),\cdots,Q_{\! X_{\! n}}(U)\right)}}$, where the random variable $U$ is uniformly distributed on $\left(0,1\right]$ and {\color{black}$Q_{\! X_{\! k}}(\cdot)$} is the quantile function of $X_k$ for {\color{black}$k\in {\color{black}\left[n\right]}$.}
	\end{itemize}
\end{definition}

As a special statistical dependence structure, comonotonicity is used to describe the phenomenon that several random variables always {\color{black}vary} in the same direction, i.e., simultaneously non-decreasing or non-increasing.

\subsection{$\left(\mathbf{h},\bm{\alpha}\right)$-Decomposability}
We first introduce the following notation. For any index set $\mathcal{J}\subseteq {\color{black}\left[\nt\right]}$, denote the partial sum of channel gains as
\begin{flalign}
	{\mathsf{H}}_{\mathcal{J}}\triangleq \sum_{i\in \mathcal{J}}h_i,
\end{flalign}
and {\color{black}hence,} the cumulative sum of channel gains {\color{black}can be expressed as}
\begin{flalign}
	\mathsf{H}_{{\color{black}[k]}}\triangleq \sum_{i=1}^{k}h_i,~\forall \, k\in {\color{black}\left[\nt\right]},
\end{flalign}
with $\mathsf{H}_{{\color{black}[0]}}=0$. 

Furthermore, denote the weighted average {\color{black}intensity} of the last $\nt-k$ transmitters as 
\begin{eqnarray}\label{eqn:constraint3}
	\bar{\alpha}_k=\frac{\sum_{i={k+1}}^{\nt}h_i\alpha_i}{1-\mathsf{H}_{\color{black}[k]}},~\forall k\in {\color{black}\left[\nt-1\right]},
\end{eqnarray}
with $\bar{\alpha}_{\nt}=0$. Clearly, we have $\bar{\alpha}_0 > \bar{\alpha}_1 > \cdots > \bar{\alpha}_{\nt-1}=a_{\nt}>\bar{\alpha}_{\nt}$.

Next, we define the $\left(\mathbf{h},\bm{\alpha}\right)$-decomposability of a random variable, which is a crucial concept in this paper.

\begin{definition}\label{def:decomposable}
	A random variable $S$ is said to be $\left(\mathbf{h},\bm{\alpha}\right)$-decomposable if there exists a random vector $\mathbf{W}=\trans{\left( W_1,\cdots,W_{\nt}\right)}$ satisfying  $\mathsf{supp}\, \mathbf{W} \subseteq [0,1]^{\nt}$ and $\mathbb{E}\left[\mathbf{W}\right]=\bm{\alpha}$ such that
	$S= \trans{\mathbf{h}}\mathbf{W}$.
\end{definition}

It is clear that any equivalent input signal $S$ (given by~\eqref{eqn:model2}) of the {\color{black}EC-OIC} must be $\left(\mathbf{h},\bm{\alpha}\right)$-decomposable. Conversely, if a random variable $S$ is $\left(\mathbf{h},\bm{\alpha}\right)$-decomposable, there exists a channel input $\mathbf{W}$ feasible to the {\color{black}EC-OIC} such that $S$ is the corresponding equivalent input. Thus, we conclude that a random variable $S$ is a feasible equivalent input for the {\color{black}EC-OIC} \textit{if and only if} $S$ is $\left(\mathbf{h},\bm{\alpha}\right)$-decomposable. {\color{black}We remind the reader that the above statement does not necessarily hold for the {\color{black}BC-OIC}. Instead, for the {\color{black}BC-OIC}, $S$ is a feasible equivalent input \textit{if and only if} there exists a vector $\mathbf{a}$ satisfying $\bm{0} \preccurlyeq \mathbf{a} \preccurlyeq \bm{\alpha}$ such that $S$ is $\left(\mathbf{h},\mathbf{a}\right)$-decomposable.}

A particular subclass of $\left(\mathbf{h},\bm{\alpha}\right)$-decomposability defined as follows will be also useful in our subsequent analysis.

\begin{definition}\label{def:comonotonic_decom}
	A random variable $S$ is said to be \textit{comonotonically $\left(\mathbf{h},\bm{\alpha}\right)$-decomposable} if there exists a {\color{black}random vector $\mathbf{X}^{\rm c}=\trans{ \left(Q_{\! X_{\! 1}}\left(U\right),\cdots,Q_{\! X_{\! \nt}}\left(U\right)\right)}$}, satisfying  $\mathsf{supp}\, \mathbf{X}^{\rm c} \subseteq [0,1]^{\nt}$ and $\mathbb{E}\left[\mathbf{X}^{\rm c}\right]=\bm{\alpha}$, such that
	\begin{flalign}\label{eq:comonotonic_decom}
		S=\trans{\mathbf{h}}\mathbf{X}^{\rm c}= \sum_{k=1}^{\nt} h_k {\color{black}Q_{\! X_{\! k}}}(U),
	\end{flalign}
	where $U$ is a random variable uniformly distributed on $\left(0,1\right]$ and {\color{black}$Q_{\! X_{\! k}}(\cdot)$} is the quantile function of $X_k$ for $k\in {\color{black}\left[\nt\right]}$.
\end{definition}

\subsection{Maximally Convex Distribution}
We develop the definition and properties of maximally convex distributions, which {\color{black}will be very useful} throughout this paper.
\begin{definition} \label{def:mcd}
	A discrete random variable $\bar{S}_{\mathbf{h},\bm{\alpha}}$ is said to {\color{black}obey} a \textit{maximally convex distribution} 
	if $\mathsf{supp}\,\bar{S}_{\mathbf{h},\bm{\alpha}} = \left\{ 0, \mathsf{H}_{\color{black}[1]},\, \cdots, \mathsf{H}_{\color{black}[\nt-1]},\, 1\right\}$ and its corresponding probability masses are $\left\{ 1-\alpha_1,\, \alpha_1-\alpha_2,\, \cdots, \, \alpha_{\nt-1}-\alpha_{\nt},\,\alpha_{\nt}\right\}$. 
\end{definition}

The following properties of the maximally convex distribution can be easily verified.
\begin{proposition}
	The SLT of $\bar{S}_{\mathbf{h},\bm{\alpha}}$, {\color{black}denoted by $\bar{\pi}_{\mathbf{h},\bm{\alpha}}(t)$}, is the piecewise linear function successively joining $\nt+1$ breakpoints $ \left(\mathsf{H}_{\color{black}[i]},\left(1-\mathsf{H}_{\color{black}[i]}\right)\bar{\alpha}_i\right)$ for all $i\in\{0\}\cup {\color{black}\left[\nt\right]} $.
\end{proposition}
\begin{IEEEproof} 
{\color{black}Directly from Definition~\ref{def:mcd}.}
\end{IEEEproof}

\begin{proposition}\label{Prob:MCD}
	$\bar{S}_{\mathbf{h},\bm{\alpha}}$ is comonotonically $\left(\mathbf{h},\bm{\alpha}\right)$-decomposable. 
\end{proposition}

\begin{IEEEproof} 
	Let $R_0=\bar{S}_{\mathbf{h},\bm{\alpha}}$ and
	\begin{flalign}
		R_k
		&= \left(\bar{S}_{\mathbf{h},\bm{\alpha}}-\mathsf{H}_{\color{black}[k]}\right)_+
	\end{flalign} 
	for $k\in {\color{black}\left[\nt\right]}$.
	{\color{black}It can be checked that} 
	\begin{flalign}
		\mathbb{E}\left[R_k\right]=\sum_{m=k+1}^{\nt} h_m\alpha_m
	\end{flalign}
	for $k\in {\color{black}\left[\nt-1\right]}$.
	Then we let
	\begin{flalign}
		X_k&=\frac{R_{k-1}-R_{k}}{h_k}
	\end{flalign}
	for $k\in {\color{black}\left[\nt\right]}$. Notice that the function $\left(s-\mathsf{H}_{\color{black}[k-1]}\right)_+-\left(s-\mathsf{H}_{\color{black}[k]}\right)_+$ is monotonically increasing with $s$. Hence, the so-constructed random vector $\bar{\mathbf{X}}_{\bm{\alpha}}=\trans{\left(X_1,\cdots,X_{\nt}\right)}$ is comonotonic and satisfies $\mathsf{supp}\, \bar{\mathbf{X}}_{\bm{\alpha}} \subseteq [0,1]^{\nt}$, $\mathbb{E}\left[\bar{\mathbf{X}}_{\bm{\alpha}}\right]=\bm{\alpha}$ and $\bar{S}_{\mathbf{h},\bm{\alpha}}= \trans{\mathbf{h}}\bar{\mathbf{X}}_{\bm{\alpha}}$.
\end{IEEEproof}
\vspace{0.2cm} 

{\color{black}
	It can be checked that the probability mass function of $\bar{\mathbf{X}}_{\bm{\alpha}}$ is given by $\mathbb{P}\left\{\bar{\mathbf{X}}_{\bm{\alpha}}=\bm{0} \right\}=1-\alpha_1$, $\mathbb{P}\left\{\bar{\mathbf{X}}_{\bm{\alpha}}=\bm{1} \right\}=\alpha_{\nt}$, and $\mathbb{P}\left\{\bar{\mathbf{X}}_{\bm{\alpha}}=\sum_{i=1}^k \mathbf{e}_i \right\}=\alpha_{k}-\alpha_{k+1}$ for $k\in [\nt-1]$, where $\mathbf{e}_i$ denotes the $i$-th column vector of the $\nt\times \nt$ identity matrix. The constructed random vector $\bar{\mathbf{X}}_{\bm{\alpha}}$ is indeed a \textit{maximally correlated $\nt$-variate binary distribution} proposed in \cite[Definition $1$]{chaabanrezkialouini18_2},} which is solely determined by $\bm{\alpha}$ and has the largest variance over all probability laws feasible to the {\color{black}EC-OIC}.

\section{{\color{black}Conditions for Decomposability and Equivalent Capacity Expressions}}
\label{sec:decomposability}

{\color{black}This section establishes the main results in the paper, which consist of two decomposition theorems and the resulting capacity expressions for the EC-OIC and the BC-OIC.

}


\subsection{{\color{black}Decomposition Results}}
{\color{black}We first present necessary and sufficient} conditions for $(\mathbf{h},\bm{\alpha})$-decomposability.

\begin{theorem}\label{main}
	Let $S$ be a random variable {\color{black}satisfying} $\mathsf{supp}\, S\subseteq [0,1]$ and {\color{black}$\mathbb{E}[S]=\trans{\mathbf{h}}\bm{\alpha}$}. Then the following claims are equivalent:
	\begin{enumerate}
		\item $S$ is $\left(\mathbf{h},\bm{\alpha}\right)$-decomposable;
		\item For each $k\in {\color{black}\left[\nt-1\right]}$,
		\begin{flalign}\label{eq:62}
			\pi_{\scriptscriptstyle \! S}(\mathsf{H}_{\color{black}[k]})\le \left(1-\mathsf{H}_{\color{black}[k]}\right)\bar{\alpha}_k;
		\end{flalign}
		\item $ S \le_{\textnormal{cx}}  \bar{S}_{\mathbf{h},\bm{\alpha}} $;
		\item $S$ is comonotonically $\left(\mathbf{h},\bm{\alpha}\right)$-decomposable;
	\end{enumerate} 
\end{theorem}

\begin{IEEEproof} 
	{\color{black}See Sec.~\ref{sec:proof_main}.}
\end{IEEEproof}
\vspace{0.2cm}	

{\color{black}Theorem~\ref{main} reveals the equivalence between $(\mathbf{h},\bm{\alpha})$-decomposability and comonotonic $(\mathbf{h},\bm{\alpha})$-decomposability. Hence, an equivalent input feasible to the {\color{black}EC-OIC} must be comonotonically $\left(\mathbf{h},\bm{\alpha}\right)$-decomposable as well. This fact suggests that it is sufficient to take comonotonic random vectors into account when we seek for the capacity-achieving input for the EC-OIC at any SNR.}

\begin{figure}[h]
	\centering
	\resizebox{13cm}{!}{\includegraphics{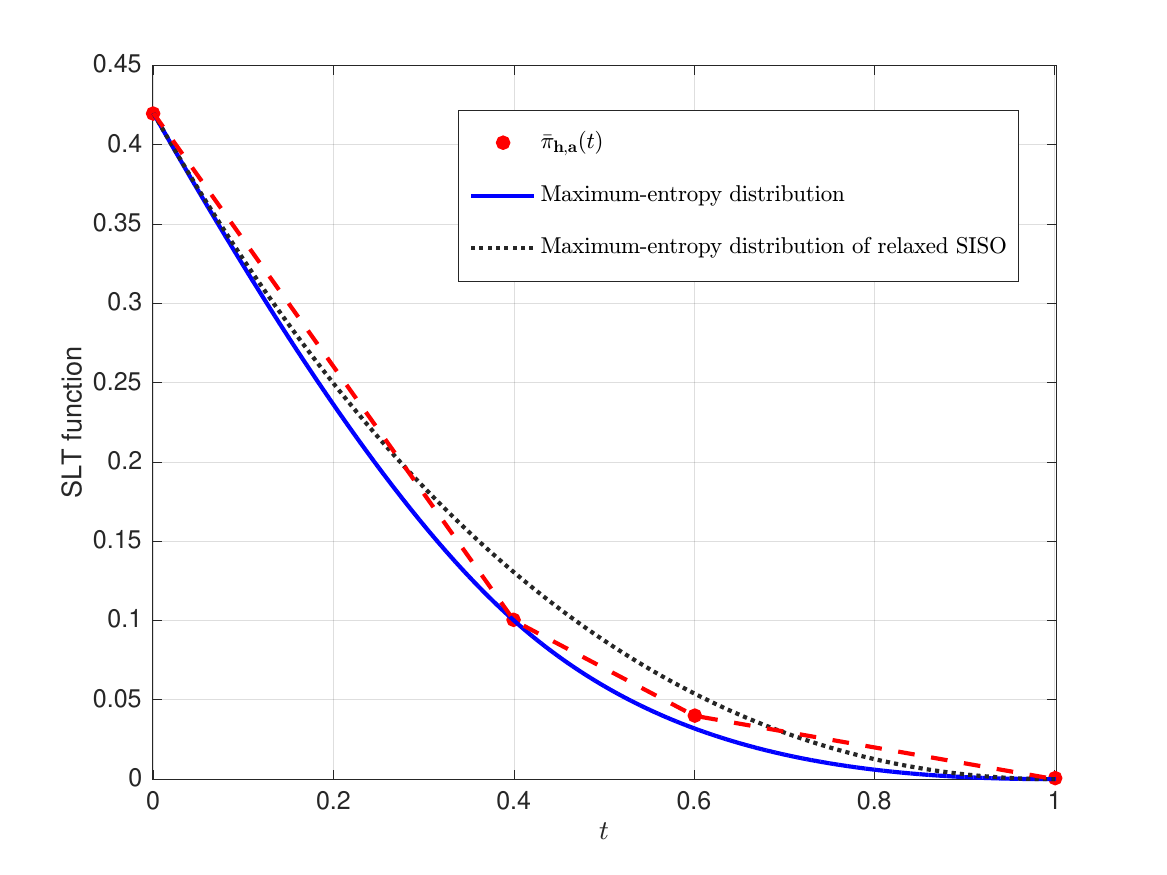}}
	\centering \caption{SLT functions of {\color{black}the maximally convex distribution and the maximum-entropy distribution for a $3\times 1$ {\color{black}EC-OIC with $\mathbf{h}=\trans{\left(0.4,0.2,0.4\right)}$ and $\bm{\alpha}=\trans{\left(0.8,0.3,0.1\right)}$, and the maximum-entropy distribution for a SISO EC-OIC with the average intensity being $0.42$.}}}
	\label{fig:SLT}
\end{figure}

Theorem~\ref{main} also demonstrates that, among all $\left(\mathbf{h},\bm{\alpha}\right)$-decomposable random variables, the maximally convex distribution $\bar{S}_{\mathbf{h},\bm{\alpha}}$ is the largest one in the sense of convex order. Hence, the best upper bound on the SLT {\color{black}over all equivalent inputs feasible to the EC-OIC} is exactly the SLT of $\bar{S}_{\mathbf{h},\bm{\alpha}}$, {\color{black}i.e., $\bar{\pi}_{\mathbf{h},\bm{\alpha}}(t)$}. In Figure~\ref{fig:SLT}, we plot the SLT of the maximally convex distribution and the maximum-entropy distribution of a $3\times 1$ {\color{black}EC-OIC} with $\mathbf{h}=\trans{\left(0.4,0.2,0.4\right)}$ and $\bm{\alpha}=\trans{\left(0.8,0.3,0.1\right)}$, which indeed follow the convex ordering as stated in Theorem~\ref{main}. We also plot the SLT of the maximum-entropy distribution of the corresponding amplitude-limited SISO {\color{black}OIC}~\eqref{model:siso} with a relaxed {\color{black}intensity} constraint~\eqref{eq:ap1}, which is not always below $\bar{\pi}_{\mathbf{h},\bm{\alpha}}(t)$, and hence, not feasible to the {\color{black}above-mentioned} {\color{black}EC-OIC}. {\color{black}We also remark that, as a consequence of Corollary~\ref{var_order},} $\bar{S}_{\mathbf{h},\bm{\alpha}}$ maximizes the variance of the equivalent input for the {\color{black}EC-OIC}. This is consistent with the low-SNR result in~\cite{chaabanrezkialouini18_2}.

\begin{remark}
	{\color{black} In the constructive proof of Theorem~\ref{main}, a decomposition method~\eqref{eq:112}, called \textit{greedy decomposition}, is provided to generate a comonotonic random vector $\mathbf{X}^{\rm c}$ satisfying $\mathbf{X}^{\rm c}\in [0,1]^{\nt}$, $\mathbb{E}\left[\mathbf{X}^{\rm c}\right]=\bm{\alpha}$ and $S=\trans{\mathbf{h}}\mathbf{X}^{\rm c}$. A more detailed description of decomposition algorithms as well as the signaling procedure is postponed until Sec.~\ref{sec: signaling}. The lack of knowledge of the greedy decomposition will not impede the reader in understanding our capacity results. 
	
	We also point out that} the greedy decomposition is just one type of the comonotonic decomposition. For an $(\mathbf{h},\bm{\alpha})$-decomposable random variable, there may exist other decompositions. Suppose the inputs $X_1,\cdots,X_{\nt}$ are independent, log-concave and satisfy {\color{black}intensity} constraints (\ref{eqn:ecc}), e.g., each $X_k$ being a truncated exponential distribution. Then we can rewrite the equivalent input $S=\trans{\mathbf{h}}\mathbf{X}$ as $S=\trans{\mathbf{h}}\mathbf{W}$ with $W_k=g_k(S)$ for $k\in {\color{black}\left[\nt\right]}$, where the functions $g_k(s)$ are given by
		\begin{flalign}\label{decomp2}
			g_k(s)=\mathbb{E}\left[X_k|\trans{\mathbf{h}}\mathbf{X}=s\right].
		\end{flalign} 
	Due to the law of total expectation, {\color{black}the} so-constructed random variables $W_k$ must satisfy the {\color{black}constraints} (\ref{eqn:ecc}) as well. Furthermore, based on Efron’s monotonicity theorem~\cite{Saumard2014Log}, $(W_1,\cdots,W_{\nt})$ is comonotonic.\footnote{Note that the decomposition (\ref{decomp2}) needs the distributions of $\nt$ inputs, while the greedy decomposition only needs the distribution of the equivalent input.}
\end{remark}

{\color{black}
Based on Theorem~\ref{main} and convex ordering, the necessary and sufficient conditions for another type of decomposition are given as follows.


\begin{theorem}\label{main2}
	Let $S$ be a random variable satisfying $\mathsf{supp}\, S\subseteq [0,1]$ and $\mathbb{E}[S]\le\trans{\mathbf{h}}\bm{\alpha}$. Then the following claims are equivalent:
	\begin{enumerate}
		\item There exists a vector $\mathbf{a}$ satisfying $\bm{0} \preccurlyeq \mathbf{a} \preccurlyeq \bm{\alpha}$ such that $S$ is $\left(\mathbf{h},\mathbf{a}\right)$-decomposable;
		\item For each $k\in \left[\nt-1\right]$,
		\begin{flalign} 
			\pi_{\scriptscriptstyle \! S}(\mathsf{H}_{\color{black}[k]})\le \left(1-\mathsf{H}_{\color{black}[k]}\right)\bar{\alpha}_k;
		\end{flalign}
		\item $S$ is $\left(\mathbf{h},\mathbf{a}^{\dagger} \right)$-decomposable, where the vector  $	\mathbf{a}^{\dagger} =   \min\left\{\beta \bm{1},\bm{\alpha}\right\} $ with some $\beta \in \left[0,\alpha_1 \right]$ such that  $\trans{\mathbf{h}}\mathbf{a}^{\dagger}=\mathbb{E}[S]$.
	\end{enumerate} 
\end{theorem}

\begin{IEEEproof} 
 See Sec~\ref{app:exp-bcmc}.
\end{IEEEproof}
}

\begin{remark}[Optimal {\color{black}intensity} allocation for {\color{black}BC-OIC}]\label{opt_power} 
	As illustrated in {\color{black}Sec.~\ref{subsec:problem}}, when $\alpha_{\nt}\ge \frac{1}{2}$ the capacity-achieving {\color{black}average intensity} of the {\color{black}BC-OIC} is $\mathbf{a}^{\star}=\frac{1}{2}\bm{1}$ so that the {\color{black}BC-OIC} is degenerated to a SISO intensity channel under a peak- and an average-{\color{black}intensity} constraint without inducing any capacity loss. 
	
	In the case of $\alpha_{\nt}< \frac{1}{2}$, {\color{black}it follows from Theorem~\ref{main2} that the capacity-achieving {\color{black}average intensity} $\mathbf{a}^\star$ belongs to the set $\left\{   \min\{\beta \bm{1},\bm{\alpha}\}:\beta\in [0,\alpha_1]	
		\right\}$.}
	Hence, the capacity of the {\color{black}BC-OIC} can be also formulated as a maximization over the capacities of a specific class of {\color{black}EC-OICs}, i.e.,
	\begin{flalign}
		\mathsf{C}_{\textnormal{B}}\left(\mathbf{h},\bm{\alpha},\sigma\right)=\max_{\beta\in [0,\alpha_1]} \, \mathsf{C}_{\textnormal{E}}\left(\mathbf{h}, \min\{\beta \bm{1},\bm{\alpha}\},\sigma \right).
	\end{flalign}
	Note that when $\beta\in [0,\alpha_{\nt}]$ the {\color{black}average intensity} vector $\min\{\beta \bm{1},\bm{\alpha}\}$ {\color{black}is exactly $\beta\bm{1}$} so that the induced {\color{black}EC-OIC} is equivalent to a SISO {\color{black}OIC} under a peak- and an average-{\color{black}intensity constraint} as well. Hence, we have $\mathsf{C}_{\textnormal{E}}\left(\mathbf{h}, \min\{\beta \bm{1},\bm{\alpha}\},\sigma \right)=\mathsf{C}_{\textnormal{E}}\left(1,\beta,\sigma\right)$ for $\beta\in [0,\alpha_{\nt}]$.
	Due to the monotonicity of the SISO capacity  $\mathsf{C}_{\textnormal{E}}\left(1,\beta,\sigma\right)$~\cite{lapidothmoserwigger09_7}, the capacity-achieving {\color{black}average intensity} can be further restricted to $\mathbf{a}^\star \in \left\{   \min\{\beta \bm{1},\bm{\alpha}\}:\beta\in [\alpha_{\nt},\alpha_1]	
	\right\} $.
\end{remark}

\subsection{Equivalent Capacity {\color{black}Expressions}}

{\color{black}
 
The above two theorems actually state necessary and sufficient conditions under which a random variable $S$ is a feasible equivalent input for the EC-OIC or the BC-OIC, respectively. These naturally lead to alternative expressions of the capacities $\mathsf{C}_{\textnormal{E}}\left(\mathbf{h},\bm{\alpha},\sigma\right)$ and $\mathsf{C}_{\textnormal{B}}\left(\mathbf{h},\bm{\alpha},\sigma\right)$ as follows.

}



\begin{corollary}[Equivalent capacity expression {\color{black}for EC-OIC}]
	\label{corollary:ecc} The capacity $\mathsf{C}_{\textnormal{E}}\left(\mathbf{h},\bm{\alpha},\sigma\right)$ can be equivalently expressed as 
	\begin{flalign}\label{ecc:capacity}
		\mathsf{C}_{\textnormal{E}}\left(\mathbf{h},\bm{\alpha},\sigma \right)=\sup_{{\color{black}\mathscr{P}_{\!  S}}} \II \left(S;S+Z\right),
	\end{flalign}
	where the supremum is over all probability laws ${\color{black}\mathscr{P}_{ \! S}}$ on $S$ with $\mathsf{supp}\, S\subseteq [0,1]$ and under {\color{black}stop-loss mean} constraints:
	\begin{subequations}\label{eq:equiv_ecmc}
		\begin{align}
			&\mathbb{E}\left[S\right]=\trans{\mathbf{h}}\bm{\alpha};\label{eq:equiv_ecmc1}\\
			&\mathbb{E}\left[\left(S-\mathsf{H}_{\color{black}[k]}\right)_+\right]\le \left(1-\mathsf{H}_{\color{black}[k]}\right) \bar{\alpha}_k, ~\forall\, k\in {\color{black}\left[\nt-1\right]}.\label{eq:equiv_ecmc2}
		\end{align}	
	\end{subequations}
\end{corollary}
\begin{IEEEproof}
Theorem~\ref{main} implies that an input {\color{black}$\mathbf{X}$} is feasible to {the \color{black}EC-OIC} if and only if {\color{black}the corresponding} equivalent input $S=\trans{\mathbf{h}}\mathbf{X}$ satisfies $\mathsf{supp}\, S\subseteq [0,1]$, $\mathbb{E}[S]=\trans{\mathbf{h}}\bm{\alpha}$ and $\pi_{\scriptscriptstyle \! S}(\mathsf{H}_{\color{black}[k]})\le \left(1-\mathsf{H}_{\color{black}[k]}\right)\bar{\alpha}_k$ for $k\in {\color{black}\left[\nt-1\right]}$. Corollary \ref{corollary:ecc} thus immediately follows.
\end{IEEEproof}
\vspace{0.2cm}

{\color{black}
For the BC-OIC, we exhibit a similar expression of $\mathsf{C}_{\textnormal{B}}\left(\mathbf{h},\bm{\alpha},\sigma\right)$ in the following.}

\begin{corollary}[Equivalent capacity expression {\color{black}for BC-OIC}]
	\label{corollary:bcmc}
	The  capacity $\mathsf{C}_{\textnormal{B}}\left(\mathbf{h},\bm{\alpha},\sigma\right)$ can be equivalently written as
	\begin{flalign}\label{ecc:capacity}
		\mathsf{C}_{\textnormal{B}}\left(\mathbf{h},\bm{\alpha},\sigma\right)=\max_{{\color{black}\mathscr{P}_{\!  S}}}\, \II \left(S;S+Z\right),
	\end{flalign}
	where the supremum is over all probability laws ${\color{black}\mathscr{P}_{\!  S}}$ on $S$ with $\mathsf{supp}\, S\subseteq [0,1]$ and under {\color{black}stop-loss mean} constraints:
	\begin{flalign}\label{eq:70}
		\mathbb{E}\left[\left(S-\mathsf{H}_{\color{black}[k]}\right)_+\right]\le \left(1-\mathsf{H}_{\color{black}[k]}\right) \bar{\alpha}_k,~\forall\, k\in {\color{black}\{0\}\cup\left[\nt-1\right]}. 
	\end{flalign}
\end{corollary}

\begin{IEEEproof}
{\color{black}
	Accroding to Theorem~\ref{main2}, an input {\color{black}$\mathbf{X}$} is feasible to the BC-OIC if and only if the corresponding equivalent input $S=\trans{\mathbf{h}}\mathbf{X}$ satisfies $\mathsf{supp}\, S\subseteq [0,1]$, $\mathbb{E}[S]\le\trans{\mathbf{h}}\bm{\alpha}$ and $\pi_{\scriptscriptstyle \! S}(\mathsf{H}_{\color{black}[k]})\le \left(1-\mathsf{H}_{\color{black}[k]}\right)\bar{\alpha}_k$ for $k\in \left[\nt-1\right]$, from which Corollary \ref{corollary:ecc} follows.
}
\end{IEEEproof}

{\color{black} Corollaries~\ref{corollary:ecc} and~\ref{corollary:bcmc} enlighten the fact that either a MISO EC-OIC or a MISO BC-OIC is equivalent to a SISO channel under an amplitude constraint and $\nt$ stop-loss mean constraints, and this significantly simplifies subsequent capacity analysis.}

\section{Capacity {\color{black}Bounds and Asymptotic Characteristics}}
\label{sec:cap-res-ecmc}
{\color{black}In this section, we present capacity results for the EC-OIC and the BC-OIC based on the derived capacity expressions.}

\subsection{Lower Bounds}
The following capacity lower bounds are obtained by applying the EPI, combined with maximizing the differential entropy over all feasible inputs. We first present a lower bound {\color{black}for the EC-OIC}.
\begin{theorem}[EPI-based lower bound {\color{black}for EC-OIC}]
\label{ecmc:lbd}
	 The capacity of the {\color{black}EC-OIC} is lower-bounded as
	\begin{flalign}
		\mathsf{C}_{\textnormal{E}}\left(\mathbf{h},\bm{\alpha},\sigma\right)\ge \frac{1}{2}\log \left(1+\frac{\exp(2 	\gamma_{ \textnormal{E}})}{2\pi e \sigma^2}\right),
	\end{flalign}
	where 
	\begin{flalign}\label{eq:gammaE}
		\gamma_{ \textnormal{E}} = &\min_{\nu_0,\lambda_0 \in \mathbb{R} \atop \lambda_1,\cdots,\lambda_{\nt   -   1}\ge 0} \sum_{i=0}^{\nt-1}\lambda_i\left(1-\mathsf{H}_{\color{black}[i]}\right)\bar{\alpha}_i-1-\nu_0\nonumber\\
		&+\exp({\color{black}\nu}_0)\left\{\sum_{k=1}^{\nt}h_{k} \exp\left[-\lambda_0\mathsf{H}_{\color{black}[k-1]}+\sum_{i=1}^{k-2}\lambda_i\left(\mathsf{H}_{\color{black}[i]}-\mathsf{H}_{\color{black}[k-1]}\right) \right]\zeta\left[ \left(-\sum_{i=0}^{k-1}\lambda_i \right)h_{k}\right]  \right\}
	\end{flalign}
	with $\zeta \left(x\right)=\frac{e^x-1}{x}$.\footnote{{\color{black}By letting the derivative of order $j$ be} $\zeta^{(j)} \left(0\right)=\frac{1}{j+1}$ for $j\in \mathbb{N}$, $\zeta \left(x\right)$ is continuously differentiable and strictly convex over the real line.}
\end{theorem}
\begin{IEEEproof}
	{\color{black}See Appendix~\ref{app:lbd}.}
\end{IEEEproof}
\vspace{0.2cm}

Similarly, we have the following lower bound {\color{black}for the BC-OIC}.
	\begin{theorem}[EPI-based lower bound {\color{black}for BC-OIC}]
	\label{bcmc:lbd}
		The capacity of the {\color{black}BC-OIC} is lower-bounded as 
			\begin{flalign}
				\mathsf{C}_{\textnormal{B}}\left(\mathbf{h},\bm{\alpha},\sigma\right)\ge \frac{1}{2}\log \left(1+\frac{\exp(2 \gamma_{\textnormal{B}})}{2\pi e \sigma^2}\right),
			\end{flalign}
			where 
			\begin{flalign}\label{eq:gammaB}
				\gamma_{\textnormal{B}} \triangleq &\min_{\nu_0 \in \mathbb{R} \atop \lambda_0,\cdots,\lambda_{\nt-1}\ge 0} \sum_{i=0}^{\nt-1}\lambda_i\left(1-\mathsf{H}_{\color{black}[i]}\right)\bar{\alpha}_i-1-\nu_0\nonumber\\
				&+\exp({\color{black}\nu}_0)\left\{\sum_{k=1}^{\nt}h_{k} \exp\left[-\lambda_0\mathsf{H}_{\color{black}[k-1]}+\sum_{i=1}^{k-2}\lambda_i\left(\mathsf{H}_{\color{black}[i]}-\mathsf{H}_{\color{black}[k-1]}\right) \right]\zeta\left[ \left(-\sum_{i=0}^{k-1}\lambda_i \right)h_{k}\right]  \right \}.
			\end{flalign} 
	\end{theorem}
\begin{IEEEproof}
	{\color{black}See Appendix~\ref{app:lbd}.}
\end{IEEEproof}
\vspace{0.2cm}

{\color{black} 
	The above two lower bounds are derived by choosing an equivalent input distribution so as to maximize the differential entropy $\hh (S)$. Since either the {\color{black}EC-OIC or the BC-OIC} can be transformed into a SISO channel with {\color{black}stop-loss mean} constraints, we can use convex programming to obtain the equivalent input distribution with a maximized differential entropy. The maximum-entropy distribution of the equivalent input $S$ for the {\color{black}EC-OIC} (or {\color{black}BC-OIC}) is given by}
		\begin{flalign}\label{eq:maxent}
			p_{\scriptscriptstyle \! S}^*(s) = \exp \left(\nu_0^*-\lambda_0^* s-\left( \sum_{i=1}^{\nt -1} \lambda_i^* \left(s-\mathsf{H}_{\color{black}[i]} \right)_+ \right)\right),~ s\in [0,1],
		\end{flalign}
		where $\nu_0^*$, $\lambda_0^*$, and $\lambda_1^*,\cdots,\lambda_{\nt -1}^*$ {\color{black}constitute} the optimal solution to~\eqref{eq:gammaE} (or~\eqref{eq:gammaB}).

\subsection{Upper Bounds}
In this subsection, we present several upper bounds on capacities of the {\color{black}EC-OIC and the BC-OIC}.
\subsubsection{Upper Bounds by SISO and MISO Capacity Expression}
Capacities of both the {\color{black}EC-OIC and the BC-OIC} can be upper-bounded by utilizing SISO or MISO capacity expression{\color{black}s} in the literature. 
\begin{proposition}[Upper bound by SISO capacity] \label{BCMC:SISO_Ubd}
The capacity of the {\color{black}EC-OIC} is upper-bounded as
	\begin{flalign}\label{eq:56}
		\mathsf{C}_{\textnormal{E}}\left(\mathbf{h},\bm{\alpha},\sigma\right)\le \mathsf{C}_{\textnormal{E}}\left(1,{\color{black}\min\{\trans{\mathbf{h}}\bm{\alpha},1-\trans{\mathbf{h}}\bm{\alpha} \}},\sigma\right),
	\end{flalign}
{\color{black}while} the capacity of the {\color{black}BC-OIC} is upper-bounded as
	\begin{flalign}\label{eq:57}
		\mathsf{C}_{\textnormal{B}}\left(\mathbf{h},\bm{\alpha},\sigma\right)\le \mathsf{C}_{\textnormal{B}}\left(1,\min\left\{ \trans{\mathbf{h}}\bm{\alpha},\frac{1}{2}\right\},\sigma\right).
	\end{flalign}
\end{proposition}
\begin{IEEEproof}
	{\color{black}We sketch the proof as follows. The upper bound~\eqref{eq:56} for the {\color{black}EC-OIC} is derived by first relaxing the intensity constraint~\eqref{eq:equiv_ecmc} on the equivalent input to~\eqref{eq:equiv_ecmc1} (i.e., ignoring the stop-loss mean inequality constraints~\eqref{eq:equiv_ecmc2}) and then using the symmetry of the {\color{black}EC-OIC} capacity (shown in \eqref{eq:symmetry}). Similarly, we can obtain the upper bound \eqref{eq:57} for the {\color{black}BC-OIC}, where the minimization operator follows from~\eqref{eq:12}.}
\end{IEEEproof}
\vspace{0.2cm}

In~\cite{moserwangwigger18_3}, the authors {\color{black}have studied the capacity of the peak-limited MISO OIC with an inequality constraint on the total average intensity}. We denote the capacity in {\color{black}this scenario as $\mathsf{C}_{\textnormal{B-TA}}(\mathbf{h},\alpha,\sigma^2)$, where $\alpha$ denotes the ratio between the maximum allowed total average intensity and
the maximum allowed peak intensity,} and here it can serve as {\color{black}an} upper bound as well.

\begin{proposition}[Upper bound by MISO capacity]\label{BCMC:MISO_Ubd}
{\color{black}The capacity of the {\color{black}EC-OIC} is upper-bounded as
	\begin{flalign}\label{eq:50}
		\mathsf{C}_{\textnormal{E}}\left(\mathbf{h},\bm{\alpha},\sigma\right)\le \mathsf{C}_{\textnormal{B-TA}}\left(\vect{h},\min\left\{ \|\bm{\alpha}\|_1,\nt-\|\bm{\alpha}\|_1,\frac{\nt}{2}\right\},\sigma\right),
	\end{flalign}
while} the capacity of the {\color{black}BC-OIC} is upper-bounded as
\begin{flalign}\label{eq:51}
		\mathsf{C}_{\textnormal{B}}\left(\mathbf{h},\bm{\alpha},\sigma\right)\le \mathsf{C}_{\textnormal{B-TA}}\left(\vect{h},\min\left\{ \|\bm{\alpha}\|_1,\frac{\nt}{2}\right\},\sigma\right).
\end{flalign}
\end{proposition}
\begin{IEEEproof}
{\color{black}
For the EC-OIC,	the upper bound~\eqref{eq:50} follows from $\sum_{k=1}^{\nt}\mathbb{E}[X_k]\le \|\bm{\alpha}\|_1$
	and the symmetry property \eqref{eq:symmetry}. For the BC-OIC, we let the intensity constraint \eqref{eq:70} be relaxed to \eqref{eq:ap2} and note that $\mathsf{C}_{\textnormal{B-TA}}(\mathbf{h},\alpha,\sigma^2)$ is maximized at $\alpha=\nt/2$. Then \eqref{eq:51} immediately follows, and \eqref{eq:50} follows from $\mathsf{C}_{\textnormal{E}}\left(\mathbf{h},\bm{\alpha},\sigma\right) \le \mathsf{C}_{\textnormal{B}}\left(\mathbf{h},\bm{\alpha},\sigma\right)$, \eqref{eq:symmetry}, and \eqref{eq:51}.  }
\end{IEEEproof}

\subsubsection{Upper Bounds by Maximum-Variance {\color{black}Argument}}
The following upper bounds are obtained by maximizing the variance of the equivalent input. In~\cite{chaabanrezkialouini18_2}, it has been shown that the maximally correlated $\nt$-variate binary distribution has the maximum variance for the {\color{black}EC-OIC}, {\color{black}which immediately leads to Theorem~\ref{ecmc:max-var-bnd}.} Here we present an alternative proof from the perspective of stochastic ordering, which can also be directly extended to the {\color{black}BC-OIC}. 

\begin{theorem}[Maximum-variance upper bound {\color{black}for EC-OIC}~{\cite{chaabanrezkialouini18_2}}]
\label{ecmc:max-var-bnd}
 The capacity of the {\color{black}EC-OIC} is upper-bounded as
	\begin{flalign}
		\mathsf{C}_{\textnormal{E}}\left(\mathbf{h},\bm{\alpha},\sigma\right)\le \frac{1}{2}\log \left(1+\frac{\const{V}_{\textnormal{max}}^{\textnormal{E}}\left(\mathbf{h},\bm{\alpha}\right)}{\sigma^2}\right),
	\end{flalign} 
where $\const{V}_{\textnormal{max}}^{\textnormal{E}}\left(\mathbf{h},\bm{\alpha}\right)$ denotes the maximum variance of $S$, {\color{black}given by}
	\begin{flalign}\label{eq:maxvar_ecmc}
		\const{V}_{\textnormal{max}}^{\textnormal{E}}\left(\mathbf{h},\bm{\alpha}\right)=\sum_{i=1}^{\nt}\sum_{j=1}^{\nt} h_ih_j \left(\min\{\alpha_i,\alpha_j\} -\alpha_i\alpha_j\right).
	\end{flalign}
\end{theorem}

\begin{IEEEproof}
	{\color{black}It immediately follows from the proof of {\cite[Theorem 1]{chaabanrezkialouini18_2}}. Alternatively, see~Appendix~\ref{app:max-var-bnd} for another proof.}
\end{IEEEproof}
\vspace{0.2cm}

\begin{theorem}[Maximum-variance upper bound {\color{black}for BC-OIC}]
\label{bcmc:max-var-bnd}
The capacity of the {\color{black}BC-OIC} is upper-bounded as
	\begin{flalign}
		\mathsf{C}_{\textnormal{B}}\left(\mathbf{h},\bm{\alpha},\sigma\right)\le \frac{1}{2}\log \left(1+\frac{\const{V}_{\textnormal{max}}^{\textnormal{B}}\left(\mathbf{h},\bm{\alpha}\right)}{\sigma^2}\right),
	\end{flalign}
where 	
		\begin{flalign}\label{eq:maxvar_bcmc}
\const{V}_{\textnormal{max}}^{\textnormal{B}}\left(\mathbf{h},\bm{\alpha}\right)&=\mathsf{H}^2_{\color{black}[{k_{\beta^*}}]}{\beta^*} \left(1-{\beta^*}\right)+2\mathsf{H}_{\color{black}[{k_{\beta^*}}]}\left(1-\mathsf{H}_{\color{black}[{k_{\beta^*}}]}\right)\bar{\alpha}_{k_{\beta^*}}(1-{\beta^*}) \nonumber \\
			&~~~+\sum_{i=k_{\beta^*}+1}^{\nt}\sum_{j=k_{\beta^*}+1}^{\nt} h_ih_j \left(\min\{\alpha_i,\alpha_j\} -\alpha_i \alpha_j\right),
		\end{flalign}
		with $\beta^*=\inf \left\{ \beta\in\left(\alpha_{\nt},\alpha_1\right]:\mathsf{H}_{\color{black}[{k_{\beta}}]} \left(1-2\beta\right)-2\left(1-\mathsf{H}_{\color{black}[{k_{\beta}}]}\right)\bar{\alpha}_{k_{\beta}} \le 0 \right\}$ and {\color{black}$k_\beta=\{k\in \left[\nt\right]:\alpha_k\ge \beta\}$}.
\end{theorem}
\begin{IEEEproof}
	{\color{black}See~Appendix~\ref{app:max-var-bnd}.}
\end{IEEEproof}

{\color{black}As seen in Theorem~\ref{bcmc:max-var-bnd}, for the BC-OIC, the maximum-variance of the equivalent input $S$ can be easily determined by comparing solutions of $\nt-1$ linear equations.}

\subsubsection{Upper Bounds by Duality Capacity Expression}
We present the following upper bounds by using the duality upper-bounding technique. These bounds are later proved to be tight at high SNR in {\color{black}Sec.}~\ref{sec:5-3}.

\begin{theorem}[Duality upper bounds {\color{black}for EC-OIC}]\label{ubd:ecc}
	The capacity of the {\color{black}EC-OIC} is upper-bounded as
	\begin{flalign}\label{eq1:ubd_ecmc}
		\mathsf{C}_{\textnormal{E}}\left(\mathbf{h},\bm{\alpha},\sigma\right)
		\leq  
		\log \left(1+\frac{  \mathsf{P}}{\sqrt{2\pi e} \sigma}\right) +  \sum_{i=0}^{\nt-1}\lambda_i \left(1-\mathsf{H}_{\color{black}[i]}\right)\bar{\alpha}_i 
		+ \sum_{i=0}^{\nt-1}\frac{\lambda_i\sigma}{\sqrt{2\pi}}\left(1-e^{-\frac{\left(1+\delta\right)^2}{2\sigma^2}}\right) 
	\end{flalign}
	for real numbers $\lambda_0,\cdots,\lambda_{\nt-1}\ge 0$
	and
	\begin{flalign}\label{eq2:ubd_ecmc}
		\mathsf{C}_{\textnormal{E}}\left(\mathbf{h},\bm{\alpha},\sigma\right)
		\leq  
		& \log \left(1+\frac{ \mathsf{P}}{\sqrt{2\pi e} \sigma}\right) +  \sum_{i=0}^{\nt-1}\lambda_i \left(1-\mathsf{H}_{\color{black}[i]}\right)\bar{\alpha}_i  
		+ \sum_{i=1}^{\nt-1}\frac{\lambda_i\sigma}{\sqrt{2\pi}}\left(1-e^{-\frac{\left(1+\delta\right)^2}{2\sigma^2}}\right)
		\nonumber \\
		&+\lambda_0 \sigma \left( \phi(\frac{1}{\sigma})-\phi\left(\frac{\delta}{\sigma}\right)\right)
		-\lambda_0  \left(\mathcal{Q}\left(\frac{1}{\sigma}\right)+\mathcal{Q}\left(\frac{\delta}{\sigma}\right)\right)
	\end{flalign}
	for real numbers $\lambda_0\le 0$ and $\lambda_1,\cdots,\lambda_{\nt-1}\ge0$,
	where
	\begin{flalign}\label{eq:P}
		\mathsf{P}=	\int_0^{1+\delta}\exp \left( -\lambda_0 y-\sum_{i=1}^{\nt-1} \lambda_i \left(y-\mathsf{H}_{\color{black}[i]} \right)_+ \right) \dd y,
	\end{flalign}
	{\color{black}$\mathcal{Q}(\cdot)$ denotes Gaussian Q-function,} and $\delta>0$ is a free parameter. 
\end{theorem}
\begin{IEEEproof}
	{\color{black}See Appendix~\ref{app:dual}.}
\end{IEEEproof}
\vspace{0.2cm}

Similarly, we get the following upper bound {\color{black}for the BC-OIC}.

\begin{theorem}[Duality upper bound {\color{black}for BC-OIC}]
	\label{ubd:bcc}
		The capacity of the {\color{black}BC-OIC} is upper-bounded as
		\begin{flalign}\label{BCMC:duality}
			\mathsf{C}_{\textnormal{B}}\left(\mathbf{h},\bm{\alpha},\sigma\right)
			\leq  
			\log \left(1+\frac{  \mathsf{P}}{\sqrt{2\pi e} \sigma}\right) 
			\!+ \!  \sum_{i=0}^{\nt-1}\lambda_i \left(1-\mathsf{H}_{\color{black}[i]}\right)\bar{\alpha}_i  	\!+ \!  \sum_{i=0}^{\nt-1}\frac{\lambda_i\sigma}{\sqrt{2\pi}}\left(1-e^{-\frac{\left(1+\delta\right)^2}{2\sigma^2}}\right) 
		\end{flalign}
		for real numbers $\lambda_0,\cdots,\lambda_{\nt-1}\ge 0$, where $\mathsf{P}$ is defined in~\eqref{eq:P}
		and $\delta>0$ is a free parameter. 
\end{theorem}
\begin{IEEEproof}
	{\color{black}See Appendix~\ref{app:dual}.}
\end{IEEEproof}

\subsection{Asymptotic Capacit{\color{black}ies}}\label{sec:5-3}
In this subsection we present low- and high-SNR asymptotic results. 

\subsubsection{Low-SNR Capacity Slope}
{\color{black}We characterize low-SNR capacities of the EC-OIC and the BC-OIC as follows.}

\begin{theorem}[Low-SNR capacity slope]\label{thm:lowsnr} The low-SNR {\color{black} capacity slope} of the {\color{black}EC-OIC} (or {\color{black}BC-OIC}) is
\begin{flalign}
	\lim_{\sigma\uparrow \infty} \bigl \{ \sigma^2\mathsf{C}_{\textnormal{E (or B)}}\left(\mathbf{h},\bm{\alpha},\sigma\right) \bigr\} = \frac{\const{V}_{\textnormal{max}}^{\textnormal{E (or B)}}\left(\mathbf{h},\bm{\alpha}\right)}{2},
\end{flalign}
where $\const{V}_{\textnormal{max}}^{\textnormal{E}}\left(\mathbf{h},\bm{\alpha}\right)$ and $\const{V}_{\textnormal{max}}^{\textnormal{B}}\left(\mathbf{h},\bm{\alpha}\right)$ are defined in~\eqref{eq:maxvar_ecmc} and~\eqref{eq:maxvar_bcmc}, respectively.
\end{theorem}

\begin{IEEEproof}
	{\color{black}Based on} \cite[Corollary $2$]{prelovverdu04_1}, the capacit{\color{black}ies of interest} can be lower-bounded as
	\begin{IEEEeqnarray}{c}
		\mathsf{C}_{\textnormal{E {\color{black}(or B)}}}\left( \mathbf{h},\bm{\alpha},\sigma \right) \geq  \frac{\const{V}_{\textnormal{max}}^{\textnormal{E {\color{black}(or B)}}}\left(\mathbf{h},\bm{\alpha}\right)}{2\sigma^2}
		+ o\left(\frac{1}{\sigma^2}\right).\label{eq:66}
	\end{IEEEeqnarray} 
	
	{\color{black}For the reverse direction, upper bounds in terms of maximum variance in Theorems~\ref{ecmc:max-var-bnd} and~\ref{bcmc:max-var-bnd} imply that
		\begin{IEEEeqnarray}{c}\label{eq:69}
			\mathsf{C}_{\textnormal{E (or B)}}\left( \mathbf{h},\bm{\alpha},\sigma \right) \leq  \frac{\const{V}_{\textnormal{max}}^{\textnormal{E (or B)}}\left(\mathbf{h},\bm{\alpha}\right)}{2\sigma^2}
			.
		\end{IEEEeqnarray} 
		Combining~\eqref{eq:66} and~\eqref{eq:69}, we complete the proof of Theorem~\ref{thm:lowsnr}.}
\end{IEEEproof}
\vspace{0.2cm}
{\color{black}We remind the reader that the maximized variances  ${\const{V}_{\textnormal{max}}^{\textnormal{E}}\left(\mathbf{h},\bm{\alpha}\right)}$ and ${\const{V}_{\textnormal{max}}^{\textnormal{B}}\left(\mathbf{h},\bm{\alpha}\right)}$ for the EC-OIC and the BC-OIC can be achieved by maximally convex distributions $\bar{S}_{\mathbf{h},\bm{\alpha}}$ and $\bar{S}_{\mathbf{h},\mathbf{a}^\dagger}$, respectively, where the vector $\mathbf{a}^\dagger$ is defined in Theorem~\ref{main2}.
}
\subsubsection{High-SNR Asymptotic Capacity}
{\color{black}By showing that the afore-mentioned EPI-based lower bounds coincide with duality upper bounds at high SNRs, we obtain the following result. }
\begin{theorem}[High-SNR asymptotic capacity]
\label{thm:highsnr}
The high-SNR asymptotic capacity of the {\color{black}EC-OIC} (or {\color{black}BC-OIC) satisfies}
		\begin{flalign}
		\lim_{\sigma\downarrow 0^+}	\mathsf{C}_{\textnormal{E (or B)}}\left(\mathbf{h},\bm{\alpha},\sigma\right)-\log\frac{1}{\sigma} = -\frac{1}{2}\log 2\pi e+\gamma_{\textnormal{E (or B)}},
	\end{flalign}
where $\gamma_{\textnormal{E}}$ and $\gamma_{\textnormal{B}}$ are defined in~\eqref{eq:gammaE} and~\eqref{eq:gammaB}, respectively.
\end{theorem}
\begin{IEEEproof}
	{\color{black}See Appendix~\ref{app:proofs-asymp}.}
\end{IEEEproof}

\subsection{Numerical Results}
\label{sec:num-res}
In this subsection, we present {\color{black}numerical evaluations} for our derived {\color{black}capacity results}, where {\color{black}the} involved duality bounds~\eqref{eq1:ubd_ecmc}, \eqref{eq2:ubd_ecmc} and \eqref{BCMC:duality} are numerically minimized over the allowed values of $\lambda_0,\cdots,\lambda_{\nt-1}$ and $\delta$.

\begin{figure}[hbtp!]
	\centering
	\resizebox{13cm}{!}{\includegraphics{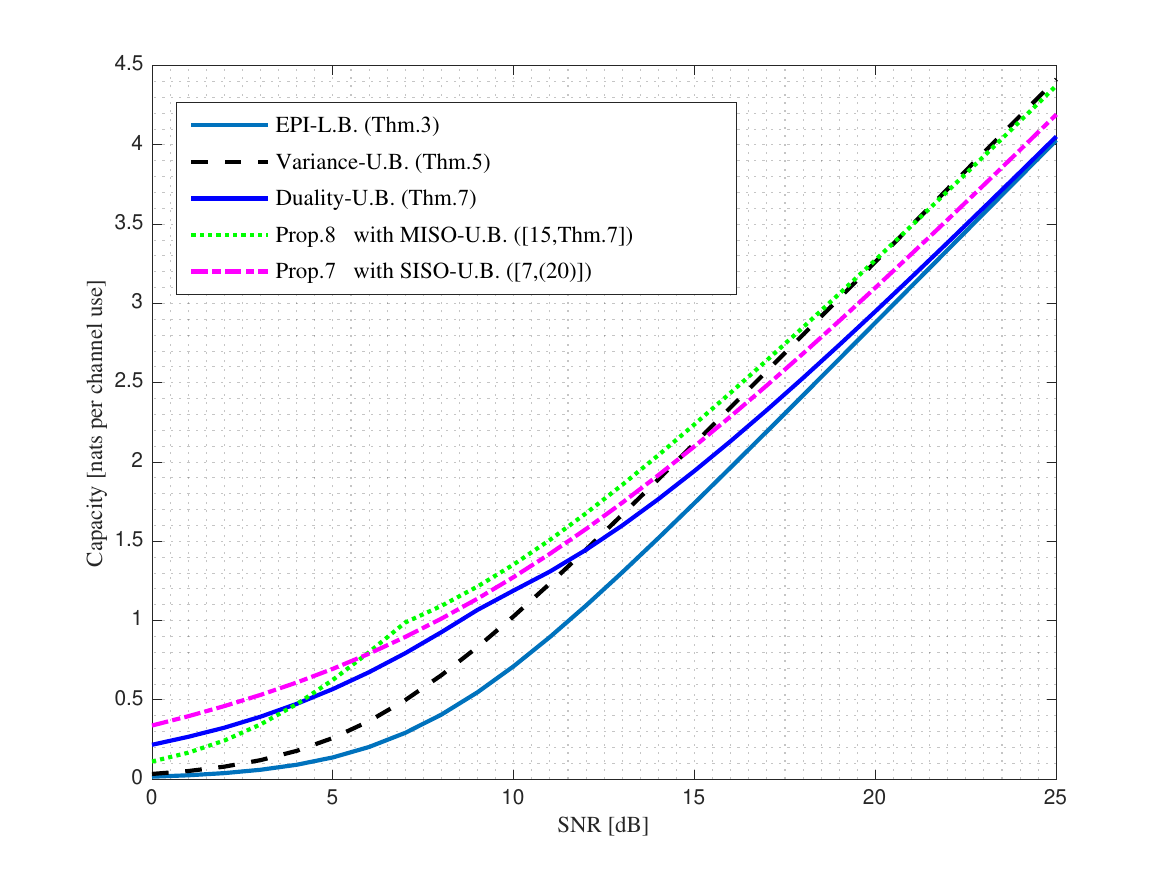}}
	\centering \caption{Capacity bounds of a $3\times1$ {\color{black}EC-OIC} with ${\mathbf{h}}=\trans{\left(0.3,0.1,0.6\right)}$ and ${\bm{\alpha}}=\trans{\left(0.8,0.3,0.1\right)}$. The maximum gap between upper and lower bounds is about $0.35$ nats.}
	\label{fig:ECMC_capacity}
\end{figure}

\begin{figure}[hbtp!]
	\centering
	\resizebox{13cm}{!}{\includegraphics{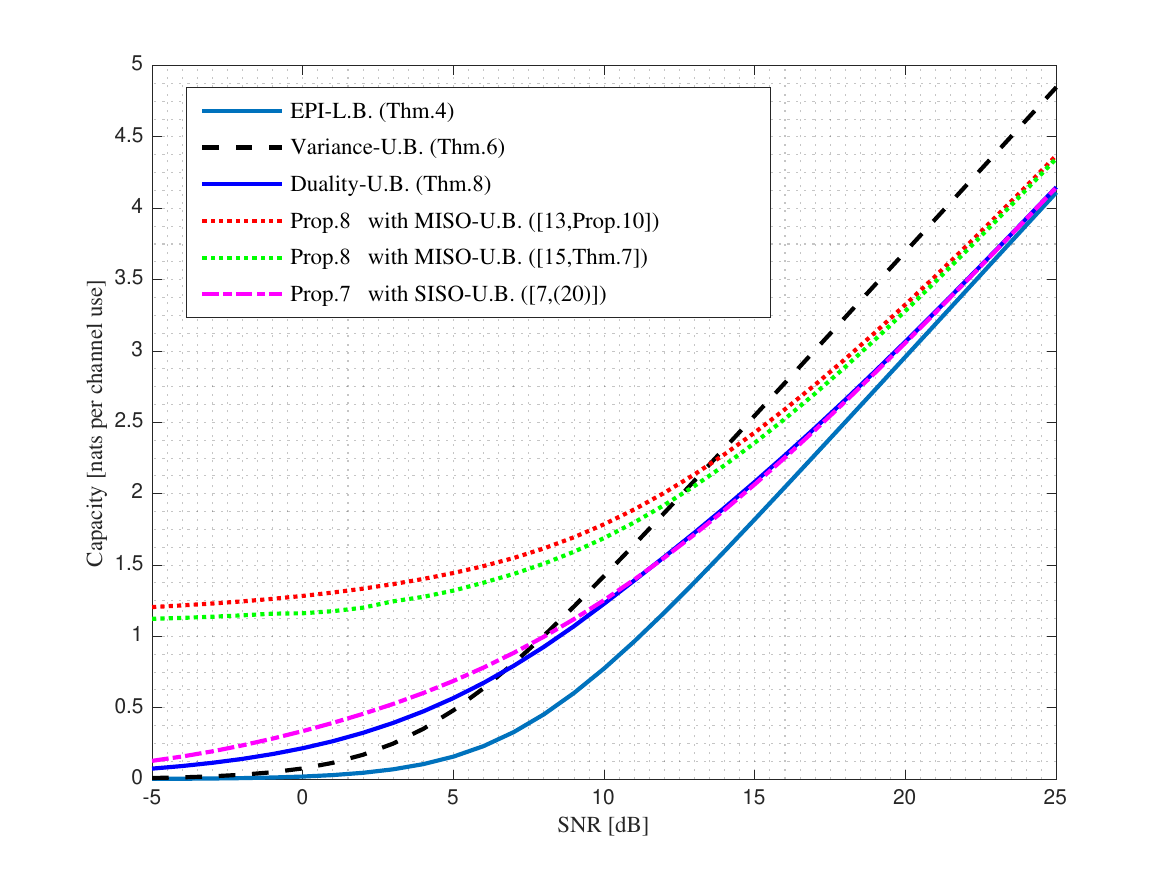}}
	\centering \caption{Capacity bounds of {\color{black}a $3\times 1$ BC-OIC} with ${\mathbf{h}}=\trans{\left(0.3,0.2,0.5\right)}$ and ${\bm{\alpha}}=\trans{\left(0.5,0.3,0.2\right)}$. The maximum gap between upper and lower bounds is about $0.47$ nats.}
	\label{fig:BCMC_capacity}
\end{figure}

Figure~\ref{fig:ECMC_capacity} depicts the derived lower and
upper bounds for a $3 \times 1$ {\color{black}EC-OIC} with ${\mathbf{h}}=\trans{\left(0.3,0.1,0.6\right)}$ and $\bm{\alpha}=\trans{\left(0.8,0.3,0.1\right)}$, {\color{black}while} Figure~\ref{fig:BCMC_capacity} depicts the lower and
upper bounds for a $3 \times 1$ {\color{black}BC-OIC} with $\mathbf{h}=\trans{\left(0.3,0.2,0.5\right)}$ and $\bm{\alpha}=\trans{\left(0.5,0.3,0.2\right)}$.
{\color{black}In both figures, we present capacity upper bounds obtained by combining Propositions~\ref{BCMC:SISO_Ubd} and \ref{BCMC:MISO_Ubd} and existing upper bounds for the SISO OIC under a peak- and an average-intensity constraint~\cite{lapidothmoserwigger09_7} and the MISO OIC with a total average-intensity and per-antenna peak-intensity constraints~\cite{moserwangwigger18_3,li2018miso}.}
Compared with existing capacity bounds, the duality-based upper bound gives better approximation on capacity at high SNR, and 
matches the EPI-based lower bound asymptotically as SNR tends to infinity. A noteworthy observation in Figure~\ref{fig:BCMC_capacity} is that the upper bound obtained by the SISO capacity (Proposition~\ref{BCMC:SISO_Ubd}) coincides with our duality upper bound~\eqref{BCMC:duality} at high SNR. This {\color{black}is} due to the fact that there exists some $\beta\in [\alpha_{\nt},\alpha_1]	$ such that the maximum-entropy distribution of the relaxed SISO channel is $(\vect{h},\min\{\beta \bm{1},\bm{\alpha}\})$-decomposable, i.e., feasible to the considered {\color{black}BC-OIC} as well. 

To better verify our derived asymptotic results at {\color{black}low and high SNRs}, we further investigate $2 \times 1$ {\color{black}EC-OICs and BC-OICs} with equal channel gains ${\vect{h}} = \trans{(\frac{1}{2},\frac{1}{2})}$ in Figure~4, whose results exhibit the symmetry property in terms of changing the order of $\bm{\alpha}$. As we see, {\color{black}for the EC-OIC both asymptotic capacities at low and high SNRs} decrease as $\bm{\alpha}$ tends to corner, while the results of the {\color{black}BC-OIC} are Schur convex.

\begin{figure}[!htbp]
	\centering
	\label{fig:asym} 
	\begin{minipage}[!htbp]{0.5\textwidth} 
		\centering 
		\includegraphics[width=7cm]{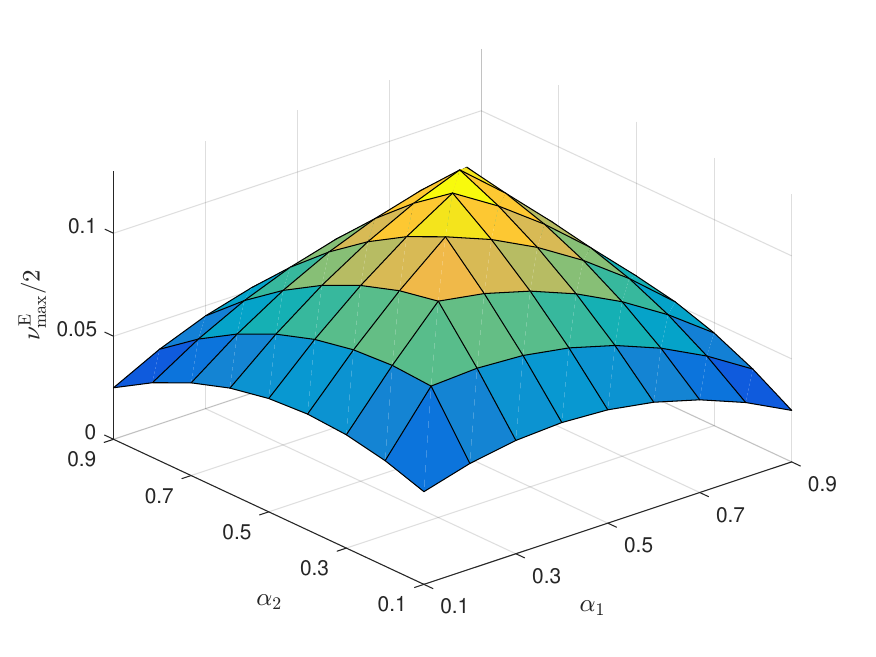} 
	\end{minipage}%
	\begin{minipage}[!htbp]{0.5\textwidth} 
		\centering 
		\includegraphics[width=7cm]{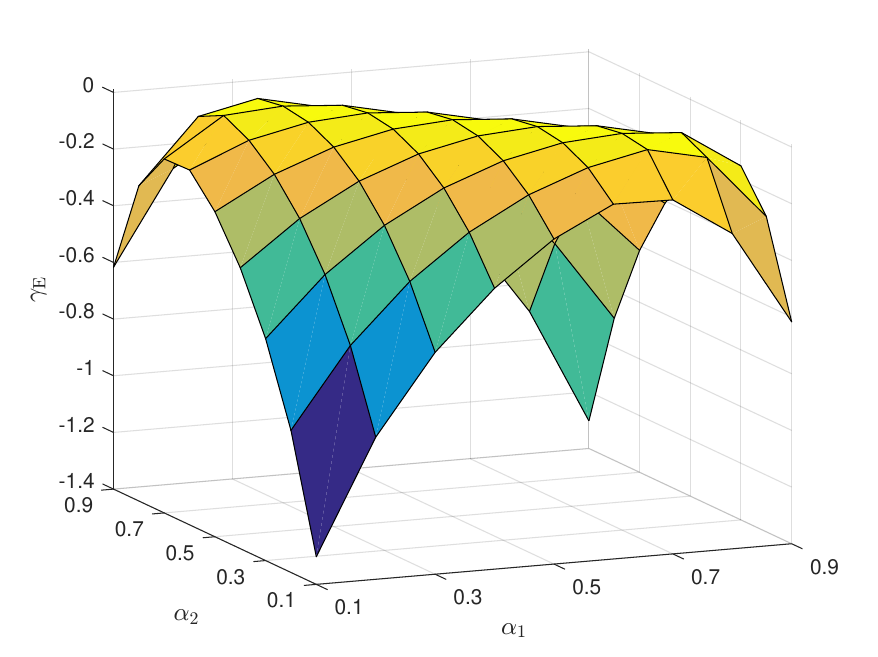} 
	\end{minipage}%

	\begin{minipage}[!htbp]{0.5\textwidth} 
		\centering 
		\includegraphics[width=7cm]{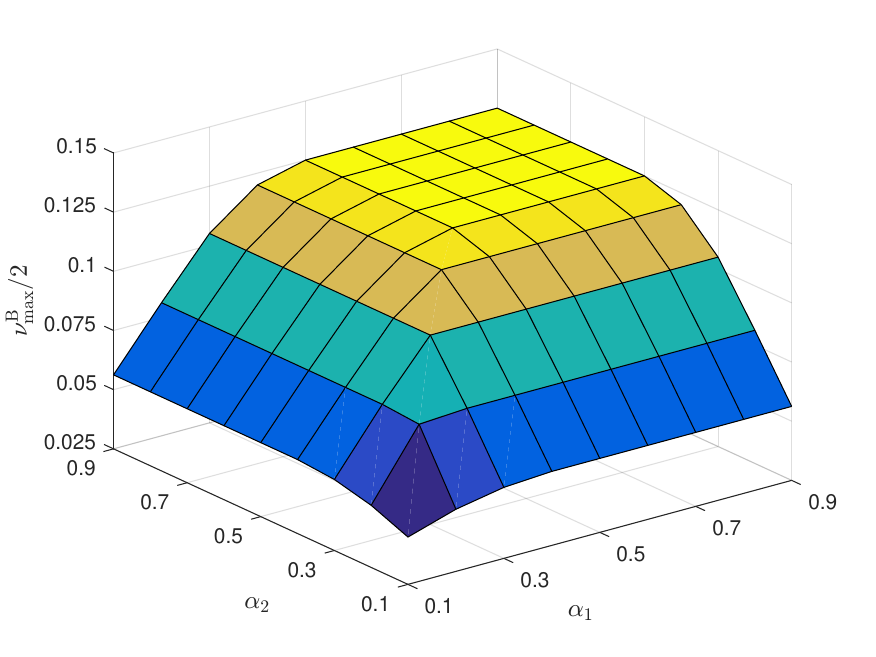} 
	\end{minipage}%
	\begin{minipage}[!htbp]{0.5\textwidth} 
		\centering 
		\includegraphics[width=7cm]{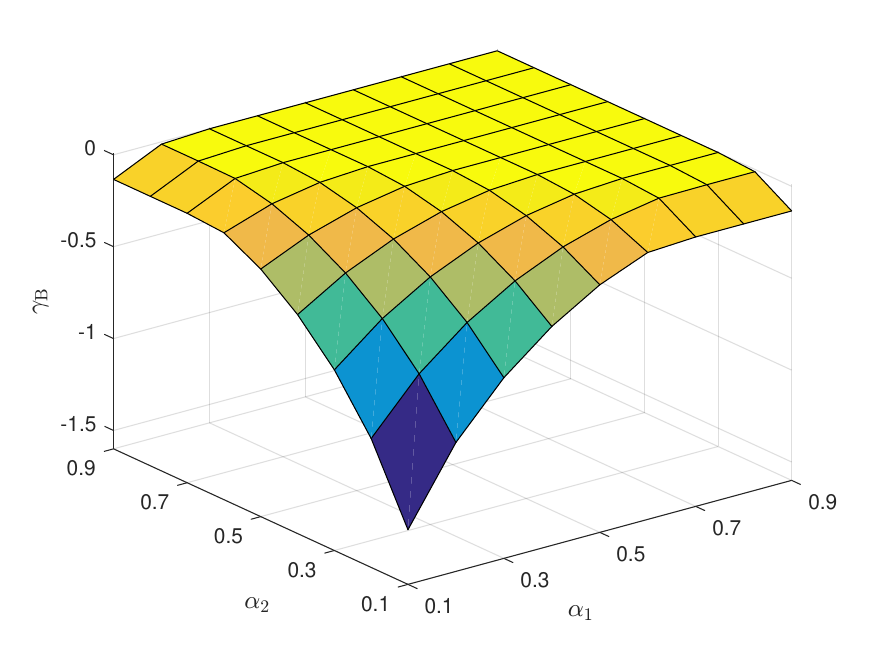} 
	\end{minipage}%
	\centering \caption{Asymptotic capacity results of $2\times 1$ {\color{black}EC-OICs} (above) and {\color{black}BC-OICs} (below) with {\color{black}equal channel gains} $h_1=h_2=0.5$.} 
	\vspace{-0.3cm}
\end{figure}

\section{Derivation of Decomposition Results}\label{sec:derivation}
{\color{black}This section deals with proofs of Theorems~\ref{main} and \ref{main2}, and the former is much more involved. A crucial part in the proof of Theorem~\ref{main} is a class of parametric functions, named \textit{greedily-constructed quantiles}, which will be introduced first in the following subsection.}

\subsection{Greedily-Constructed Quantile}
{\color{black}
Let us begin with a short literature review.} As shown in~\cite{moserwangwigger18_3,limoserwangwigger20_1}, the capacity of a MISO OIC under a peak-{\color{black}intensity} and a total average-{\color{black}intensity constraint} can be achieved by a minimum-energy signaling strategy, which prioritizes the input with a larger channel coefficient and enables it to grow first due to the nature of linear programming. For {\color{black}the EC-OIC}, a maximally correlated multivariate binary input has been shown to maximize the variance of the equivalent input in~\cite{chaabanrezkialouini18_2}, which lets the transmitter with the maximum average {\color{black}intensity} grow prior to the others. We remark that both the minimum-energy signaling and the maximally correlated multivariate binary input can be regarded as special classes of comonotonic distributions. For the above reasons, it is natural to investigate the behaviors of comonotonic inputs in MISO OICs of our interest.

\begin{figure}[hbtp!]
	\centering
	\resizebox{13cm}{!}{\includegraphics{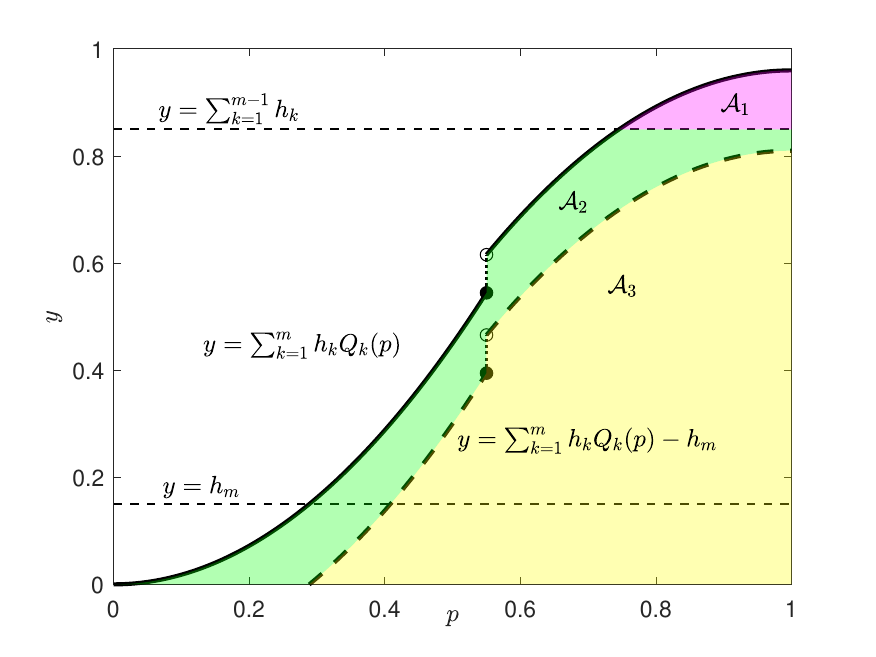}}
	\centering \caption{{\color{black}Decomposition of the quantile of a nonnegative and bounded random variable}. }
	\label{fig:quantile}
\end{figure}

{\color{black}Given a comonotonic input $\mathbf{X}$ with mean $\bm{\alpha}$, the equivalent input $S=\trans{\mathbf{h}}\mathbf{X}$ must be comonotonically $\left(\mathbf{h},\bm{\alpha}\right)$-decomposable. According to Definition~\ref{def:comonotonic_decom},} the quantile {\color{black}$Q_{\!  S}(p)$} of $S$ is a mixture of quantiles as {\color{black}$Q_{\!  S}(p)=\sum_{k=1}^{\nt} h_k Q_{\! X_{\! k}}(p)$}. Hence, a random variable $S$ with the quantile $Q_{\! S}(p)$ is comonotonically $\left(\mathbf{h},\bm{\alpha}\right)$-decomposable if and only if there exist $\nt$ quantiles {\color{black}$Q_{\! X_{\! k}}(p)$} ($k\in {\color{black}\left[\nt\right]}$) satisfying the following three conditions:
\begin{enumerate}
	\item ${\color{black}Q_{\! X_{\! k}}}\!:\!(0,1]\!\to\![0,1]$ is non-decreasing and left-continuous;
	\item $\int_{0}^1 {\color{black}Q_{\! X_{\! k}}}(p) \dd p=\alpha_k$;
	\item $Q_{\! S} (p )=\sum_{k=1}^{\nt} h_k {\color{black}Q_{\! X_{\! k}}}(p)${\color{black}, $\forall p \in(0,1]$}.
\end{enumerate}
Thus, the core problem {\color{black}in} comonotonic decomposition of a random variable is how to decompose its quantile function $Q_{\! S}(p)$. In Figure~\ref{fig:quantile}, we present a heuristic geometrical interpretation of the strategy for constructing desired quantiles. Suppose that $\sum_{k=1}^{m} h_k {\color{black}Q_{\! X_{\! k}}}(p)$ is known and is the quantile of $R_m=\sum_{k=1}^{m}h_k {\color{black}Q_{\! X_{\! k}}(U)}$. The curve of quantile sum $\sum_{k=1}^{m-1} h_k{\color{black}Q_{\! X_{\! k}}}(p)$ must lie in the green region due to boundedness of quantiles, i.e., larger than zero, less than $\sum_{k=1}^{m} h_k {\color{black}Q_{\! X_{\! k}}}(p)$ and $\sum_{k=1}^{m-1} h_k$, and within a curved strip of width $h_{m}$ below $\sum_{k=1}^{m} h_k {\color{black}Q_{\! X_{\! k}}}(p)$. In addition, to meet the integral requirement of ${\color{black}Q_{\! X_{\! m}}}(p)$, the sum quantile $\sum_{k=1}^{m} h_k{\color{black}Q_{\! X_{\! k}}}(p)$ must satisfy $\textnormal{Area}~ \mathcal{A}_1 \le h_{m}\alpha_{m}$ and $\textnormal{Area}~ \mathcal{A}_1+ \textnormal{Area}~\mathcal{A}_2 \ge h_{m}\alpha_{m}$. Note that the magenta region $\mathcal{A}_1=\mathcal{R}^{  \scriptscriptstyle R_m}_{\mathsf{H}_{\color{black}[m-1]}}$ (the right-hand side is defined in Proposition~\ref{prop:area}) and the yellow region $\mathcal{A}_3$ is a vertical translate of $\mathcal{R}^{  \scriptscriptstyle R_m}_{h_{m}}$. Then we have
\begin{flalign}
	&~~~~\textnormal{Area}~ \mathcal{A}_1+ \textnormal{Area}~\mathcal{A}_2\nonumber \\
	&=\int_0^1 \left(\sum_{k=1}^{m} h_k{\color{black}Q_{\! X_{\! k}}}(p)\right) \dd p -\textnormal{Area}~\mathcal{A}_3 \nonumber \\
	&= \sum_{k=1}^{m} h_k\alpha_k  -\textnormal{Area}~\mathcal{A}_3 \nonumber \\
	&= \sum_{k=1}^{m} h_k\alpha_k  -  \textnormal{Area}~ \mathcal{R}^{  \scriptscriptstyle R_m}_{h_{m}}.
\end{flalign}
Thus, to guarantee the existence of $Q_{m}(p)$ with a required definite integration value, the most greedy approach in constructing $\sum_{k=1}^{m}h_k{\color{black}Q_{\! X_{\! k}}}(p)$ is to let the areas of $\mathcal{R}^{  \scriptscriptstyle R_m}_{h_{m}}$ and $\mathcal{R}^{  \scriptscriptstyle R_m}_{\mathsf{H}_{\color{black}[m-1]}}$ {\color{black}be} as small as possible, i.e., the area of $\sum_{k=1}^{m}h_k{\color{black}Q_{\! X_{\! k}}}(p)$ should be preempted at the bottom. From above considerations, {\color{black}in Proposition \ref{prop:SLT}} we construct and analyze the ``most greedy'' class of quantile functions, i.e., so-called greedily-constructed quantiles, which coincide with $Q_{\! S} (p )$ before a threshold and otherwise coincide with a downward translate of $	Q_{\! S} (p )$.

\begin{proposition}\label{prop:SLT}
	Let $S$ be a random variable on $[0,1]$ and define {\color{black}the greedily-constructed quantile as the following} nonnegative parametric function 
	\begin{flalign}\label{eq:parametric1}
		\psi_{\scriptscriptstyle \!S} ( p;v,z ) {\color{black}\triangleq}	
		\begin{cases}
			Q_{\! S} (p )&,~\mbox{$0 < p \le F_{\! S}(v )$};\\
			v&,~ \mbox{$F_{\! S}(v ) < p \le F_{\! S}(v+z)$};\\
			Q_{\! S} (p  )-z&,~\mbox{$  F_{\! S}(v+z ) < p \le1$},\\
		\end{cases}
	\end{flalign}
	where the variable $p\in(0,1]$, and parameters  $v \in [0,1]$ and $z\in\left[0,1-v\right]$. Then
	\begin{enumerate}
		\item $\psi_{\scriptscriptstyle \!S} ( p;v,z)$ is the quantile function of $R=\varphi \left( S;v,z\right)$, where 	\begin{flalign}\label{eq:parametric2}
			\varphi \left( s;v,z\right) =
			\begin{cases}
				s&,~\mbox{$s \le v$},\\
				v&,~\mbox{$v<s \le v+z$},\\
				s-z&,~\mbox{$ s > v+z$}.
			\end{cases}	
		\end{flalign}
		\item The SLT of $R$ is
		\begin{flalign}
			\pi_{\scriptscriptstyle \! R} (t) =
			\begin{cases}
				\pi_{\scriptscriptstyle \! S}(t)-\pi_{\scriptscriptstyle \! S}(v)+\pi_{\scriptscriptstyle \! S}(v+z)&,~\mbox{$t \le v$},\\
				\pi_{\scriptscriptstyle \! S}(t+z)&,~\mbox{$ t > v$}.
			\end{cases}	
		\end{flalign}
	\end{enumerate}
	
\end{proposition}
\begin{IEEEproof}
	We first rewrite
	\begin{flalign}
		R&=\varphi \left( S;v,z\right) \nonumber \\
		&=\varphi \left( Q_{\! S}(U);v,z\right)  \nonumber \\
		&=
		\begin{cases}
			Q_{\! S}(U)&,~\mbox{$Q_{\! S}(U) \le v$}\\
			v&,~\mbox{$v<Q_{\! S}(U) \le v+z$}\\
			Q_{\! S}(U)-z&,~\mbox{$ Q_{\! S}(U) > v+z$}
		\end{cases}	 \nonumber \\
		&=\begin{cases}
			Q_{\! S}(U)&,~\mbox{$U \le F_{\! S}(v )$}\\
			v&,~\mbox{$F_{\! S}(v )<U \le F_{\! S}(v+z )$}\\
			Q_{\! S}(U)-z&,~\mbox{$ U >F_{\! S}(v+z )$}
		\end{cases}	\label{eq:43}\\
		&=\psi_{\scriptscriptstyle \!S} ( U;v,z),
	\end{flalign}
	where~\eqref{eq:43} follows from the Galois inequality~\eqref{eq:Galois}. 
	
	Since $Q_{\! S} (p )$ is the quantile of $S$, the function $\psi_{\scriptscriptstyle \!S} ( p;v,z )$ is left-continuous on the interval $p\in(0,1]$ and {\color{black}piecewise} non-decreasing on the intervals $\left(0,F_{\! S}(v )\right]$, $ \left(F_{\! S}(v ) ,F_{\! S}(v+z)\right]$ and $\left(F_{\! S}(v+z),1\right]$, respectively. Recall that $Q_{\! S}\left(F_{\! S}(v)\right)=\inf \left\{
	s: F_{\! S}(s)\ge F_{\! S}(v) \right\}
	$ and hence $Q_{\! S}\left(F_{\! S}(v)\right)\le v$.
	Next, based on the contrapositive of Proposition~\ref{prop:galois}, we have $Q_{\! S}(p)>v+z$ for any real number $p\in \left(F_{\! S}(v+z),1\right]$. Hence, the parametric function $\psi_{\scriptscriptstyle \!S} ( p;v,z )$ is non-decreasing and left-continuous on the interval $p\in(0,1]$. Proposition~\ref{thm:quantile} shows that $\psi_{\scriptscriptstyle \!S} ( U;v,z)$ is the quantile function of $R$. 
	
	It is clear that $0 \le R \le S$, and hence, $\mathsf{supp}\,R \subseteq [0,1]$. We notice that if $t>v$, $\left(\varphi \left( s;v,z\right)-t\right)_+=\left(s-t-z\right)_+$. Then 
	\begin{flalign}
		\pi_{\scriptscriptstyle \! R} (t)&=\mathbb{E}\left[ \left(R-t\right)_+\right] \nonumber \\
		&=\mathbb{E}\left[ \left(\varphi \left( S;v,z\right)-t\right)_+\right]\nonumber  \\
		&=\mathbb{E}\left[ \left(S-t-z\right)_+\right] \nonumber \\
		&=\pi_{\scriptscriptstyle \! S}(t+z),~t>v.
	\end{flalign}
	
	The {\color{black}other} case is concluded similarly by noticing that the following equality
	\begin{flalign}
		&~~~~\left(\varphi \left( s;v,z\right)-t\right)_+ \nonumber \\
		&=(s-t)_+-\left(\left(s-v\right)_++v-t\right) +\left(\left(s-v-z\right)_++v-t\right)  \nonumber \\
		&=(s-t)_+-(s-v)_++(s-v-z)_+
	\end{flalign}
	holds for $t\le v$.	
\end{IEEEproof}	

{\color{black}With a slight abuse of terminology, we also refer to the expression~\eqref{eq:parametric2} as the greedily-constructed quantile. Furthermore, in} the remainder of this paper, we work with the expression~\eqref{eq:parametric2} rather than~\eqref{eq:parametric1}, which frees us from the redundant discussion about quantiles.
{\color{black}We will see later that, components of a decomposable random variable can be constructed via the greedily-constructed quantiles.} 
\subsection{Proof of Theorem~\ref{main}}\label{sec:proof_main}
We first prove a necessary condition for the $\left(\mathbf{h},\bm{\alpha}\right)$-decomposability as follows. 
\begin{lemma}\label{lemma:necessary}
	An $\left(\mathbf{h},\bm{\alpha}\right)$-decomposable random variable $S$ must satisfy:
	\begin{enumerate}
		\item $\mathsf{supp}\, S\subseteq [0,1]$;
		\item Its SLT $\pi_{\scriptscriptstyle \! S}(t)$ satisfies 
		\begin{flalign}\label{eq:necessary}
			\pi_{\scriptscriptstyle \! S} \left({\mathsf{H}}_{\mathcal{J}}\right)\le \sum_{k\in \mathcal{J}^{\rm c}} h_k\alpha_k
		\end{flalign}
		for any index set $\mathcal{J}\subseteq {\color{black}\left[\nt\right]}$.
	\end{enumerate}  
	
\end{lemma}

\begin{IEEEproof} From the definition of $\left(\mathbf{h},\bm{\alpha}\right)$-decomposability, we can rewrite $S$ as  
	$S= \trans{\mathbf{h}}\mathbf{W}$, where the random vector $\mathbf{W}=\trans{\left( W_1,\cdots,W_{\nt}\right)}$ {\color{black}satisfies}  $\mathsf{supp}\, \mathbf{W} \subseteq [0,1]^{\nt}$ and $\mathbb{E}\left[\mathbf{W}\right]=\bm{\alpha}$. Due to nonnegativity of $\mathbf{h}$ and $\trans{\bm{1}}\mathbf{h}=1$, we immediately know $\mathsf{supp}\, S\subseteq [0,1]$. Then note that the function with respect to $s\in[0,1]$ satisfies the following inequality for any index set $\mathcal{J}\subseteq  {\color{black}\left[\nt\right]}$
	\begin{flalign}
		\mathbb{E}\left[\sum_{k\in\mathcal{J}^{\rm c}}h_kW_k\Big|S=s\right]\ge \left(s-{{\mathsf{H}}}_{\mathcal{J}} \right)_+,
	\end{flalign}
	due to $\mathsf{supp}\, \mathbf{W} \subseteq [0,1]^{\nt}$. By using the law of total expectation, we get
	\begin{flalign}
		\pi_{\scriptscriptstyle \! S} \left({{\mathsf{H}}}_{\mathcal{J}}\right)&= \mathbb{E}\left[\left(S-{{\mathsf{H}}}_{\mathcal{J}} \right)_+\right] \nonumber \\
		&\le \mathbb{E}_S\left[\mathbb{E}\left[\sum_{k\in\mathcal{J}^{\rm c}} h_kW_k\Big|S\right]\right] \nonumber \\
		& = \mathbb{E}\left[\sum_{k\in\mathcal{J}^{\rm c}}h_kW_k\right] \nonumber  \\
		&=\sum_{k\in\mathcal{J}^{\rm c}} h_k\alpha_k.
	\end{flalign}
	This {\color{black}completes the} proof of Lemma~\ref{lemma:necessary}.
\end{IEEEproof}
\vspace{0.2cm}

Then we show the equivalence among the four statements of Theorem \ref{main} by proving $1\ \Longrightarrow \ 2$,   $2 \ \Longrightarrow \ 3$, $3 \ \Longrightarrow \ 4$, and $4 \ \Longrightarrow \ 1$ all hold. 

The definition of comonotonic $(\mathbf{h},\bm{\alpha})$-decomposability immediately leads to $4 \ \Longrightarrow \ 1$.

By Lemma~\ref{lemma:necessary}, we can directly get 1 $\Longrightarrow$ 2. 

{\color{black}N}ote that $\bar{\pi}_{\mathbf{h},\bm{\alpha}}(0)=\pi_{\scriptscriptstyle \! S}\left( 0 \right)$, $\bar{\pi}_{\mathbf{h},\bm{\alpha}}(1)=\pi_{\scriptscriptstyle \! S}\left( 1 \right)=0$, and
\begin{equation}
	\bar{\pi}_{\mathbf{h},\bm{\alpha}}(\mathsf{H}_{\color{black}[k]})=\left(1-\mathsf{H}_{\color{black}[k]}\right)\bar{\alpha}_k\ge \pi_{\scriptscriptstyle \! S}\left(\mathsf{H}_{\color{black}[k]}\right).
\end{equation}
Furthermore, due to the convexity of the SLT and the piecewise linearity of $\bar{\pi}_{\mathbf{h},\bm{\alpha}}(t)$, we have
\begin{equation}
	\pi_{\scriptscriptstyle \! S}\left(t\right)\le\bar{\pi}_{\mathbf{h},\bm{\alpha}}\left(t\right), \ \forall t\in[0,1]. 
\end{equation}
Hence $2 \ \Longrightarrow \ 3$.

A bit involved part is to show  $3 \ \Longrightarrow \ 4$. Let $R_{\nt}^*=S$, $v_1^*=0$, and the $\nt$-dimensional real vector $\hat{\bm{\alpha}}^{(i)}$ be determined by $\hat{\alpha}_{k}^{(i)}=\alpha_k$ for $1 \le k \le i$ and others zero. We now use an iterative decomposition method to prove this.  It is sufficient to show
for each $i\in\{\nt,\cdots,2\}$, if $R_{i}^*$ satisfies  $\mathsf{supp}\, R_{i}^* \subseteq [0,\mathsf{H}_{\color{black}[i]}] $ and $R_{i}^* \le_{\textnormal{cx}} \bar{S}_{\mathbf{h},\hat{\bm{\alpha}}^{(i)}}$, then
\begin{enumerate}
	\item there exists a number $v_i^*\in\left[0,\mathsf{H}_{\color{black}[i-1]}\right]$ such that $R_{i-1}^*=\varphi\left(R_{i}^*;v_i^*,h_{i}\right)$ has the expectation $\mathbb{E}\left[R_{i-1}^*\right]=\sum_{m=1}^{i-1}h_m\alpha_m$;
	\item so-constructed  $R_{i-1}^*$ satisfies $\mathsf{supp}\, R_{i-1}^* \subseteq [0,\mathsf{H}_{\color{black}[i-1]}] $ and $R_{i-1}^* \le_{\textnormal{cx}} \bar{S}_{\mathbf{h},\hat{\bm{\alpha}}^{(i-1)}}$;
\end{enumerate}

Now we prove the above two statements. Denote the SLT of $R_{i}^*$ by $\pi_{\scriptscriptstyle\! {\color{black}R_{i}^*}}(t)$. Using Proposition ~\ref{prop:SLT}, we obtain
\begin{flalign}\label{eqn:expectation}
	\mathbb{E}\left[\varphi\left(R_{i}^*,v,h_{i}\right)\right]=\pi_{\scriptscriptstyle\! R_{i}^*}(0)-\pi_{\scriptscriptstyle\! R_{i}^*}(v)+\pi_{\scriptscriptstyle\! R_{i}^*}(v+h_{i})
\end{flalign}
which is continuous and nondecreasing on the closed interval $v\in\left[0,\mathsf{H}_{\color{black}[i-1]}\right]$.

Noting that $R_{i}^* \le_{\textnormal{cx}} \bar{S}_{\mathbf{h},\hat{\bm{\alpha}}^{(i)}}$, we have 
\begin{flalign}\label{comb_inequality}
	\pi_{\scriptscriptstyle\! R_{i}^*} \left({{\mathsf{H}}}_{\mathcal{J}}\right)\le \bar{\pi}_{\mathbf{h},\hat{\bm{\alpha}}^{(i)}} \left({{\mathsf{H}}}_{\mathcal{J}}\right) \le  \sum_{k\in \mathcal{J}^{\rm c}}h_k \hat{\alpha}_{k}^{(i)}
\end{flalign}
for any index set $\mathcal{J}\subseteq {\color{black}\left[\nt\right]}$, 
where the second inequality follows from the $(\mathbf{h},\hat{\bm{\alpha}}^{(i)})$-decomposability of $\bar{S}_{\mathbf{h},\hat{\bm{\alpha}}^{(i)}}$.
Substituting $v=0$ into (\ref{eqn:expectation}), we have
\begin{flalign}
	\mathbb{E}\left[\varphi\left(R_{i}^*,0,h_{i}\right)\right] 
	&=\pi_{\scriptscriptstyle\! R_{i}^*}\left(h_{i}\right) \nonumber \\
	&\le  \sum_{m=1}^{i-1}h_m\hat{\alpha}_{m}^{(i)}+\sum_{m=i+1}^{n}h_m\hat{\alpha}_{m}^{(i)} \label{eqn:87}\\
	&=\sum_{m=1}^{i-1}h_m \alpha_m ,\label{eqn:88}
\end{flalign}
where (\ref{eqn:87}) follows from the inequality (\ref{comb_inequality}) by letting $\mathcal{J}=\{i\}$ and (\ref{eqn:88}) follows from the definition of $\hat{\bm{\alpha}}^{(i)}$. In the same way, we have
\begin{flalign}
	\mathbb{E}\left[\varphi\left(R_{i}^*,\mathsf{H}_{\color{black}[i-1]},h_{i}\right)\right] 
	&=\pi_{\scriptscriptstyle\! R_{i}^*}(0)-\pi_{\scriptscriptstyle\! R_{i}^*}(\mathsf{H}_{\color{black}[i-1]})+\pi_{\scriptscriptstyle\! R_{i}^*}(\mathsf{H}_{\color{black}[i]}) \nonumber \\
	&=\pi_{\scriptscriptstyle\! R_{i}^*}(0)-\pi_{\scriptscriptstyle\! R_{i}^*}(\mathsf{H}_{\color{black}[i-1]})\label{eqn:91}  \\
	&\ge \sum_{m=1}^{i}h_m\alpha_m-h_{i} \alpha_{i}\\
	&= \sum_{m=1}^{i-1}h_m\alpha_m,
\end{flalign}
where (\ref{eqn:91}) follows from
 $\mathsf{supp}\, R_{i}^* \subseteq [0,\mathsf{H}_{\color{black}[i]}]$. Hence, by the continuity of $\mathbb{E}\left[\varphi\left(R_{i}^*,v,h_{i}\right)\right]$ on the closed interval $\left[0,\mathsf{H}_{\color{black}[i-1]}\right]$, there must exist a solution $v_i^*\in\left[0,\mathsf{H}_{\color{black}[i-1]}\right]$ {\color{black}(possibly non-unique)} to the following equation
\begin{flalign}
	&\mathbb{E}\left[\varphi\left(R_{i}^*,v,h_{i}\right)\right]=\sum_{m=1}^{i-1}h_m\alpha_m,
\end{flalign}
which is equivalent to
\begin{flalign}\label{eq:145}
	\pi_{\scriptscriptstyle\! R_{i}^*}(v)-\pi_{\scriptscriptstyle\! R_{i}^*}(v+h_{i})=h_{i}\alpha_{i}.
\end{flalign}
Then we let $R_{i-1}^*=\varphi\left(R_{i}^*;v_i^*,h_{i}\right)$ and the SLT of  $R_{i-1}^*$ is given by Proposition \ref{prop:SLT} as follows 
\begin{flalign}\label{eqn:94}
	\pi_{\scriptscriptstyle\! R_{i-1}^*}(t)=	
	\begin{cases}
		\pi_{\scriptscriptstyle\! R_{i}^*}(t)-h_{i}\alpha_{i} &,~\mbox{$t \le v_i^*$},\\
		\pi_{\scriptscriptstyle\! R_{i}^*}(t+h_{i})&,~\mbox{$ t > v_i^*$},
	\end{cases}	
\end{flalign}
where the first equality follows from $ \pi_{\scriptscriptstyle\! R_{i}^*}(v_i^*)-\pi_{\scriptscriptstyle\! R_{i}^*}(v_i^*+h_{i})=h_{i}\alpha_{i}$.

Clearly, there exists a unique $\tau\in\{ 1,2,\cdots, i-1\}$ such that $\mathsf{H}_{\color{black}[{\tau-1}]}<v_i^{*} \le \mathsf{H}_{\color{black}[{\tau}]}$.
To prove that so-constructed $R_{i-1}^*$ satisfies $\mathsf{supp}\, R_{i-1}^* \subseteq [0,\mathsf{H}_{\color{black}[{i-1}]}] $ and $R_{i-1}^* \le_{\textnormal{cx}} \bar{S}_{\mathbf{h},\hat{\bm{\alpha}}^{(i-1)}}$, we will alternatively provide a case-by-case proof of $\pi_{\scriptscriptstyle\! R_{i-1}^*}\left(\mathsf{H}_{\color{black}[{k}]}\right)\le \bar{\pi}_{\mathbf{h},\hat{\bm{\alpha}}^{(i-1)}} \left(\mathsf{H}_{\color{black}[{k}]}\right)$ for all $k\in {\color{black}\left[\nt\right]}$ as follows.

\begin{itemize}
	\item Case 1: $k\ge \tau$. Note that  $\mathsf{H}_{\color{black}[{k}]}\ge \mathsf{H}_{\color{black}[{\tau}]}\ge v_i^*$. If $k\ge i-1$, then, plugging $t=\mathsf{H}_{\color{black}[{k}]}$ into (\ref{eqn:94}) and utilizing the {\color{black}monotonicity} of SLT, we get
	\begin{flalign}
		\pi_{\scriptscriptstyle\! R_{i-1}^*}\left(\mathsf{H}_{\color{black}[{k}]}\right)&=\pi_{\scriptscriptstyle\! R_{i}^*}\left(\mathsf{H}_{\color{black}[{k}]}+h_{i}\right)
		\nonumber \\
		&\le  \pi_{\scriptscriptstyle\! R_{i}^*}\left(\mathsf{H}_{\color{black}[{i}]}\right)
		\nonumber \\
		&=0.
	\end{flalign}
	Due to the nonnegativity of SLT, we have $\pi_{\scriptscriptstyle\! R_{i-1}^*}\left(\mathsf{H}_{\color{black}[{k}]}\right)=0$ if $k\ge i-1$.
	Otherwise, we have
	\begin{flalign}
		\pi_{\scriptscriptstyle\! R_{i-1}^*}\left(\mathsf{H}_{\color{black}[{k}]}\right)&=\pi_{\scriptscriptstyle\! R_{i}^*}\left(\mathsf{H}_{\color{black}[{k}]}+h_{i}\right)
		\nonumber \\
		&\le  \sum_{m=k+1}^{i-1}h_{m}\hat{\alpha}_{m}^{(i)}+\sum_{m=i+1}^{\nt}h_{m}\hat{\alpha}_{m}^{(i)} \label{eq:149}\\
		&\le  \sum_{m=k+1}^{i-1}h_{m}\hat{\alpha}_{m}^{(i)} \nonumber  \\
		&=\sum_{m=k+1}^{\nt}h_m\hat{\alpha}_{m}^{(i-1)},
	\end{flalign}
	where \eqref{eq:149} follows from (\ref{comb_inequality}) by letting $\mathcal{J}=\left\{1,2,\cdots,k,i \right\}$.
	\item Case 2: $k\le \tau-1$. We have $\mathsf{H}_{\color{black}[{k}]}< v_i^*$, and hence,
	\begin{flalign}
		\pi_{\scriptscriptstyle\! R_{i-1}^*}\left(\mathsf{H}_{\color{black}[{k}]}\right)&=\pi_{\scriptscriptstyle\! R_{i}^*}\left(\mathsf{H}_{\color{black}[{k}]}\right)-h_{i}\alpha_{i} \nonumber \\
		&\le  \sum_{m=k+1}^{\nt}h_{m}\hat{\alpha}_{m}^{(i)}-h_{i}\alpha_{i} \nonumber \\
		&=\sum_{m=k+1}^{\nt}h_m\hat{\alpha}_{m}^{(i-1)}, \label{eqn:117}
	\end{flalign}
	where (\ref{eqn:117}) follows from the fact that $k+1\le \tau \le i-1$.
\end{itemize}

In summary, we have derived that
\begin{flalign}
	&\pi_{\scriptscriptstyle\! R_{i-1}^*}(0)= \bar{\pi}_{\mathbf{h},\hat{\bm{\alpha}}^{(i-1)}}(0)=\sum_{m=1}^{i-1}h_i\alpha_i,\\
	&	\pi_{\scriptscriptstyle\! R_{i-1}^*}(1)= \bar{\pi}_{\mathbf{h},\hat{\bm{\alpha}}^{(i-1)}}(1)=0,\\
	&	\pi_{\scriptscriptstyle\! R_{i-1}^*}\left(\mathsf{H}_{\color{black}[{k}]}\right)\le \bar{\pi}_{\mathbf{h},\hat{\bm{\alpha}}^{(i-1)}}\left(\mathsf{H}_{\color{black}[{k}]}\right),~\forall \,k\in {\color{black}\left[\nt\right]}.
\end{flalign} 
Then, in view of the convexity of SLT and piecewise linearity of $\bar{\pi}_{\mathbf{h},\hat{\bm{\alpha}}^{(i-1)}}\left(t\right)$, we have
\begin{flalign}
	\pi_{\scriptscriptstyle\! R_{i-1}^*}(t)\le \bar{\pi}_{\mathbf{h},\hat{\bm{\alpha}}^{(i-1)}}(t),~t\in [0,1],
\end{flalign}
and thus $R_{i-1}^* \le_{\textnormal{cx}} \bar{S}_{\mathbf{h},\hat{\bm{\alpha}}^{(i-1)}}$. Considering that $\pi_{\scriptscriptstyle\! R_{i-1}^*}(\mathsf{H}_{\color{black}[{i-1}]})\le \pi_{\mathbf{h},\hat{\bm{\alpha}}^{(i-1)}}\left(\mathsf{H}_{\color{black}[{i-1}]}\right)=0$ and $R_{i-1}^*\ge 0$, we have $\mathsf{supp}\,R_{i-1}^*\subseteq [0,\mathsf{H}_{\color{black}[{i-1}]}]$.  

Conduct the above iterative construction of $R_{i-1}^*$ from $i=\nt$ to $i=2$ and  let
\begin{flalign}
	X_k^{\rm c}&\triangleq  \frac{R_{k}^*-R_{k-1}^*}{h_{k}}\nonumber \\
	&=\frac{R_{k}^*-\varphi\left(R_{k}^*;v_k^*,h_{k}\right)}{h_{k}}\label{eq:112}
\end{flalign} 
for $k\in {\color{black}\left[\nt\right]}$.

Considering the definition of $\varphi \left( x;v,z\right)$ and the fact $\mathsf{supp}\,R_{{\color{black}k}}^*\subseteq [0,\mathsf{H}_{\color{black}[{{\color{black}k}}]}]$, we have $\mathsf{supp}\, X_k^{\rm c} \subseteq [0,1]$ and $\mathbb{E}\left[X_k^{\rm c}\right]=\alpha_k$ for all $k\in {\color{black}\left[\nt\right]}$. Furthermore, we note that: 1) (\ref{eq:112}) implies the comonotonicity between $X_k^{\rm c}$ and $R_{k}^*$; 2) $R_{k-1}^*=\varphi\left(R_{k}^*;v_k^*,h_{k}\right)$ implies the comonotonicity between $R_{k}^*$ and $R_{k-1}^*$. Thus, we conclude that $\mathbf{X}^{\rm c}=\trans{\left(X_1^{\rm c},\cdots,X_{\nt}^{\rm c}\right)}$ is comonotonic and $S=\sum_{k=1}^{\nt} h_kX_k^{\rm c}$ is comonotonically $\left(\mathbf{h},\bm{\alpha}\right)$-decomposable.

\subsection{Proof of Theorem~\ref{main2}}\label{sec:proof_main2}
\label{app:exp-bcmc}
{\color{black}We first show 
	$1\ \Longrightarrow \ 2$. Let $\mathbf{a}$ be an arbitrary vector satisfying $\bm{0} \preccurlyeq \mathbf{a} \preccurlyeq  \bm{\alpha}$ and $S$ can be decomposed as $S=\sum_{k=1}^{\nt}h_kX_k$ with $\mathsf{supp}\, X_k \subseteq [0,1]$ and $\mathbb{E}[X_k]=a_i$ for each $k\in [\nt]$. Then we have}
\begin{flalign}
	\mathbb{E}\left[\left(S-\mathsf{H}_{\color{black}[{k}]}\right)_+\right]&= 
	\pi_{\scriptscriptstyle\! S} (\mathsf{H}_{\color{black}[{k}]}) \nonumber  \\
	&\le \sum_{j=k+1}^{\nt} h_j\mathbb{E}\left[X_j\right] \label{eq:73}\\
	&\le \sum_{j=k+1}^{\nt} h_j\alpha_j \nonumber \\
	&=\left(1-\mathsf{H}_{\color{black}[{k}]}\right) \bar{\alpha}_k,~{\color{black} k\in \{0\} \cup \left[\nt-1\right]},
\end{flalign}
where~\eqref{eq:73} follows from {\color{black}Theorem~\ref{main}}. 

Next, {\color{black}we will show $2\ \Longrightarrow \ 3$. For any random variable $S$ with $\mathsf{supp}\, S\subseteq [0,1]$ and satisfying constraints~\eqref{eq:70}, we let} $\mathbb{E}\left[S\right]{\color{black}=\pi_{\scriptscriptstyle \! S}(0)}=\varepsilon$, $\varepsilon_0=0$ and $${\color{black}\varepsilon_j=\left( \mathsf{H}_{\color{black}[{\nt-j+1}]} \right) \times \left( \alpha_{\nt-j+1} \right)
	+\left(1-\mathsf{H}_{\color{black}[{\nt-j+1}]}\right) \times 
	\left( \bar{\alpha}_{\nt-j+1}\right)
	,~\forall \,} j\in {\color{black}\left[\nt\right]}.$$ 
Then there must exist a unique integer $ \tau \in {\color{black}\left[\nt\right]}$ such that $\varepsilon \in \left(\varepsilon_{\tau-1},\varepsilon_{\tau}\right]$. 
Let 
\begin{flalign}\label{eq:117}
	\mathbf{a}^{\varepsilon}={\color{black} \trans{\left(a_{1}^{\varepsilon},\cdots,a_{\nt}^{\varepsilon}\right)}}=\varepsilon \bm{1},~\text{if}~\tau=1;
\end{flalign}
and otherwise
\begin{flalign}\label{eq:118}
	a^{\varepsilon}_{m}=
	\begin{cases}
		\alpha_{m} &,\mbox{ $\nt-\tau+2 \le m\le \nt$},\\
		{\color{black}
			\left(\varepsilon- \sum_{m=\nt-\tau+2}^{\nt}h_ma^{\varepsilon}_{m}\right)/\mathsf{H}_{\color{black}[{\nt-\tau+1}]} } &,\mbox{ others}.\\
	\end{cases}
\end{flalign}
Note that the so-constructed $\mathbf{a}^{\varepsilon}$ satisfies $a_1^{\varepsilon} \ge a_2^{\varepsilon} \ge \cdots a^{\varepsilon}_{\nt}$ and $\bm{0} \preccurlyeq \mathbf{a}^{\varepsilon} \preccurlyeq  \bm{\alpha}$. For $k\ge \nt-\tau+1$, the inequalities~\eqref{eq:70} imply
\begin{flalign}
	\pi_{\scriptscriptstyle \! S}(\mathsf{H}_{\color{black}[{k}]}) &\le \sum_{j=k+1}^{\nt} h_j\alpha_j= \sum_{j=k+1}^{\nt} h_j a_j^{\varepsilon},
\end{flalign}
while for $k\le \nt-\tau$ we have
\begin{flalign}
	\pi_{\scriptscriptstyle \! S}(\mathsf{H}_{\color{black}[{k}]})  &\le \left(1-\frac{\mathsf{H}_{\color{black}[{k}]}}{\mathsf{H}_{\color{black}[{\nt-\tau+1}]}}\right)\pi_{\scriptscriptstyle \! S}(0)
	+\frac{\mathsf{H}_{\color{black}[{k}]}}{\mathsf{H}_{\color{black}[{\nt-\tau+1}]}}\pi_{\scriptscriptstyle \! S}(\mathsf{H}_{\color{black}[{\nt-\tau+1}]}) \label{eq:76}
	\\
	&\le \left(1-\frac{\mathsf{H}_{\color{black}[{k}]}}{\mathsf{H}_{\color{black}[{\nt-\tau+1}]}}\right)\pi_{\scriptscriptstyle \! S}(0)
	+\frac{\mathsf{H}_{\color{black}[{k}]}}{\mathsf{H}_{\color{black}[{\nt-\tau+1}]}}\sum_{m=\nt-\tau+2}^{\nt}h_ma_{m}^{\varepsilon} \nonumber \\
	&=\sum_{m=\nt-\tau+2}^{\nt}h_ma_{m}^{\varepsilon}+\left(1-\frac{\mathsf{H}_{\color{black}[{k}]}}{\mathsf{H}_{\color{black}[{\nt-\tau+1}]}}\right)\left(\varepsilon-\sum_{m=\nt-\tau+2}^{n}h_ma_{m}^{\varepsilon}\right) \nonumber \\
	&=\sum_{m=\nt-\tau+2}^{\nt}h_ma_{m}^{\varepsilon}+\sum_{m=k+1}^{\nt-\tau+1}h_ma_m^{\varepsilon} \nonumber \\
	&=\sum_{m=k+1}^{\nt}h_ma_{m}^{\varepsilon},
\end{flalign}
where~\eqref{eq:76} follows from the convexity of the SLT. {\color{black}Theorem~\ref{main} immediately leads to the $\left(\mathbf{h},\mathbf{a}^{\varepsilon}\right)$-decomposability of $S$. It is clear that the set $\left\{ \mathbf{a}^\varepsilon: \varepsilon \in \left[ 0 ,  \trans{\mathbf{h}}\bm{\alpha} \right] \right\} $ can be reparameterized as
	\begin{flalign}
		\left\{ \mathbf{a}^\varepsilon: 0 \le \varepsilon \le \trans{\mathbf{h}}\bm{\alpha}\right\}
		=\left\{   \min\{\beta \bm{1},\bm{\alpha}\}:\beta\in [0,\alpha_1]	
		\right\}.
	\end{flalign}
	Therefore, $S$ must be also $\left(\mathbf{h},\mathbf{a}^{\dagger} \right)$-decomposable, where $	\mathbf{a}^{\dagger} =   \min\left\{\beta \bm{1},\bm{\alpha}\right\} $ with some $\beta \in \left[0,\alpha_1 \right]$ such that  $\trans{\mathbf{h}}\mathbf{a}^{\dagger}=\mathbb{E}[S]$.

	The proof is concluded by noticing $\bm{0} \preccurlyeq \mathbf{a}^{\dagger} \preccurlyeq \bm{\alpha}$, which implies $3\ \Longrightarrow  1$.}

\section{Signaling for {\color{black}EC-OIC} and {\color{black}BC-OIC}}\label{sec: signaling}
In this section, we discuss the signaling problem for a given equivalent input $S$ feasible to the {\color{black}EC-OIC or the BC-OIC}. The optimization of equivalent inputs or channel inputs should be paid particular attention and are not studied here{\color{black}; for example, in} \cite{zhang18_1}, an optimal spatial constellation in terms of the received minimum Euclidean distance is proposed for the case with or without {\color{black}channel state information at the transmitter}.
	
We have seen that our capacity results can be derived once we prove the equivalence between a MISO OIC with per-antenna {\color{black}intensity} constraints and a SISO channel with an amplitude constraint and several {\color{black}stop-loss mean} constraints. However, in practical signaling, it is vital for each transmitter to know its corresponding component {\color{black}obtained by decomposing} an $(\mathbf{h},\bm{\alpha})$-decomposable random variable $S$ rather than $S$ itself. Hence, an algorithm to efficiently decompose $S$ into {\color{black}a} feasible channel input $\mathbf{X}^{\rm c}$ will be useful.

In the proof of Theorem~\ref{main} ({\color{black}see Sec.~\ref{sec:proof_main}}), {\color{black}a constructive method, namely greedy decomposition, iteratively computes $\nt$ comonotonic components of an $(\mathbf{h},\bm{\alpha})$-decomposable random variable, which relies on greedily-constructed quantile functions with $\nt-1$ parameters $v_{\nt}^{*}$, $v_{\nt-1}^{*}$, $\cdots$, and $v_2^*$ to be determined ($v_1^{\ast}=0$). To assist the reader, we excerpt this iterative algorithm from Sec.~\ref{sec:proof_main} as follows.
\begin{algorithm}[Iteration-based greedy decomposition] 
	\label{Algorithm:iteration}
	Conditioned on $S=s$ for any given $s\in[0,1]$, the greedy decomposition computes  instantaneous intensity signals of $\nt$ transmitters as
\begin{flalign}
		x_{\nt}^{\rm{c}}=\frac{s-\varphi\left(s;v_{\nt}^*,h_{\nt}\right)}{h_{\nt}}	
\end{flalign}
and
\begin{flalign}
	x_{k-1}^{\rm{c}}=\frac{\left(s-\sum_{i=k}^{\nt}h_ix_i^{\rm{c}}\right)-\varphi\left(s-\sum_{i=k}^{\nt}h_ix_i^{\rm{c}};v_{k-1}^*,h_{k-1}\right)}{h_{k-1}}	
\end{flalign}
for $k\in \left[\nt\right] \setminus \{1\}$.
\end{algorithm}
}

{\color{black}We remind that the parameters $v_{\nt}^{*}$, $v_{\nt-1}^{*}$, $\cdots$, and $v_2^*$ are solutions to equations~\eqref{eq:145}}, which needs the SLT $\pi_{\scriptscriptstyle\! R_{i-1}^*}(t)$ (given in~\eqref{eqn:94}) of the remainder part $R_{i-1}^*$ in each step.
In what follows, we present an alternative perspective {\color{black}on} the greedy decomposition via partitioning the interval $[0,1)$, and {\color{black}show that the solutions to equations~\eqref{eq:145} can also be acquired} only based on the SLT of $S$.\footnote{By convention, we adopt a left-closed-right-open interval $[0,1)$ {\color{black}for a unified representation of sets obtained by finite union and set difference in this section.}}


\begin{proposition}[Interval partition]
	\label{Prop:Partition}
	Let $S$ be an $(\mathbf{h},\bm{\alpha})$-decomposable random variable, $\kappa_{\nt}^*=v_{\nt}^*$, $\eta_{\nt}(\kappa)=\kappa+h_{\nt}$ and the interval $\mathcal{P}_{\nt}=\left[\kappa_{\nt}^*,\eta_{\nt}(\kappa_{\nt}^*)\right)$. For $i\in\{\nt,\cdots,2\}$, $\kappa_{i-1}^*$ is a solution to the equation 
	\begin{flalign}\label{eq:65}
		\pi_{\scriptscriptstyle \! S}(\kappa)-\pi_{\scriptscriptstyle \! S}\left(\eta_{i-1}(\kappa)\right)
		=\sum_{m=i-1}^{i-1+N_{i-1}(\kappa)}h_{m}\alpha_{m},~\kappa \in\left[0,\min \left\{\kappa_i^*,\mathsf{H}_{\color{black}[i-1]}\right\}\right],
	\end{flalign}
	where {\color{black}the quantity $\eta_{i-1}(\kappa)$ satisfies} $\kappa+h_{i-1} \le \eta_{i-1}(\kappa) \le 1$ {\color{black}and} is uniquely determined by the Lebesgue measure
	\begin{flalign}
		\mu\left(\left[\kappa,\eta_{i-1}(\kappa) \right]\setminus \bigcup_{m=i}^{\nt}\mathcal{P}_i  \right)=h_{i-1},
	\end{flalign}
	the Borel sets
	\begin{flalign}
		\mathcal{P}_{j}=\left[\kappa_{j}^*,\eta_{j}(\kappa_{j}^*) \right)\setminus \bigcup_{m=j+1}^{\nt}\mathcal{P}_m,~j\le \nt-1,
	\end{flalign}
	and 
	\begin{flalign}
		N_{i-1}(\kappa)=\left|\left\{j:\kappa_j^*\le \eta_{i-1}(\kappa), i-1< j\le \nt  \right\}\right|.
	\end{flalign}
	Then, for each $k\in {\color{black}\left[\nt\right]}$, $\kappa_k^*$ is also a solution to~\eqref{eq:145}.
\end{proposition}
{\color{black}
\begin{IEEEproof}
	See Appendix~\ref{app:Partition}.
\end{IEEEproof}

}

For any given $(\mathbf{h},\bm{\alpha})$-decomposable $S$, {\color{black}we will assign the $k$-th one of the above $\nt$ disjoint sets, i.e., $\mathcal{P}_k$ of length $h_k$, to the $k$-th transmitter}. Recall that we let $R_{k-1}^*=\varphi\left(R_{k}^*;\kappa_k^*,h_{k}\right)$ and $X_k^{\rm c}
=\left(R_{k}^*-R_{k-1}^*\right)/{h_{k}}$ in the greedy decomposition. Then the definition of the parametric function $\varphi\left(s;v,z\right)$ immediately implies that when $\kappa_k^* \le R_{k}^* <\kappa_k^*+h_{k} $ the variation of $R_{k}^*$ can be regarded as solely changing $X_k^{\rm c}$ and keeping $R_{k-1}^*$ invariant. Motivated by this fact, via mathematical induction  we can prove that the continuous variation of the random variable $S$ in the set $\mathcal{P}_k$ is solely determined by the $k$-th transmitter for each $k\in {\color{black}\left[\nt\right]}$.\footnote{We omit the induction-based proof for brevity since its involved techniques are similar to those in Appendix~\ref{app:Partition}.} Hence, we obtain the following decomposition algorithm, {\color{black}which implements the greedy decomposition as well.}

\begin{algorithm}[{\color{black}Partition-based greedy} decomposition] 
	\label{Algorithm:partition}
	Conditioned on $S=s$ for any given $s\in[0,1]$, the greedy decomposition performs $x_{k}^{\rm{c}}=0$ if $s\le  \kappa_k^*$, and otherwise
	\begin{flalign}\label{eq:68}
		x_{k}^{\rm{c}}=
		\frac{\mu \left\{p\in \mathcal{P}_k:p\le s \right \}}{h_{k}}, ~\forall \, k\in {\color{black}\left[\nt\right]}.	
	\end{flalign}
\end{algorithm}

{\color{black}From a geometric perspective, the signal sent by the $k$-th transmitter is exactly} a scaled length of the part of $\mathcal{P}_k$ with elements no larger than the realization $s$. Since each set $\mathcal{P}_k$ is a finite (at most $\nt-k+1$) union of disjoint intervals, the computational complexity involved in~Algorithm~\ref{Algorithm:partition} is low.

\begin{remark}
	Note that the sets $\mathcal{P}_k$ {\color{black}obtained in Proposition~\ref{Prop:Partition}} depend on the distribution of the equivalent input $S$. For a MISO OIC with inputs under {\color{black}a total average-intensity constraint and peak-intensity} constraints~\cite{moserwangwigger18_3}, given a feasible equivalent input $S=\trans{\mathbf{h}}\mathbf{X}$ and sorting all transmitters in descending order $h_1\ge \cdots \ge h_{n_t}$, the inputs can be directly obtained by $X_k=(S-\mathsf{H}_{\color{black}[k]})_+/h_k$ for $k\in {\color{black}\left[\nt\right]}$. Those piecewise linear functions can be regarded as a special case of~(\ref{eq:68}) with a simple and deterministic interval partition $\mathcal{P}_k=\left[\mathsf{H}_{\color{black}[k-1]},\mathsf{H}_{\color{black}[k]}\right)$ for $k\in {\color{black}[\nt]}$, regardless of the distribution of $S$. A similar phenomenon arises in its MIMO counterpart~\cite{limoserwangwigger20_1}, where the inputs can be computed according to {\color{black}a deterministic partition of the zonotope generated by all column vectors of the channel matrix.}\footnote{{\color{black}A zonotope (also named ``zonohedra'') generated by $n$ vectors $\mathbf{a}_1,~\mathbf{a}_2,\cdots,~\mathbf{a}_{n}\in \mathbb{R}^{m}$ is the Minkowski sum $\left\{\sum_{i=1}^{n}x_i \mathbf{a}_i: 0 \le x_i \le 1 \right\}$ \cite{zamirfeder98_1}.}} This observation reflects a fundamental difference between {\color{black}OICs with per-antenna average-intensity constraints and those with a total average-intensity constraint}.
\end{remark}

{\color{black}Clearly, Algorithms 1 and 2 carry out the greedy decomposition method and compute comonotonic components  of an $\left(\mathbf{h},\bm{\alpha}\right)$-decomposable random variable $S$. The difference between the two algorithms is that the former outputs in an iterative manner while the latter in a direct manner. Both algorithms} can be applied directly to the {\color{black}EC-OIC}. For the {\color{black}BC-OIC}, the signaling problem can be easily addressed after an additional step. {\color{black}Given a random variable $S$ feasible to the BC-OIC, we calculate the auxiliary average-intensity vector  $	\mathbf{a}^{\dagger} =   \min\left\{\beta \bm{1},\bm{\alpha}\right\} $ with some $\beta \in \left[0,\alpha_1 \right]$ such that  $\trans{\mathbf{h}}\mathbf{a}^{\dagger}=\mathbb{E}[S]$. As shown in Theorem~\ref{main2}, the inequality $\bm{0} \preccurlyeq \mathbf{a}^{\dagger} \preccurlyeq  \bm{\alpha}$ holds and  $S$ is $\left(\mathbf{h},\mathbf{a}^{\dagger}\right)$-decomposable so that Algorithms 1 and 2 can be applied as well}.

At the end of this section, we present several examples to illustrate {\color{black}signaling procedures for the EC-OIC and the BC-OIC.}

\begin{example}[Signaling for the maximally convex distribution] By solving~\eqref{eq:65} for $\bar{S}_{\mathbf{h},\bm{\alpha}}$, {\color{black}the} partition is simply given by
	$\mathcal{P}_k=\left[\mathsf{H}_{\color{black}[k-1]},\mathsf{H}_{\color{black}[k]}\right),~k\in {\color{black}\left[\nt\right]}$. Algorithm~\ref{Algorithm:partition} outputs {\color{black}the} maximally correlated $\nt$-variate binary distributions as well, i.e., the same {\color{black}as} the decomposition results in Proposition~\ref{Prob:MCD}.
\end{example}

\begin{figure*}[!htbp]
	\vspace{-5mm}
	\centering
		\begin{minipage}{11.5cm}
			\centering\includegraphics [width=11.5cm]{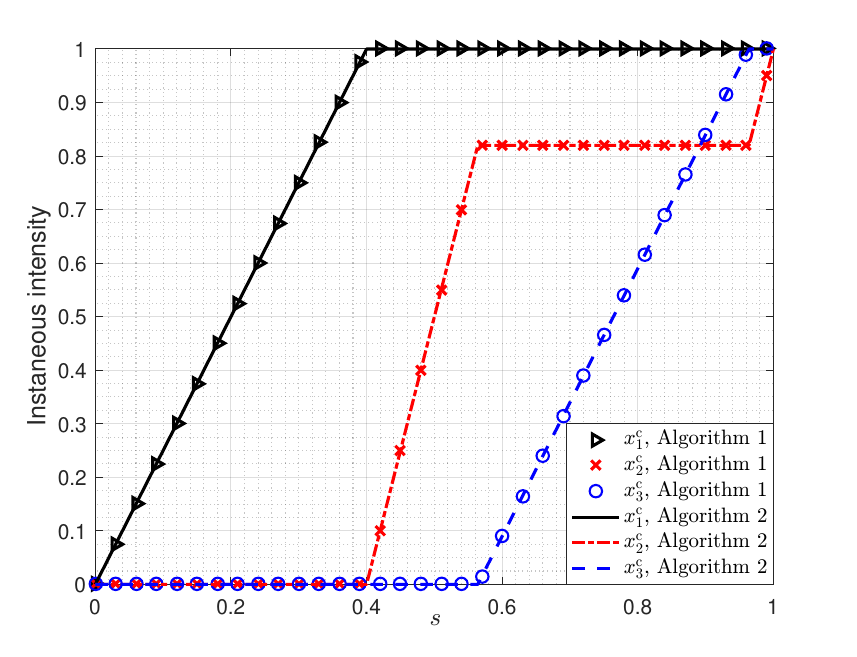}
		\end{minipage}
\vspace{-3mm}

		\begin{minipage}{9.4cm}
			\centering\includegraphics [width=9.4cm]{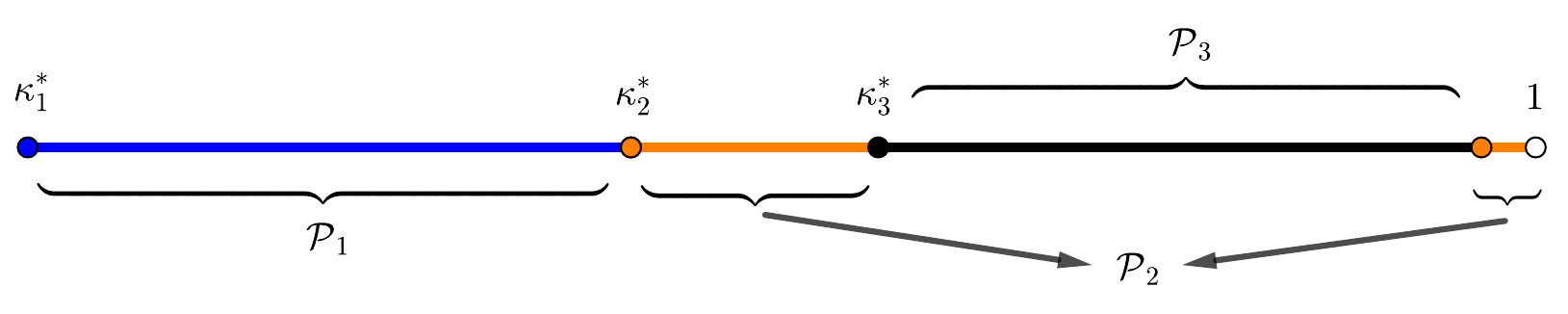}
		\end{minipage}
	\centering\caption{{\color{black}Instantaneous transmitted signals conditioned on the equivalent input signal $s$ and interval} partition for a maximum-entropy distribution.} \label{fig:partition}
\end{figure*}

\begin{example}[{\color{black}Signaling} for the maximum-entropy distribution]
	Consider a $3\times 1$ {\color{black}EC-OIC} with normalized {\color{black}channel gains $\mathbf{h}=\trans{\left(0.4,0.2,0.4\right)}$ and required average intensities} ${\color{black}\bm{\alpha}}=\trans{\left(0.8,0.3,0.1\right)}$. The maximum-entropy distribution is numerically computed as \llg{$p_{\scriptscriptstyle \! S}^*(s)=\exp ({-0.4286+2.9176s-6.5987(s-0.4)_+})$}.
		By solving~(\ref{eq:65}), we obtain relevant parameters as  $\kappa_3^*=0.564$, $\kappa_2^*=0.4$, and $\kappa_1^*=0$. Thus, $
		\mathcal{P}_3= \left[0.564,0.964\right),~
		\mathcal{P}_2= \left[0.4,0.564\right) \cup \left[(0.964,1\right)$ and $
		\mathcal{P}_1= \left[0,0.4\right)$, which are plotted {\color{black}at the bottom of} Figure~\ref{fig:partition}. {\color{black}Moreover, at the top of Figure~\ref{fig:partition}, we align the $s$-axis and plot the outputs of Algorithm 1 and Algorithm 2, which {\color{black}exactly match} each other. }
\end{example}

\begin{example}[Signaling for a mergeable $3 \times1 $ {\color{black}BC-OIC}]
		Consider a $3\times 1$ {\color{black}BC-OIC} with {\color{black}unnormalized} parameters: the channel gains  $\widetilde{\mathbf{h}}=\trans{\left(4\times 10^{-6} ,1.5\times 10^{-6},3\times 10^{-6}\right)}$, the maximum allowed peak {\color{black}intensities} $\bm{\amp}=\trans{\left(2 ,3,2.5 \right)}$, ratios of the maximum allowed average {\color{black}intensity} to corresponding maximum allowed peak {\color{black}intensity}  $\widetilde{\bm{\alpha}}=\trans{\left(0.4,0.1,0.1\right)}$, and the AWGN standard deviation $\tilde{\sigma}$. 
	
	We first compute the normalized channel gains as $h_1'=0.4$, $h_2'=0.225$ and $h_3'=0.375$ by $h_k' = \tilde{h}_k \amp_k/\left(\sum_{i=1}^{\nt}\tilde{h}_i \amp_i\right)$ for all $ k\in{\color{black}[3]}$. Correspondingly, the normalized standard deviation of AWGN is $\sigma=\tilde{\sigma}/\left(\sum_{i=1}^{\nt}\tilde{h}_i \amp_i\right)$.
	
	Next, note that $\alpha_2=\alpha_3$ and thus we can let the second and the third transmitters send identical normalized signals. Based on Proposition \ref{lemma:mergeability}, the considered channel is equivalent to a $2\times 1$ {\color{black}BC-OIC} with normalized {\color{black}channel gains} $\mathbf{h}=\trans{\left(0.4,0.6\right)}$ and {\color{black}maximum allowed average intensities} $\bm{\alpha}=\trans{\left(0.4,0.1\right)}$.
	In what follows, we consider a continuous and two finite-alphabet distributions of the equivalent input, respectively.
	
	(1) Maximum-entropy distribution: 
		$p_{\scriptscriptstyle \! S}^*(s) = \exp \left(1.466-4.270s\right),~ s\in [0,1]$. We compute the expectation as $\mathbb{E}\left[S\right]=\trans{\mathbf{h}}\bm{\alpha}$ and thus $S$ is $\left(\mathbf{h},\bm{\alpha}\right)$-decomposable.
		Applying (\ref{eq:65}), we obtain $\kappa_2^*=0.272$, $\mathcal{P}_2= \left[0.272,0.872\right)$ and $
		\mathcal{P}_1= \left[0,0.272\right)\cup \left[0.872,1\right)$.
		
		Given a realization $s=0.2$ generated by $p_{\scriptscriptstyle \! S}^*(s)$, the first transmitter sends the unnormalized signal $x_1=\frac{s}{h_1}\times \amp_1=1$ and other two {\color{black}transmitters} send zero. 
		
		Given a realization $s=0.5$, the first transmitter sends $x_1=\frac{0.272}{h_1}\amp_1=1.36$, and other two {\color{black}transmitters} send $x_2=\frac{s-0.272}{h_2}\amp_2=1.14$ and $x_3=\frac{s-0.272}{h_2}\amp_3=0.95$. 
		
		Given a realization $s=0.9\ge \sup \mathcal{P}_2$, the second and the third transmitters send their peak values, i.e., $x_2=\amp_2=3$ and $x_3=\amp_3=2.5$, and the first transmitter sends $x_1=\frac{0.272+s-0.872}{h_1}\amp_1=1.5$.
		
		(2) Nonequiprobable OOK: $\mathcal{S}_{\textnormal{OOK}}=\{0,1\}$ with $\mathbb{P}\left\{S=1\right\}=0.1$. Note that the expectation $\mathbb{E}\left[S\right]=0.15{\color{black}\,\,=\trans{\mathbf{h}}\mathbf{a}^{\dagger}  }$, {\color{black}where $\mathbf{a}^{\dagger}=\trans{\left(0.1,0.1\right)}$}. Here $S$ can be regarded as a feasible input to a mergeable $2\times 1$ {\color{black}EC-OIC} with parameters $\mathbf{h}$ and {\color{black}$\mathbf{a}^{\dagger}$}, and thus, spatial repetition can be used {\color{black}so} that the signaling for $\mathcal{S}_{\textnormal{OOK}}$ is simplified as
		$X_1=\amp_1 S$, $x_2=\amp_2 S$ and $x_3=\amp_3 S$.
		
		(3) Equally spaced and equiprobable $8$ASK: $\mathcal{S}_{8\textnormal{ASK}}=\{0,\Delta,2\Delta,\cdots,7\Delta\}$, where the minimum Euclidean distance $\Delta$ is maximized as $\Delta=0.0629$. {\color{black}Note that the expectation of the equivalent input $S$ in this case is exactly} $\trans{\mathbf{h}}\bm{\alpha}$. {\color{black}Then} $S$ is $\left(\mathbf{h},\bm{\alpha}\right)$-decomposable {\color{black}due to Theorem~\ref{main2}}. Applying (\ref{eq:65}), we obtain $\kappa_2^*=0.226$, $\mathcal{P}_2= \left[0.226,0.826\right)$ and $
		\mathcal{P}_1= \left[0,0.226\right)\cup \left[0.826,1\right)$. Given $s=i\Delta$, the decomposition can be further simplified as
		$x_1=\frac{i\Delta}{h_1}\amp_1$ and $x_2=x_3=0$ if $i\le 3$, and otherwise $x_1=\frac{\kappa_2^*}{h_1}\amp_1$, $x_2=\frac{(i\Delta-\kappa_2^*)}{h_2}\amp_2$ and $x_3=\frac{(i\Delta-\kappa_2^*)}{h_2}\amp_3$.\footnote{For a specific distribution, a carefully designed decomposition algorithm may be more suitable than the greedy {\color{black}decomposition} from the considerations of practical signaling. For example, we can alternatively let 
		$X_1=\frac{\alpha_1}{h_1\alpha_1+h_2\alpha_2}\amp_1S$, $X_2=\frac{\alpha_2}{h_1\alpha_1+h_2\alpha_2}\amp_2S$ and $X_3=\frac{\alpha_2}{h_1\alpha_1+h_2\alpha_2}\amp_3S$ for $\mathcal{S}_{8\textnormal{ASK}}$.}
\end{example}

{\color{black}Inspired by the above examples, in order to design a good spatial constellation, we can concatenate our proposed algorithms to a well-designed constellation of the equivalent input in terms of Euclidean distance. This transmitter architecture may deserve further attention.}


\section{Conclusion}
\label{sec:conclusion}
In this paper, we consider two types of MISO {\color{black}OIC}s under per-antenna {\color{black}intensity} constraints.
We first express their capacities as a maximization problem
over distributions of the sum random variable $S = \trans{\vect{h}} \vect{X}$. The main
challenge is to equivalently transform per-antenna average-{\color{black}intensity constraints} on
$\vect{X}$ to constraints on $S$, since the mapping from $\vect{x}$
onto $s=\trans{\mathbf{h}}\vect{x}$ is not injective. This problem is solved by finding the {\color{black}necessary and sufficient conditions for two types of decomposition of a bounded and nonnegative random variable.} Specifically, we show that both types of original MISO OICs with per-antenna {\color{black}intensity} constraints can be transformed into
a SISO one with an {\color{black}amplitude constraint} and several {\color{black}stop-loss mean} constraints.\

{\color{black}Then we obtain several capacity bounds for our considered MISO OICs by alternatively bounding capacities of their corresponding equivalent SISO OICs.} {\color{black}U}sing EPI and duality upper-bounding techniques, we derive new lower and upper bounds on the capacity, and they asymptotically match at high SNR.
In the low-SNR regime, the capacity slope is shown to be proportional
to the maximum variance of the equivalent input $S$ under {\color{black}stop-loss mean}
constraints. We characterize several properties of the input distribution that maximizes
the variance. {\color{black}For the EC-OIC, the equivalent input of the maximum-variance input is the maximally convex distribution, which is an upper bound (in terms of convex ordering) for all feasible equivalent inputs. For the BC-OIC, the parameter of the maximum-variance input can be easily determined by several linear equations.}

{\color{black}
\section*{Acknowledgment} 
The authors wish to thank the Associate Editor and the anonymous reviewers for their valuable comments and suggestions that improved the quality of this paper.
}
\appendices

\section{Proof of Theorems~\ref{ecmc:lbd} and~\ref{bcmc:lbd}}\label{app:lbd}
	We will prove Theorem~\ref{ecmc:lbd} for the {\color{black}EC-OIC}, and the involved techniques are applied to the case of the {\color{black}BC-OIC} as well.  
	Rewrite the original entropy-maximizing problem under {\color{black}stop-loss} equality and inequality constraints~\eqref{eq:equiv_ecmc} as follows:
	
	\begin{equation}\label{prob:original}
		\begin{aligned}
			&\textnormal{minimize} ~~&&\int_0^1 p (s)\log p (s)\dd s\\
			&\textnormal{subject to}~~ && \quad ~~ p(s)\ge 0\\
			& ~&& \int_0^1   p (s) \dd s  =1\\
			& ~&& \int_0^1 s p (s) \dd s =\trans{\mathbf{h}}\bm{\alpha}\\
			& ~&&
			\int_0^1  \left(s-\mathsf{H}_{\color{black}[k]}\right)_+ 
			p (s) \dd s
			\le \left(1-\mathsf{H}_{\color{black}[k]}\right) \bar{\alpha}_k, ~\forall\, k\in {\color{black}\left[\nt-1\right]}
		\end{aligned}
	\end{equation}
	
	Then we consider the Lagrangian functional\footnote{Here we let a Lagrange multiplier be $-({\color{black}\nu}_0+1)$ instead of the conventional ${\color{black}\nu}_0$.}
	\begin{flalign}\label{174}
		L\left(p,{\color{black}\nu}_0,\bm{\lambda}\right)= & \int_0^1 p  \log p \dd s-({\color{black}\nu}_0+1)\left(\int_0^1 p \dd s-1\right)
		+\lambda_0\left(\int_0^1 sp \dd s-\trans{\mathbf{h}}\bm{\alpha}\right) \nonumber \\ 
		&+\sum_{i=1}^{\nt-1}\lambda_i
		\left( \int_0^1 \left(s-\mathsf{H}_{\color{black}[k]}\right)_+ 
		p \dd s
		-\left(1-\mathsf{H}_{\color{black}[k]}\right) \bar{\alpha}_k
		\right),
	\end{flalign}
	where numbers ${\color{black}\nu}_0,\lambda_0\in \mathbb{R}$ and $\lambda_1,\cdots,\lambda_{\nt-1}\ge 0$. By using the Euler-Lagrange equation, we obtain the following extremum condition for the functional $L\left(p,{\color{black}\nu}_0,\bm{\lambda}\right)$ from the perspective of calculus of variations:
	\begin{flalign}
		\frac{\partial L}{\partial p}= \log p-{\color{black}\nu}_0+\lambda_0s+\sum_{i=1}^{\nt-1}\lambda_i\left(s-\mathsf{H}_{\color{black}[k]}\right)_+ =0,
	\end{flalign}
	whose solution is
	\begin{flalign}\label{eq:175}
		p (s) = \exp \left(\nu_0 -\lambda_0  s-\left( \sum_{i=1}^{\nt -1} \lambda_i  \left(s-\mathsf{H}_{\color{black}[i]} \right)_+ \right)\right)>0,~ s\in [0,1].
	\end{flalign}
	
	{\color{black}Substituting}~\eqref{eq:175} into~\eqref{174}, we obtain the dual function as
	\begin{flalign}\label{eq:176}
		g\left( {\color{black}\nu}_0,\bm{\lambda}\right)= & - \sum_{i=1}^{\nt-1}\lambda_i\left(1-\mathsf{H}_{\color{black}[i]}\right)\bar{\alpha}_i
		+1+\nu_0-\lambda_0\bar{\alpha}_0 \nonumber\\
		&-\int_0^1 \exp \left(\nu_0-\lambda_0 s-\left( \sum_{i=1}^{\nt-1} \lambda_i \left(s-\mathsf{H}_{\color{black}[i]} \right)_+ \right)\right) \dd s .
	\end{flalign}
	Then we consider the Lagrange dual problem
	\begin{flalign}\label{prob:maxent}
		\gamma_{ \textnormal{E}} = & \min_{\nu_0,\lambda_0 \in \mathbb{R} \atop \lambda_1,\cdots,\lambda_{\nt -1 }\ge 0} -g\left( {\color{black}\nu}_0,\bm{\lambda}\right).
	\end{flalign}
	
	Since the duality gap between the primary problem~\eqref{prob:original} and its Lagrange dual~\eqref{prob:maxent} is zero, i.e., strong duality holds, the maximum differential entropy {\color{black}equals} the solution to the dual problem, i.e., $\gamma_{ \textnormal{E}} $.
	
	The proof is concluded by calculating the involved integration in~\eqref{eq:176} and using EPI to obtain the lower bound on the differential entropy {\color{black}$\hh(Y)$} of the  output $Y$~\cite{coverthomas06_1}.

\section{Proof of Theorems~\ref{ecmc:max-var-bnd} and~\ref{bcmc:max-var-bnd} }
\label{app:max-var-bnd}
By the maximum entropy argument, we have 
\begin{IEEEeqnarray}{c}
  \hh(Y) \leq \frac{1}{2}\log\bigl(2\pi e (\sigma^2+ \const{V}_{\textnormal{max}}^{\textnormal{E } (\textnormal{or B})}\left(\mathbf{h},\bm{\alpha}\right) )\bigl),  
\end{IEEEeqnarray}
where {\color{black}$\const{V}_{\max}^{\textnormal{E}}(\vect{h},\bm{\alpha})$ and $\const{V}_{\max}^{\textnormal{B}}(\vect{h},\bm{\alpha})$ denote} the maximum variances {\color{black}over all feasible equivalent input $S$} for the {\color{black}EC-OIC and the BC-OIC}, respectively. Hence, we only need to show $\const{V}_{\textnormal{max}}^{\textnormal{E }}\left(\mathbf{h},\bm{\alpha}\right)$ and $\const{V}_{\textnormal{max}}^{\textnormal{B}}\left(\mathbf{h},\bm{\alpha}\right)$ are exactly the RHS of~\eqref{eq:maxvar_ecmc} and~\eqref{eq:maxvar_bcmc}, respectively.
	
In the {\color{black}EC-OIC}, for any {\color{black}feasible} input $\mathbf{X}$, we have $S \le_{\textnormal{cx}}\bar{S}_{\mathbf{h},\bm{\alpha}}$. Then by Corollary \ref{var_order} and the fact 
\begin{IEEEeqnarray}{c}
\textnormal{Var}\left(\bar{S}_{\mathbf{h},\bm{\alpha}} \right)=\sum_{i=1}^{\nt}\sum_{j=1}^{\nt} h_ih_j \left(\min\{\alpha_i,\alpha_j\} -\alpha_i\alpha_j\right),
\end{IEEEeqnarray}
{\color{black}we have $\const{V}_{\textnormal{max}}^{\textnormal{E }}\left(\mathbf{h},\bm{\alpha}\right)=\textnormal{Var}\left(\bar{S}_{\mathbf{h},\bm{\alpha}} \right)$.} The proof {\color{black}of Theorem~\ref{ecmc:max-var-bnd}} is completed.

We now turn to the {\color{black}BC-OIC} case. Under {\color{black}bounded-cost constraints}~\eqref{eqn:bcc}, finding the maximum variance can be formulated as the following {\color{black}intensity} optimization problem with respect to the variables $\mathbf{a}$
\begin{equation}\label{prob:low}
	\begin{aligned}
		&\textnormal{maximize} ~~&&\const{V}_{\textnormal{max}}^{\textnormal{E}}\left(\mathbf{h},\mathbf{a}\right)\\
		&\textnormal{subject to}~~ && \bm{0} \preccurlyeq \mathbf{a} \preccurlyeq \bm{\alpha}
	\end{aligned}
\end{equation}
i.e., maximization over all feasible average{\color{black}-intensity vectors} for the {\color{black}BC-OIC}.

Next, we show the solution to the optimization problem~\eqref{prob:low} is $\mathbf{a}^{*}=\min\{\beta^{*} \bm{1},\bm{\alpha}\}$, and the maximum variance is the RHS of~\eqref{eq:maxvar_bcmc}.

Following the same arguments as in the proof of Theorem~\ref{main2}, we can show for any feasible input $\mathbf{X}$ with the average {\color{black}intensity} $\mathbf{a}$ satisfying  {\color{black}$\bm{0} \preccurlyeq \mathbf{a} \preccurlyeq \bm{\alpha}$}, its equivalent input signal $S$ must be $(\mathbf{h},{\color{black}\mathbf{a}^\dagger})$-decomposable, {\color{black} where the vector  $	\mathbf{a}^{\dagger} =   \min\left\{\beta \bm{1},\bm{\alpha}\right\} $ with some $\beta \in \left[0,\alpha_1 \right]$ such that  $\trans{\mathbf{h}}\mathbf{a}^{\dagger}=\trans{\mathbf{h}}\mathbf{a}$.} Following Theorem~\ref{main}, we have  $S \le_{\textnormal{cx}} \bar{S}_{\mathbf{h},{\color{black}\mathbf{a}^\dagger}}$. Hence, the following order relation holds
	\begin{flalign}
		\textnormal{Var}\left(\bar{S}_{\mathbf{h},{\color{black}\mathbf{a}^\dagger}} \right)\ge \textnormal{Var}\left(S \right),
	\end{flalign}
which implies the variance-maximized {\color{black}average-intensity vector} $\mathbf{a}^{\star}$ {\color{black}can be restricted to} $\mathbf{a}^{\star}=\min\{\beta \bm{1},\bm{\alpha}\}$ for $\beta\in [0,\alpha_1]$ as well (similar to the discussion in Remark~\ref{opt_power}).
%

We only need to consider the nontrivial case {\color{black}where} $\alpha_{\nt} < \frac{1}{2}$. Let $\nu(\beta)=\const{V}_{\textnormal{max}}^{\textnormal{E}}\left(\mathbf{h},\min\{\beta \bm{1},\bm{\alpha}\}\right)$. Note that the function $\nu(\beta)$ is continuous in the interval $\beta\in[0,\alpha_{\nt}]$, and {\color{black}piecewise} smooth. It is clear that $\nu(\beta)=\beta(1-\beta)$ for $\beta\in[0,\alpha_{\nt}]$, which is maximized at $\beta=\alpha_{\nt}$.
		Letting {\color{black}$k_\beta=\{k\in \left[\nt\right]:\alpha_k\ge \beta\}$}, then $\beta\in \left(\alpha_{{k_\beta}+1},\alpha_{k_\beta} \right]$ holds. We rewrite $\nu(\beta)$ as
		\begin{flalign}\label{eq:opt-var}
			\nu(\beta)=&\mathsf{H}^2_{\color{black}[{k_\beta}]}\beta \left(1-\beta\right)+2\mathsf{H}_{\color{black}[{k_\beta}]}\sum_{m=k_\beta+1}^nh_i\alpha_i(1-\beta) \nonumber \\
			&+ \underbrace{ \sum_{i=k_\beta+1}^{\nt}\sum_{j=k_\beta+1}^{\nt} h_ih_j \left(\min\{\alpha_i,\alpha_j\} -\alpha_i \alpha_j\right)}_{\mathsf{K}}
		\end{flalign}
	where the quantity $\mathsf{K}$ is a constant in the interval $\beta \in\left(\alpha_{{k_\beta}+1},\alpha_{k_\beta} \right]$.
		Then we obtain 	
		\begin{flalign}
			\nu^\prime(\beta)=\mathsf{H}^2_{\color{black}[{k_\beta}]} \left(1-2\beta\right)-2\mathsf{H}_{\color{black}[{k_\beta}]}\left(1-\mathsf{H}_{\color{black}[{k_\beta}]}\right)\bar{\alpha}_{k_\beta},~\beta \in \left(\alpha_{{k_\beta}+1},\alpha_{k_\beta} \right].
		\end{flalign}
		 {\color{black}For any pair $(\beta_1,\beta_2)$ satisfying $ \alpha_{\nt}<\beta_1\le \beta_2$, it is clear that} $k_{\beta_1}\ge k_{\beta_2}\ge 1$, $\mathsf{H}_{\color{black}[{k_{\beta_1}}]}\ge \mathsf{H}_{\color{black}[{k_{\beta_2}}]}>0$, and $\bar{\alpha}_{k_{\beta_1}}\le\bar{\alpha}_{k_{\beta_2}}$. We have
		\begin{flalign}
			\nu^\prime(\beta_2) 
			&=\mathsf{H}_{\color{black}[{k_{\beta_2}}]}\left(\mathsf{H}_{\color{black}[{k_{\beta_2}}]} \left(1-2\beta_2\right)-2\left(1-\mathsf{H}_{\color{black}[{k_{\beta_2}}]}\right)\bar{\alpha}_{k_{\beta_2}}\right)\nonumber \\
			&\le \mathsf{H}_{\color{black}[{k_{\beta_2}}]}\left(\mathsf{H}_{\color{black}[{k_{\beta_1}}]} \left(1-2\beta_1\right)-2\left(1-\mathsf{H}_{\color{black}[{k_{\beta_1}}]}\right)\bar{\alpha}_{k_{\beta_1}}\right) \nonumber \\
			&=\frac{\mathsf{H}_{\color{black}[{k_{\beta_2}}]}}{\mathsf{H}_{\color{black}[{k_{\beta_1}}]}}	\nu^\prime(\beta_1)
		\end{flalign}
		Hence, 
		if $\nu^\prime(\beta_1)\le 0$, we have $\nu^\prime(\beta_2)\le 0$. Thus, the optimal {\color{black}average intensity} satisfies
		\begin{flalign} \label{eq:opt-beta}
			\beta^*=\inf \left\{ \beta\in\left(\alpha_{\nt},\alpha_1\right]:\mathsf{H}_{\color{black}[{k_{\beta}}]} \left(1-2\beta\right)-2\left(1-\mathsf{H}_{\color{black}[{k_{\beta}}]}\right)\bar{\alpha}_{k_{\beta}} \le 0 \right\}.
		\end{flalign}	
		Substituting~\eqref{eq:opt-beta} into~\eqref{eq:opt-var}, the proof {\color{black}of Theorem~\ref{bcmc:max-var-bnd}} is completed.


\section{Proof of Theorems~\ref{ubd:ecc} and~\ref{ubd:bcc}  }
\label{app:dual}

We only prove Theorem~\ref{ubd:ecc} for the {\color{black}EC-OIC}, and the proof of Theorem~\ref{ubd:bcc} follows essentially the same arguments.

{\color{black}Let $W(\cdot|s)$ denote the distribution of the channel output conditioned on the equivalent input $s\in[0,1]$.} We evaluate the duality-based upper bound 
		\begin{align}\label{E101}
			\mathsf{C}_{\textnormal{E}}\left(\mathbf{h},\bm{\alpha},\sigma\right) \leq \sup_{{\color{black}\mathscr{P}_{\! S}} }\mathbb{E}_{{\color{black}S\sim \mathscr{P}_{\! S}}}\left[D\left(W\left(\cdot|S\right)||R\left(\cdot\right)\right)\right]
		\end{align}
		for the auxiliary density
		\begin{flalign}
			R(y) =
			\begin{cases} 
				\frac{\beta}{\sqrt{2\pi}\sigma}
				e^{-\frac{y^2}{2\sigma^2}} 
				& \textnormal{if } y \in (-\infty, 0),   
				\\[2mm]
				\frac{(1-\beta)}{\mathsf{P}} \cdot  e^ {-\sum_{i=0}^{\nt-1} \lambda_i \left(y-\mathsf{H}_{\color{black}[i]} \right)_+ } &\textnormal{if }  y\in [0,1+\delta],
				\\[1mm]
				\frac{\beta}{\sqrt{2\pi}\sigma}
				e^{-\frac{(y-1-\delta)^2}{2\sigma^2}}  
				&  \textnormal{if } y \in (1+\delta, \infty), 
			\end{cases}
		\end{flalign}
		with free parameters $\delta>0$,  $\lambda_0\in \mathbb{R}$, $\lambda_1,\cdots,\lambda_{\nt-1}\ge 0$, and $\beta\in(0,1)$ which will be specified later.

		We notice that 
		\begin{IEEEeqnarray}{rCl}
		\IEEEeqnarraymulticol{3}{l}{%
				-\int_{-\infty}^{0} W(y|s)\log R(y)   \dd y 
			}\nonumber\\*\quad%
			& = &  -\int_{-\infty}^{0}
			\frac{1}{\sqrt{2\pi}\sigma}
			e^{-\frac{(y-s)^2}{2\sigma^{2}}} 
			\left(\log{\frac{\beta}{\sqrt{2\pi}\sigma}} -
			\frac{y^2}{2\sigma^{2}}\right)   \dd y  \nonumber \\
			& =  &-\log\left(\frac{\beta}{\sqrt{2\pi}\sigma}\right)
			\mathcal{Q}\left({\frac{s}{\sigma}}\right)
			+ \frac{1}{2}\mathcal{Q}\left({\frac{s}{\sigma}}\right)
		        +\> \frac{1}{2}\left(\frac{s}{\sigma}\right)^2
			\mathcal{Q}\left({\frac{s}{\sigma}}\right)
			- \frac{s}{2\sigma}
			\phi\left(\frac{s}{\sigma}\right) 
			\IEEEeqnarraynumspace
			\nonumber \\
			& \leq &  -\left(\log
			\frac{\beta}{\sqrt{2\pi}{\sigma}}-
			\frac{1}{2}\right)\mathcal{Q}\left({{\frac{s}{\sigma}}}\right)
			\label{eq:25}
			\\
			& = &  - \log \frac{\beta}{\sqrt{2\pi e}{\sigma}}
			\cdot \mathcal{Q}\left({{\frac{s}{\sigma}}}\right),
			\label{eq:e207}
		\end{IEEEeqnarray}
		where 
		\begin{IEEEeqnarray}{c}
			\phi(s) = \frac{1}{\sqrt{2\pi}} e^{-\frac{s^2}{2}},
		\end{IEEEeqnarray}
		{\color{black}$\mathcal{Q}(\cdot)$ denotes Gaussian Q-function}, and \eqref{eq:25} holds because of 
		\begin{IEEEeqnarray}{c}\label{eq:78}
		0 \le \phi(\eta)-	\eta \mathcal{Q}(\eta) \le \frac{1}{\sqrt{2\pi}}, \quad \eta \ge 0.
		\end{IEEEeqnarray}
	Similarly,
		\begin{IEEEeqnarray}{rCl}
			-\int_{1+\delta}^{\infty} {W(y|s)\log{{R(y)} }}
		  \dd y
			& \leq &   - \log \frac{\beta}{\sqrt{2\pi e}\sigma}
			\cdot \mathcal{Q}\left({\frac{1+\delta-s}{\sigma}}\right).
			\label{eq:e208} 
		\end{IEEEeqnarray}		
		Moreover,
		\begin{IEEEeqnarray}{rCl}
			\IEEEeqnarraymulticol{3}{l}{%
				-\int_{0}^{1+\delta} {W(y|s)\log{{R(y)} }} \dd y
			}\nonumber\\*\quad%
			& = & -\int_{0}^{1+\delta}
			\frac{1}{\sqrt{2\pi}\sigma}
			e^{-\frac{(y-s)^2}{2\sigma^2}} 
		 \log\left(\frac{1-\beta}{\mathsf{P}}  \right)
			\dd y \nonumber\\
                    &&\hspace{0.1cm} +\int_{0}^{1+\delta}
			\frac{1}{\sqrt{2\pi}\sigma}
			e^{-\frac{(y-s)^2}{2\sigma^2}} 
			\left(\sum_{i=0}^{\nt-1} \lambda_i \left(y-\mathsf{H}_{\color{black}[i]} \right)_+ 
			\right) \dd y.
			\label{eq:E2061} 
			\IEEEeqnarraynumspace
		\end{IEEEeqnarray}	
		Rewrite the first term in the RHS of \eqref{eq:E2061} as 
		\begin{IEEEeqnarray}{rCl}
		\IEEEeqnarraymulticol{3}{l}{%
				-\int_{0}^{1+\delta}
				\frac{1}{\sqrt{2\pi}\sigma}
				e^{-\frac{(y-s)^2}{2\sigma^2}} 
				\log\left(\frac{1-\beta}{\mathsf{P}} 
				\right) \dd y
				}\nonumber\\*\quad%
			& = & -\log \left(\frac{1-\beta}{\mathsf{P}} \right)
			\left(1-\mathcal{Q}{\left(\frac{s}{\sigma}\right)} -
			\mathcal{Q}{\left(\frac{1+\delta-s}{\sigma}\right)} \right).
			\label{eq:E2062} 
			\IEEEeqnarraynumspace
		\end{IEEEeqnarray}	
		The second term in the RHS of \eqref{eq:E2061} can further be 
		upper-bounded by
		\begin{IEEEeqnarray}{rCl}
			\IEEEeqnarraymulticol{3}{l}{%
				\int_{0}^{1+\delta}
				\frac{1}{\sqrt{2\pi}\sigma}
				e^{-\frac{(y-s)^2}{2\sigma^2}} 
				\left(\sum_{i=0}^{\nt-1} \lambda_i \left(y-\mathsf{H}_{\color{black}[i]} \right)_+ 
				\right) \dd y
			}\nonumber\\*\quad%
			& = &  \sum_{i=0}^{\nt-1}\lambda_i\sigma  \left(
			\phi\left(\frac{s-\mathsf{H}_{\color{black}[i]}}{\sigma}\right)
			- \phi\left(\frac{1+\delta-s}{\sigma}\right) \right)
			\nonumber\\
			&&+ \sum_{i=0}^{\nt-1}\lambda_i (s-\mathsf{H}_{\color{black}[i]}) \left(1 - \mathcal{Q}\left(\frac{s-\mathsf{H}_{\color{black}[i]}}{\sigma}\right)
			- \mathcal{Q}\left(\frac{1+\delta-s}{\sigma}\right) \right)
			\nonumber \\
			& \leq &  \sum_{i=1}^{\nt-1}\lambda_i\sigma  \left(
			\phi\left(0\right)
			- \phi\left(\frac{1+\delta}{\sigma}\right) \right)
			+ \sum_{i=1}^{\nt-1}\lambda_i (s-\mathsf{H}_{\color{black}[i]})_+ +\lambda_0 s \nonumber\\
			&&+\lambda_0 \sigma \left( \phi(\frac{s}{\sigma})-\phi\left(\frac{1+\delta-s}{\sigma}\right)\right)
			-\lambda_0 s \left(\mathcal{Q}\left(\frac{s}{\sigma}\right)+\mathcal{Q}\left(\frac{1+\delta-s}{\sigma}\right)\right)	\label{eq:E2063}  \\
			& \leq & \sum_{i=0}^{\nt-1}\lambda_i (s-\mathsf{H}_{\color{black}[i]})_+ + \sum_{i=0}^{\nt-1}\frac{\lambda_i\sigma}{\sqrt{2\pi}}\left(1-e^{-\frac{\left(1+\delta\right)^2}{2\sigma^2}}\right) , ~\textnormal{if}~\lambda_0\ge 0,\label{eq:85}
	  \IEEEeqnarraynumspace
		\end{IEEEeqnarray}
	while 
	\begin{IEEEeqnarray}{rCl}
		\IEEEeqnarraymulticol{3}{l}{%
			\int_{0}^{1+\delta}
			\frac{1}{\sqrt{2\pi}\sigma}
			e^{-\frac{(y-s)^2}{2\sigma^2}} 
			\left(\sum_{i=0}^{\nt-1} \lambda_i \left(y-\mathsf{H}_{\color{black}[i]} \right)_+ 
			\right) \dd y
		}\nonumber\\*%
		&\le&~\sum_{i=0}^{\nt-1}\lambda_i (s-\mathsf{H}_{\color{black}[i]})_+  + \sum_{i=1}^{\nt-1}\frac{\lambda_i\sigma}{\sqrt{2\pi}}\left(1-e^{-\frac{\left(1+\delta\right)^2}{2\sigma^2}}\right)\nonumber \\
		&&+\lambda_0 \sigma \left( \phi(\frac{1}{\sigma})-\phi\left(\frac{\delta}{\sigma}\right)\right)
		-\lambda_0  \left(\mathcal{Q}\left(\frac{1}{\sigma}\right)+\mathcal{Q}\left(\frac{\delta}{\sigma}\right)\right), ~\textnormal{if}~\lambda_0\le 0. \label{eq:86}
		  \IEEEeqnarraynumspace
	\end{IEEEeqnarray}
		Here \eqref{eq:E2063} follows from \eqref{eq:78}  and the fact that, for $s \in[0,1]$,
		\begin{IEEEeqnarray}{rCl}
				1 - \mathcal{Q}\left(\frac{s-\mathsf{H}_{\color{black}[i]}}{\sigma}\right)
				- \mathcal{Q}\left(\frac{1+\delta-s}{\sigma}\right)
			\geq 1 - \mathcal{Q}\left(\frac{s-1}{\sigma}\right)
			- \mathcal{Q}\left(\frac{1+\delta-s}{\sigma}\right)
			&>&0,
		\end{IEEEeqnarray}
		and \eqref{eq:85} and \eqref{eq:86} follow from the fact that
	$
			\sigma \left( \phi(\frac{s}{\sigma})-\phi\left(\frac{1+\delta-s}{\sigma}\right)\right)
			-s \left(\mathcal{Q}\left(\frac{s}{\sigma}\right)+\mathcal{Q}\left(\frac{1+\delta-s}{\sigma}\right)\right)$ is strictly decreasing with $s\in [0,1]$.
	
		By choosing
		\begin{IEEEeqnarray}{c}
			\beta = \frac{\sqrt{2\pi e }\sigma }{ \mathsf{P}+\sqrt{2\pi
					e}\sigma}
		\end{IEEEeqnarray}
		and combining \eqref{E101}, \eqref{eq:e207}, \eqref{eq:e208}, \eqref{eq:E2062}, \eqref{eq:85} and \eqref{eq:86}, we conclude the proof. 

\section{Proof of {\color{black}Theorem}~\ref{thm:highsnr}}
\label{app:proofs-asymp}

{\color{black}We only prove Theorem~\ref{thm:highsnr} for the {\color{black}EC-OIC}, and the involved techniques can be applied directly to the {\color{black}BC-OIC}.}

The maximum-entropy distribution $p_{\scriptscriptstyle \! S}^*(s)$ of the equivalent input for the {\color{black}EC-OIC} is defined in~\eqref{eq:maxent}. For the case of $\lambda_0^*\ge 0$, by substituting {\color{black}a suboptimal choice} $\delta=\sqrt{\sigma}$ and $\lambda_i=\lambda_i^*$ for $i\in\{0,\cdots,\nt-1\}$ into~\eqref{eq1:ubd_ecmc} in {\color{black}Theorem}~\ref{ubd:ecc}, we get {\color{black}an} upper bound as
	\begin{flalign*}
	\mathsf{C}_{\textnormal{E}}\left(\mathbf{h},\bm{\alpha},\sigma\right)
	\leq   
	\log \left(1+\frac{  \mathsf{P}(\sigma)}{\sqrt{2\pi e} \sigma}\right) +  \sum_{i=0}^{\nt-1}\lambda_i^* \left(1-\mathsf{H}_{\color{black}[i]} \right)\bar{\alpha}_i 
	+ \sum_{i=0}^{\nt-1}\frac{\lambda_i\sigma}{\sqrt{2\pi}}\left(1-e^{-\frac{\left(1+\delta\right)^2}{2\sigma^2}}\right) .
	\end{flalign*}
Note that
	\begin{flalign}
	\lim_{\sigma\downarrow 0^+} \mathsf{P}(\sigma)&=\lim_{\sigma\downarrow 0^+}	\int_0^{1+{\color{black}\sqrt{\sigma}}}\exp \left( -\lambda_0^* y-\sum_{i=1}^{\nt-1} \lambda_i^* \left(y-\mathsf{H}_{\color{black}[i]} \right)_+ \right) \dd y \nonumber\\
	&=	\int_0^{1}\exp \left( -\lambda_0^* y-\sum_{i=1}^{\nt-1} \lambda_i^* \left(y-\mathsf{H}_{\color{black}[i]} \right)_+ \right)\dd y \nonumber \\
	&=e^{-{\color{black}\nu}_0^*}	\int_0^{1}p_{\scriptscriptstyle \! S}^*(y) \dd y \nonumber \\
	&=e^{-{\color{black}\nu}_0^*}
\end{flalign}
and
\begin{flalign}
	0 \le  \left(1-e^{-\frac{\left(1+\delta\right)^2}{2\sigma^2}}\right)  \le 1.
\end{flalign}
Then we have
	\begin{flalign}
	& {\color{black}\limsup \limits_{\sigma\downarrow 0^+} } \,\, \mathsf{C}_{\textnormal{E}}\left(\mathbf{h},\bm{\alpha},\sigma\right)-\log\frac{1}{\sigma} \nonumber\\
	\le& \lim_{\sigma\downarrow 0^+}   \log \left(1+\frac{ e^{-{\color{black}\nu}_0^*}}{\sqrt{2\pi e} \sigma}\right) -\log\frac{1}{\sigma} +  \sum_{i=0}^{\nt-1}\lambda_i^* \left(1-\mathsf{H}_{\color{black}[i]}\right)\bar{\alpha}_i \nonumber \\
	=& {\color{black}\lim_{\sigma\downarrow 0^+}   \log \left(\sigma+\frac{ e^{-{\color{black}\nu}_0^*}}{\sqrt{2\pi e} }\right)+  \sum_{i=0}^{\nt-1}\lambda_i^* \left(1-\mathsf{H}_{\color{black}[i]}\right)\bar{\alpha}_i } \nonumber \\
	=& -\frac{1}{2}\log 2\pi e-{\color{black}\nu}_0^* +  \sum_{i=0}^{\nt-1}\lambda_i^* \left(1-\mathsf{H}_{\color{black}[i]}\right)\bar{\alpha}_i \label{eq:136} \\
	=& -\frac{1}{2}\log 2\pi e+\gamma_{\textnormal{E}}, \label{eq:137}
\end{flalign}
{\color{black}where~\eqref{eq:136} follows from the continuity of a logarithmic function and~\eqref{eq:137} follows from the fact that 
\begin{flalign}
	\gamma_{ \textnormal{E}} &=  \sum_{i=0}^{\nt-1}\lambda_i^{*}\left(1-\mathsf{H}_{\color{black}[i]}\right)\bar{\alpha}_i
	-1-\nu_0^*+\int_0^1 p^{*}_{\scriptscriptstyle \! S} (s) \dd s \nonumber \\
	&= \sum_{i=0}^{\nt-1}\lambda_i^{*}\left(1-\mathsf{H}_{\color{black}[i]}\right)\bar{\alpha}_i
	-\nu_0^* \nonumber.
\end{flalign}

}

By using the lower bound in {\color{black}Theorem}~\ref{ecmc:lbd}, the reverse part of the proposition can be obtained {\color{black}as follows}
	\begin{flalign}
		&{\color{black}\liminf \limits_{\sigma\downarrow 0^+} } \,\, \mathsf{C}_{\textnormal{E}}\left(\mathbf{h},\bm{\alpha},\sigma\right)-\log\frac{1}{\sigma} \\
	\ge& \lim_{\sigma\downarrow 0^+}\frac{1}{2}\log \left(1+\frac{\exp(2 	\gamma_{ \textnormal{E}})}{2\pi e \sigma^2}\right)-\log\frac{1}{\sigma} \nonumber \\
	\ge& -\frac{1}{2}\log 2\pi e+\gamma_{\textnormal{E}}.
\end{flalign}

{\color{black}Due to} the boundedness of $ \phi(\frac{1}{\sigma})-\phi\left(\frac{\delta}{\sigma}\right)$ and 
$
\lim_{x\to +\infty}\mathcal{Q}(x)=0
$,
 the proof for the case of  $\lambda_0^*\le 0$ can be {\color{black}accomplished} similarly.
 
 \section{Proof of Proposition~\ref{Prop:Partition}}\label{app:Partition}
 Before proving Proposition~\ref{Prop:Partition}, we first need the following lemmas:
 \begin{lemma}
 	Suppose both $v_i^*$ and $\tilde{v}_i^*\in \left[0,\mathsf{H}_{\color{black}[i-1]}\right]$ are solutions to the equation \eqref{eq:145} and let $R_{i-1}^*=\varphi\left(R_{i}^*;v_i^*,h_{i}\right)$ and $\tilde{R}_{i-1}^*=\varphi\left(R_{i}^*;\tilde{v}_i^*,h_{i}\right)$. Then $\pi_{\scriptscriptstyle\! R_{i-1\!}^*}(t)=\pi_{\scriptscriptstyle\! \tilde{R}_{ i-1\! }^*}(t)$.	
 \end{lemma}

\begin{IEEEproof}
	Without loss of generality, we assume that	$v_i^* \le \tilde{v}_i^*$. Applying \eqref{eqn:94}, it is clear that $\pi_{\scriptscriptstyle\! R_{i-1\!}^*}(t)=\pi_{\scriptscriptstyle\! \tilde{R}_{ i-1\! }^*}(t)$ for $t\in \left[ 0,v_i^*\right]$ and $t\in \left[ \tilde{v}_i^*,1\right]$, respectively. 
	
	For $t\in \left[v_i^*,\tilde{v}_i^*\right]$, we notice that
	\begin{flalign}
		\pi_{\scriptscriptstyle\! R_{i-1\!}^*}(t)-\pi_{\scriptscriptstyle\! \tilde{R}_{ i-1\! }^*}(t) =\pi_{\scriptscriptstyle\! R_{i\!}^*}(t+h_{i})-\pi_{\scriptscriptstyle\! R_{i\!}^*}(t)+h_{i}\alpha_{i},
	\end{flalign}
	which is continuous and nonincreasing with $t$. Note that 
	\begin{flalign}
		\pi_{\scriptscriptstyle\! R_{i-1\!}^*}(v_i^*)-\pi_{\scriptscriptstyle\! \tilde{R}_{ i-1\! }^*}(\tilde{v}_i^*)
		&=\pi_{\scriptscriptstyle\! R_{i\!}^*}(v_i^*+h_{i})-\pi_{\scriptscriptstyle\! R_{i\!}^*}(v_i^*)+h_{i}\alpha_{i} \nonumber \\
		&=\pi_{\scriptscriptstyle\! R_{i\!}^*}(\tilde{v}_i^*+h_{i})-\pi_{\scriptscriptstyle\! R_{i\!}^*}(\tilde{v}_i^*)+h_{i}\alpha_{i} \nonumber \\
		&=0,~\label{eq:181}
	\end{flalign}
	where~\eqref{eq:181} follows from the assumption that $v_i^*$ and $\tilde{v}_i^*$ are {\color{black}solutions} to \eqref{eq:145}. Thus, we conclude that $\pi_{\scriptscriptstyle\! R_{i-1\!}^*}(t)-\pi_{\scriptscriptstyle\! \tilde{R}_{ i-1\! }^*}(t) = 0$ for all $t\in[0,1]$.
\end{IEEEproof}
 
\begin{lemma}
$v_{i}^*\ge v_{i-1}^*$  for $i\in\{n,\cdots,2\}$.
\end{lemma}
 \begin{IEEEproof}
 	Notice that $v_2^*\ge 0$ and $v_1^*=0$. We will prove $v_{i}^*\ge v_{i-1}^*$ for any $i \in \{\nt,\cdots,3\}$ (when $\nt \ge 3$) by contradiction. 
 	
 	Assuming $v_{i-1}^*>v_i^*$, we get $v_i^* \le \mathsf{H}_{\color{black}[i-2]}$. Substituting $v=v_{i}^*$ into $\varphi\left(R_{i-1}^*,v,h_{i-1}\right)$, we have
 	\begin{flalign}
 		\varphi\left(R_{i-1}^*,v_{i}^*,h_{i-1}\right) 
 		&=\varphi\left(\varphi\left(R_{i}^*,v_i^*,h_{i}\right),v_{i}^*,h_{{\color{black}i-1}}\right) \nonumber \\
 		&=\varphi\left(R_{i}^*,v_{i}^*,h_{i}+h_{i-1}\right),
 	\end{flalign}
 	which follows from the definition of $\varphi \left( x;v,z\right)$.
 	Then we consider
 	\begin{flalign}
 		&~~~~\mathbb{E}\left[\varphi\left(R_{i-1}^*,v_{i}^*,h_{i-1}\right)\right]   \nonumber \\ 
 		&=\mathbb{E}\left[\varphi\left(R_{i}^*,v_{i}^*,h_{i}+h_{i-1}\right)\right]  \nonumber \\
 		&=\pi_{\scriptscriptstyle\! R_{i}^*}(0)-\pi_{\scriptscriptstyle\! R_{i}^*}(v_{i}^*)+\pi_{\scriptscriptstyle\! R_{i}^*}(v_{i}^*+h_{i}+h_{i-1}) \nonumber \\
 		&=\pi_{\scriptscriptstyle\! R_{i-1}^*}(0)-\pi_{\scriptscriptstyle\! R_{i}^*}(v_{i}^*+h_{i})+\pi_{\scriptscriptstyle\! R_{i}^*}(v_{i}^*+h_{i}+h_{i-1}), \label{eqn:113}
 	\end{flalign}
 	where (\ref{eqn:113}) follows from $\pi_{\scriptscriptstyle\! R_{i-1}^*}(0)=\pi_{\scriptscriptstyle\! R_{i}^*}(0)-\pi_{\scriptscriptstyle\! R_{i}^*}(v_{i}^*)+\pi_{\scriptscriptstyle\! R_{i}^*}(v_{i}^*+h_{i})$.
 	Thus, we have
 	\begin{flalign}
 		&~~~~\frac{\mathbb{E}\left[R_{i-1}^*\right]-\mathbb{E}\left[\varphi\left(R_{i-1}^*,v_{i}^*,h_{i-1}\right)\right]}{h_{i-1}} \nonumber \\
 		&=\frac{\pi_{\scriptscriptstyle\! R_{i}^*}(v_{i}^*+h_{i})-\pi_{\scriptscriptstyle\! R_{i}^*}(v_{i}^*+h_{i}+h_{i-1})}{h_{i-1}} \nonumber \\
 		&=\frac{1}{h_{i-1}}\int_{v_{i}^*+h_{i}}^{v_{i}^*+h_{i}+h_{i-1}} \left( 1-F_{\scriptscriptstyle\! R_{i}^*}(x) \right) \dd x \nonumber \\
 		& \le 1-F_{\scriptscriptstyle\! R_{i}^*}(v_{i}^*+h_{i}) \nonumber  \\
 		&\le \frac{1}{h_{i}}\int_{v_{i}^*}^{v_{i}^*+h_{i}} \left(1-F_{\scriptscriptstyle\! R_{i}^*}(x)\right) \dd x \nonumber \\
 		&=\frac{\pi_{\scriptscriptstyle\! R_{i}^*}(v_{i}^*)-\pi_{\scriptscriptstyle\! R_{i}^*}(v_{i}^*+h_{i})}{h_{i}}\nonumber  \\
 		&= \alpha_{i} \nonumber \\
 		&< \alpha_{i-1},
 		\IEEEeqnarraynumspace
 	\end{flalign}
 	where $F_{\scriptscriptstyle\! R_{i}^*}(x)$ denotes the cumulative distribution function of $R_{i}^*$.
 	Hence, we know
 	\begin{flalign}
 		\mathbb{E}\left[\varphi\left(R_{i-1}^*,v_{i}^*,h_{i-1}\right)\right]>\mathbb{E}\left[\varphi\left(R_{i-1}^*,v_{i-1}^*,h_{i-1}\right)\right],
 	\end{flalign}
 	which implies $v_{i}^*\ge v_{i-1}^*$ and contradicts the assumption  due to the monotonicity of  $\mathbb{E}\left[\varphi\left(R_{i}^*,v,h_{{\color{black}i}}\right)\right]$. 	
 \end{IEEEproof}
 \vspace{0.2cm}
 Now we prove Proposition~\ref{Prop:Partition} via induction on $k$. The existence of the solution to~\eqref{eq:65} can be verified by letting $\kappa=v_{k-1}^*$, which is omitted here. 
 \begin{itemize}
 	\item The proposition in the base case $k=\nt$ is clearly true.
 	\item Suppose $\kappa_k^*$ is the solution to 
 	\begin{flalign}\label{eq:suppose}
 		\pi_{\scriptscriptstyle\! R_{k}^*}(\kappa_k^*)-\pi_{\scriptscriptstyle\! R_{k}^*}(\kappa_k^*+h_{k})=h_{k}\alpha_{k}
 	\end{flalign}
 	for each $k\ge i$ ($2\le i\le \nt$) and consider the case of $k=i-1$. 
 	
 	Due to the monotonicity of the solutions $v_k^*$, we have $\kappa_i^*\le \cdots \le \kappa_{\nt}^*$. Then using~\eqref{eqn:94}, for $\kappa \le \kappa_{i}$ we have
 	\begin{flalign}
 		\pi_{\scriptscriptstyle \! S}(\kappa)&=\pi_{\scriptscriptstyle\! R_{\nt-1}^*}(\kappa)+h_{\nt}\alpha_{\nt} \nonumber \\
 		&=\pi_{\scriptscriptstyle\! R_{i-1}^*}(\kappa)+\sum_{m=i}^{\nt}h_m\alpha_m,\label{eq:180}
 	\end{flalign}
 	where~\eqref{eq:180} follows from~\eqref{eq:suppose}. 
 	
 	It is clear that the Borel sets $\mathcal{P}_i,\cdots, \mathcal{P}_{\nt}$ are disjoint and satisfy $\mu\left( \mathcal{P}_j\right)=h_j$ for $i \le j\le \nt$. In addition, we have
 	\begin{flalign}\label{eq:189}
 		\eta_{j}(\kappa_{j}^*) \le \eta_{i-1}(\kappa_{i-1}^*), ~\textnormal{if}~\kappa_{j}^* < \eta_{i-1}(\kappa_{i-1}^*),
 	\end{flalign}
 	which follows from the following equality
 	\begin{flalign}\label{eq:185}
 		\mu\left(\left[\kappa_{i-1}^*,\eta_{i-1}(\kappa_{i-1}^*)\right)\setminus \bigcup_{m=i}^{\nt}\mathcal{P}_i  \right)=h_{i-1}.
 	\end{flalign}
 	We let $\mathcal{J}= \left\{j:\kappa_j^* < \eta_{i-1}(\kappa_{i-1}^*) , i-1< j\le \nt \right\}$.
 	
 	If $\mathcal{J}=\emptyset$, then we have $\eta_{i-1}(\kappa_{i-1}^*)\le \kappa_i^*$ and similarly obtain
 	\begin{flalign}
 		\pi_{\scriptscriptstyle \! S}\left(\eta_{i-1}(\kappa_{i-1}^*)\right)&=\pi_{\scriptscriptstyle\! R_{\nt-1}^*}\left(\eta_{i-1}(\kappa_{i-1}^*)\right)+h_{\nt}\alpha_{\nt}  \nonumber \\
 		&=\pi_{\scriptscriptstyle\! R_{i-1}^*}\left(\eta_{i-1}(\kappa_{i-1}^*)\right)+\sum_{m=i}^{\nt}h_m\alpha_m.
 	\end{flalign}
 	If $\mathcal{J} \neq \emptyset$, we denote $M=\max \mathcal{J}$. Due to $\kappa_i^*\le \cdots \le \kappa_{\nt}^*$, we have $\kappa_j^* \ge \eta_{i-1}(\kappa_{i-1}^*)$ if and only if $j>M$. Using~\eqref{eq:189}, we conclude that the disjoint sets $\mathcal{P}_j$ satisfy
 	\begin{align}
 		&\mathcal{P}_j \subseteq	\left[\kappa_{i-1}^*,\eta_{i-1}(\kappa_{i-1}^*) \right),~i\le j \le M,\\
 		&\mathcal{P}_j   \cap 	\left[\kappa_{i-1}^*,\eta_{i-1}(\kappa_{i-1}^*)\right)=\emptyset,~j>M.
 	\end{align}
 	Applying the equality~\eqref{eq:185}, we immediately have
 	\begin{flalign}
 		\eta_{i-1}(\kappa_{i-1}^*)=\kappa_{i-1}^*+\sum_{ j=i}^{M} h_j\alpha_j
 	\end{flalign} 
 	Then we rewrite $\pi_{\scriptscriptstyle \! S}\left(\eta_{i-1}(\kappa_{i-1}^*)\right)$ as
 	\begin{flalign}
 		\pi_{\scriptscriptstyle \! S}\left(\eta_{i-1}(\kappa_{i-1}^*)\right)&=\pi_{\scriptscriptstyle\! R_{\! \scriptscriptstyle M}^*}(\kappa)+\sum_{M<m\le \nt}h_m\alpha_m
 		\nonumber \\
 		&=\pi_{\scriptscriptstyle\! R_{\! \scriptscriptstyle M-1\!}^*}\left(\eta_{i-1}(\kappa_{i-1}^*)-h_M\right)+\sum_{M<m\le \nt}h_m\alpha_m
 		\nonumber \\
 		&=\pi_{\scriptscriptstyle\! R_{\! \scriptscriptstyle i-1\!}^*}\left(\eta_{i-1}(\kappa_{i-1}^*)-\sum_{j=i}^Mh_j\right)+\sum_{M<m\le \nt}h_m\alpha_m
 		\nonumber \\
 		&=\pi_{\scriptscriptstyle\! R_{\! \scriptscriptstyle i-1\!}^*}(\kappa_{i-1}^*+h_{i-1})+\sum_{M<m\le \nt}h_m\alpha_m.
 	\end{flalign}
 	Hence, we have
 	\begin{flalign}
 		\pi_{\scriptscriptstyle \! S}(\kappa_{i-1}^*)-\pi_{\scriptscriptstyle \! S}\left(\eta_{i-1}(\kappa_{i-1}^*)\right)=\sum_{m=i-1}^{i-1+N_{i-1}(\kappa_{i-1}^*)}h_{m}\alpha_{m},
 	\end{flalign}
 	which implies
 	\begin{flalign}
 		\pi_{\scriptscriptstyle\! R_{i-1}^*}(\kappa_{i-1}^*)-\pi_{\scriptscriptstyle\! R_{i-1}^*}(\kappa_{i-1}^*+h_{i-1})=h_{i-1}\alpha_{i-1}.
 	\end{flalign}
 	The proof is completed.
 \end{itemize}
 
\bibliographystyle{./myIEEEtran}
\normalem
\bibliography{./defshort1,./biblio1}
\end{document}